\documentclass[pre,aps,10pt,floatfix,showpacs,twocolumn,superscriptaddress]{revtex4}
\usepackage{enumerate,bbm,stmaryrd}
\usepackage{amssymb}
\usepackage[tbtags]{amsmath}
\usepackage{epsfig}
\usepackage{graphicx}
\usepackage[dvips]{hyperref}
\usepackage{textcomp}
\usepackage[OT1,T1]{fontenc}
\DeclareMathOperator{\sgn}{sgn}
\DeclareMathOperator{\const}{const}
\newcommand{\dd}{\mathrm{d}}
\newcommand{\ee}{\mathrm{e}}
\newcommand{\ii}{\mathrm{i}}
\newcommand{\pp}{\pi}
\newcommand{\llb}{\llbracket}
\newcommand{\rrb}{\rrbracket}
\newcommand{\blparen}{{\boldsymbol{(}}}
\newcommand{\brparen}{{\boldsymbol{)}}}
\newcommand{\blbracket}{{\boldsymbol{[}}}
\newcommand{\brbracket}{{\boldsymbol{]}}}

\newcommand{\intx}[1]{\!\int#1\,}
\newcommand{\intxlim}[3]{\!\int_{#1}^{#2}#3\,}
\newcommand{\subs}[1]{_{\rm #1}}
\newcommand{\sups}[1]{^{(#1)}}
\newcommand{\Epsilon}{\mathrm{E}}
\newcommand{\Jvec}{{\boldsymbol{J}}}
\newcommand{\evec}{{\boldsymbol{e}}}
\newcommand{\Hcal}{{\mathcal{H}}}
\newcommand{\Kcal}{\mathcal{K}}
\newcommand{\Fcal}{\mathcal{F}}
\newcommand{\Vcal}{\mathcal{V}}
\newcommand{\Scal}{\mathcal{S}}
\newcommand{\Ucal}{\mathcal{U}}
\newcommand{\Lcal}{\mathcal{L}}
\newcommand{\Ecal}{\mathcal{E}}
\newcommand{\Acal}{\mathcal{A}}
\newcommand{\Gcal}{\mathcal{G}}
\newcommand{\Ncal}{\mathcal{N}}
\newcommand{\Zcal}{\mathcal{Z}}
\newcommand{\Ehat}{\hat{E}}
\newcommand{\Ebar}{\tilde{E}}
\newcommand{\SigmaBar}{\tilde{\varSigma}}
\newcommand{\hBar}{\bar{h}}
\newcommand{\uBar}{\bar{u}}
\newcommand{\omegaBar}{\bar{\omega}}
\newcommand{\Pbar}{\tilde{P}}
\newcommand{\Qbar}{\tilde{Q}}
\newcommand{\Rbar}{\tilde{R}}
\newcommand{\Phat}{\hat{P}}
\newcommand{\Qhat}{\hat{Q}}
\newcommand{\p}{$p$\nobreakdash}
\newcommand{\K}{$K$\nobreakdash}
\newcommand{\N}{$N$\nobreakdash}

\begin{document}

\preprint{paper submitted to Physical Review E}

\title{Statistical Mechanics of the Quantum $K$-Satisfiability problem.}

\author{Sergey Knysh}
\affiliation{ELORET Corporation, NASA Ames Research Center, MS 229-1, Moffett Field, CA 94035-1000.}
\email{Sergey.I.Knysh@nasa.gov}
\author{Vadim N. Smelyanskiy}
\affiliation{NASA Ames Research Center, MS 269-1, Moffett Field, CA 94035-1000.}
\email{Vadim.N.Smelyanskiy@nasa.gov}

\begin{abstract}
We study the quantum version of the random $K$-Satisfiability problem in the presence of the external
magnetic field $\Gamma$ applied in the transverse direction. We derive the replica-symmetric free energy
functional within static approximation and the saddle-point equation for the order parameter: the
distribution $P[h(m)]$ of functions of magnetizations. The order parameter is interpreted as the histogram
of probability distributions of individual magnetizations. In the limit of zero temperature and small
transverse fields, to leading order in $\Gamma$ magnetizations $m\approx 0$ become relevant in addition to
purely classical values of $m \approx \pm 1$. Self-consistency equations for the order parameter are
solved numerically using Quasi Monte Carlo method for $K=3$. It is shown that for an arbitrarily small
$\Gamma$ quantum fluctuations destroy the phase transition present in the classical limit $\Gamma=0$,
replacing it with a smooth crossover transition. The implications of this result with respect to the
expected performance of quantum  optimization algorithms via adiabatic evolution are discussed. The
replica-symmetric solution of the classical random $K$-Satisfiability problem is briefly revisited. It is
shown that the phase transition at $T=0$ predicted by the replica-symmetric theory is of continuous type
with atypical critical exponents.
\end{abstract}


\maketitle

\section{Introduction\label{sec:Intro}}

Quantum phase transition (QPT) is a transition between different ground states driven by quantum
fluctuations and controlled by certain parameters, for example, an external magnetic field. Study of QPTs
in systems with strongly interacting spins attracted attention in the field of quantum computing due to
the possibility of creating massively entangled states at the quantum critical point \cite{Osterloh:2002}
and the relevance of QPTs to the analysis of the performance of quantum algorithms for solving classical
combinatorial optimization problems (COPs)
\cite{Brooke:2001,Schutzhold:2006,Santoro:2002,Banuls:2006,Caneva:2007}. Quantum mechanics offers an
alternative to the mechanism of thermal fluctuations for the transitions between the states, which can be
exploited in the optimization procedures \cite{Kadowaki:1998,Brooke:2001}. QPT in this paper will be
studied  in the context of a general-purpose quantum adiabatic algorithm (QAA)  proposed by Farhi and
coworkers \cite{Farhi:2001}. In its simplest form the algorithm is defined via a quantum \N-spin
Hamiltonian that is a sum of two terms
\begin{equation}
  \hat{H} = \Hcal\subs{cl}(\hat\sigma_{i}^{z},\dots,\hat\sigma_{N}^{z})- \Gamma \sum_{i=1}^N
  \hat{\sigma}_i^x.\label{H}
\end{equation}
The first operator term is derived from a cost (energy) function of classical spins
$\Hcal\subs{cl}(s_1,\dots,s_N)$ by replacing each classical spin $s_i=\pm $1 with a Pauli matrix,
$\hat\sigma_i^z$. The ground state of this operator encodes the solution of a classical COP described by
$\Hcal\subs{cl}$. The second term describes spin coupling  to the external magnetic field $\propto \Gamma$
applied in the transverse direction (e.g. along the positive $x$ axis). At the start of the algorithm,
$\Gamma$ is made very large and the ground state of $\hat H(0)$ is prepared with all the spins pointing in
$\hat x$ direction. Then $\Gamma$=$\Gamma(t)$ is slowly reduced to zero while the state of the quantum
system remains close to the instantaneous adiabatic ground state of $H(t)$ --- provided that the condition
$\langle \Psi_0|\frac{\partial}{\partial t} \hat H|\Psi_0\rangle\ll (E_1-E_0)^2$ is satisfied. Here $\hat
H(t)|\Psi_n(t)\rangle=E_n(t)|\Psi_n(t)\rangle$. At the end of the algorithm at $\Gamma=0$, the system is
found in a state which is a superposition of spin configurations corresponding to all degenerate global
minima of $\Hcal\subs{cl}$. The runtime of the algorithm is proportional to $1/g\subs{min}^{2}$, where
$g\subs{min}=\min_{\Gamma}(E_1-E_0)$ is a minimum  of the energy gap \cite{Messiah} taken over the range
of $\Gamma$.

It has been noticed several decades ago that properties of the solution space of complex COPs are closely
related to those of spin glass systems \cite{Mezard:1985,Fu:1986}. It has been also recognized
\cite{Mezard} that many of the spin glass models are in almost one-to-one correspondence with
computationally hard COPs encountered in practice and forming a class of NP-hard \cite{Garey} problems.

Whereas theoretical computer science is mostly concerned with the worst-case complexity,
from the statistical physics perspective the main interest lies in the typical running time of algorithms
over the random ensemble of problem instances
 (or samples of spin glass system) \cite{Fu:1986,Stein:89}.  When this expected runtime scales
exponentially with the number of spins, the COP is considered intractable. This intractability was linked
to so-called threshold phenomena \cite{Kirkpatrick:1994,Hogg,Dubois} in NP-complete problems.
In physics community, these threshold phenomena were recognized as phase transitions in models of
classical spin glasses \cite{Monasson:1999}.
Many NP-complete problems, including the most basic of them
--- random \K-Satisfiability (or \K-SAT) --- correspond to infinite-range \emph{dilute}
spin glass models with \K-local interactions, i.e. $\Hcal\subs{cl}(s_1,\dots,s_N)$ is given by a sum of
interaction terms; each involving a set of $K$ spins, chosen at random from a set of size $N$. In contrast
to finite-dimensional models, the topology of links corresponding to spin couplings is completely random,
with no non-trivial correlations. The random ensembles of instances are described by a single parameter
the connectivity $\gamma$ which is the number of interaction terms per spin, $\gamma=M/N$. The probability
for a given spin to be involved in $d$ interactions is Poisson with the finite mean value of $d$ equal to
$K\gamma$. This is different from infinite-range fully-connected spin models such as the
Sherrigton-Kirkpatrick model \cite{Sherrington:1975}, where the value of $d=N-1$ scales with the number of
spins.
\begin{figure}[!h]
  \includegraphics[width=0.99\linewidth,clip=true]{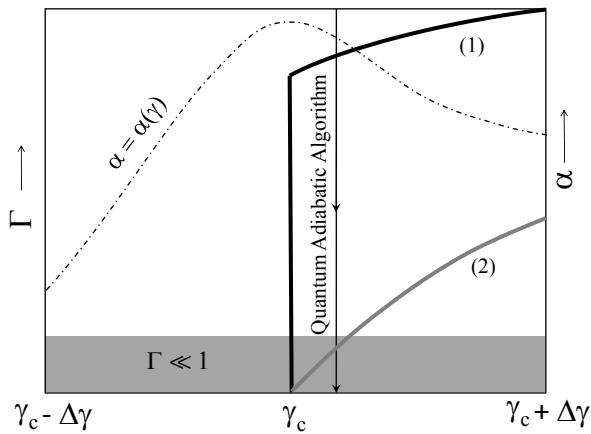}
  \caption{  Thick black (1) and gray (2) lines show two possible forms of quantum phase diagrams on  transverse field $\Gamma$ \emph{vs}
  connectivity $\gamma$ plane for  random $K$-SAT problem. Black line (1) corresponds to the quantum
  dilute ferromagnet. Gray filled rectangle shows
   the  region of interest in this paper with small transverse fields,  $\Gamma\ll$ 1. Dot-dashed
   line depicts the scaled exponent $\alpha=\alpha(\gamma)$ of the median runtime ${\cal T}$ of a classical
   algorithm, ${\cal T}\sim \exp(-\alpha N)$, over an ensemble of problem instances with the same $\gamma$.
\label{fig:quantum}}
\end{figure}

Classical infinite-range spin glass models in the dilute limit have been studied, though recent results
concentrate on the zero-temperature limit \cite{Monasson:1996,Monasson:1997,Mezard:2002Sci,Mezard:2002PRE,Mezard:2003Parisi,Krzakala:2007,Montanari:2008}. However, very little
is known about their respective quantum versions [described by the Hamiltonian (\ref{H}) with $H_{\rm cl}$
corresponding to an infinite-range dilute spin glass] despite a lot of interest in these models from the
perspective of quantum computing. Recently, quantum versions of random Exact Cover and other related
optimization problems have been studied \cite{Smelyanskiy:2004} using a generalized annealing
approximation \cite{Knysh:2004}. At the same time, fully connected infinite-range quantum spin models have
been analyzed in the  literature using various approximations. This includes quantum versions of the
Sherrington-Kirkpatrick \cite{Thirumalai:1989,Yamamoto:1987,Goldschmidt:1990PRL,Miller:1993,Ye:1993},
random Heisenberg \cite{Bray:1980}, \p-spin and random energy models \cite{Goldschmidt:1990PRB}. Exact
solutions in quantum spin glasses are mainly limited to one-dimensional models
\cite{McCoy:1968,Fisher:1994}.

Numerical studies  \cite{Hogg} have demonstrated that the typical runtime of known classical algorithms
applied to ensembles of randomly generated instances of \K-SAT and similar models, as a function of
$\gamma$, peaks at the point of static transition, which is a major bottleneck of classical optimization
algorithms (see Fig.~\ref{fig:quantum}). This can be understood by analogy with the critical slowing down
of the dynamics in the vicinity of phase transitions in problems without disorder.
Similarly, we expect that the dynamics of QAA for random \K-SAT could be governed by the corresponing
quantum phase tranisiton (QPT). If the system underwent a QPT as the value of $\Gamma$ is lowered from a large
value to $0$, the gap would attain its minimum value at the point of the transition. The critical exponent
associated with the singularity of the free energy would determine the scaling of the minimum gap
(which would have the form of an exponential or stretched exponential \cite{Fisher:1994}).

In this paper we concentrate on static transition that corresponds to satisfiability transition at zero
temperature. We concentrate on $K=3$ as the most interesting case. It is the smallest value of $K$ for
which $K$-SAT is NP-complete. Moreover, random $K$-SAT undergoes random first-order phase transition
for $K \geqslant 3$. As it is the case with all random first-order transitions, the static transition
is preceded by the dynamic transition. Results for similar model --- \K-XOR-SAT, or dilute \p-spin glass
--- at finite temperature indicate \cite{Montanari:2006} that the free energy remains analytic across
dynamic transition, which would imply that the static transition is the real bottleneck of simulated annelaing
algorithm. While giving credence to the idea of the analysis of static transition, this picture may not necessarily
apply to \K-SAT for $K=3$, where the dynamic transition is accompanied by another,
condensation, transition \cite{Krzakala:2007,Montanari:2008}.
Due to difficulties of replica-symmetry-breaking analysis in quantum case, we have only performed
the replica-symmetric analysis. Although replica-symmetric approximation is capable of correctly capturing
the existence and qualitative properties of static transition, it fails to describe the dynamic transition
and overestimates the critical threshold $\gamma_c$.

In Fig.~\ref{fig:quantum} we sketch two conjectured forms of QPT line $\Gamma=\Gamma_c(\gamma)$.
One possiblity $\Gamma_c(\gamma)$ changes continuously from the value of $0$ at $\gamma=\gamma_c$.
Alternatively, it may exhibit a finite jump (i.e. $\Gamma_c(\gamma_c)=\Gamma_{c0}>0$) as in dilute transverse Ising
models without frustration \cite{Harris:1974,Sachdev}. Another (third) possibility is that the phase transition
at $\Gamma=0$ disappears for any finite $\Gamma>0$.
One may distinguish between these
cases by setting $\Gamma\ll$1 and studying the free energy  for a range
of values of $\gamma$ containing
$\gamma_c$, as shown in Fig.~\ref{fig:quantum}. In QAA the parameter
$\Gamma(t)$ decreases with time,
corresponding to a vertical line in the $(\gamma,\Gamma)$ plane as shown
in Fig.~\ref{fig:quantum}.
The central result of this paper is that it is the third possibility that takes place:
quantum effects (the transverse field $\Gamma$) in the QAA Hamiltonians (\ref{H}) make the static
phase transition disappear; the free energy becomes analytical in the vicinity of $\gamma_c$ for
small but finite $\Gamma$.

It should be  mentioned in passing that certain highly symmetric examples of COPs have been constructed
\cite{VanDam:2001,Farhi:2002}, where a total spin is an exact quantum number of the  Hamiltonian $H$
of Eq.~(\ref{H}) and QAA  fails due to the onset of a large spin tunneling  through a broad, order $N$,
semiclassical barrier with amplitude that scales down exponentially with $N$
\cite{Farhi:2002,Boulatov:2003}. However, in spin glasses, quantum evolution does not correspond to large
spin dynamics. Instead, an exponentially large (in $N$) number of deep local minima of the classical
energy are connected by an extremely large number of tunneling paths with amplitudes proportional to high
powers of $\Gamma$. This picture as well as the analysis of QPTs is more relevant for understanding  the
typical complexity of QAA for NP-hard problems such as \K-SAT.

This paper is organized as follows. Section~\ref{sec:MZ} presents a brief overview of important results
for the classical version of random \K-SAT and discusses the relationship between the present work's
replica-symmetric analysis of quantum \K-SAT and that of the classical \K-SAT corresponding to the limit
of $\Gamma=0$. We formulate the quantum version of \K-SAT and analyze it using replica-symmetric theory in
Sec.~\ref{sec:Replica}. This is followed by the analysis of small magnetic fields $\Gamma$ in
Sec.~\ref{sec:smallG}. In Sec.~\ref{sec:CT0} we revisit the classical $T=0$ random \K-SAT to demonstrate
that the replica-symmetric analysis predicts a continuous phase transition; it was previously thought to
be of random first-order type. In Sec.~\ref{sec:Numerical} we present the numerical results for both
finite-temperature classical \K-SAT and zero-temperature quantum \K-SAT. We concentrate on $K=3$,
which is the most intersting case. Since we utilize the replica-symmetric approximation in the analysis
of quantum \K-SAT, we compare these results with those predicted by the replica-symmetric theory for
finite temperature classical \K-SAT (despite the fact tools to study replica symmetry breaking in classical
\K-SAT have appeared recently).
In the Conclusion we discuss our results, especially in relation to quantum adiabatic algorithm and
describe possible extensions of the present work. The mathematical details of the calculation of the
replica free energy functional are relegated to Appendix~\ref{app:Replica}. Appendix~\ref{app:BP}
discusses correspondence between the replica-symmetric ansatz and the Bethe-Peierls approximation.
A novel Quasi Monte Carlo algorithm used in numerical calculations is described in Appendix~\ref{app:QMC}.

\section{Classical statistical mechanics of random $K$-SAT:  Monasson-Zecchina replica symmetric
solution and its connection to the present work.\label{sec:MZ}}

An instance of random \K-SAT is a system of $N$ classical spins with the
energy function that is written as a sum of $M$ terms:

\begin{equation}
\Hcal\subs{cl}(s_1,\dots,s_N)=\sum_{\evec=(i_1\dotsc i_K) < \Ecal}
  E(s_{i_1},\dots,s_{i_K};\Jvec_\evec).
\label{Hcl0}
\end{equation}
Each term is associated with a \K-tuple $\evec=(i_1,\dots,i_K)$. If spins labeled by $i=1,\dots,N$ are
viewed as vertices of some graph, \K-tuples $\evec$ correspond to its hyperedges.
The set of all hyperedges for a given instance is labeled $\Ecal$. Hyperedges corresponding to each term
are chosen independently and uniformly at random; hence with each instance of random \K-SAT we may
associate a realization of a random hypergraph. This represent the geometric part of disorder.

Each term defines a constraint involving spin variables $s_{i_1},\dots,s_{i_K}$. The cost function
$\Hcal\subs{cl}(s_1,\dots,s_K)$ can be either zero or some positive value representing the energy penalty
for those combinations $(s_1,\dots,s_K)$ that violate the constraint.

For \K-SAT the constraints penalize exactly one out $2^K$ assignments. The cost function
is chosen in the following form
\begin{equation}
  E_\Jvec(s_1,\dots,s_K) = 2 \prod_{\ell=1}^K \frac{1+J_\ell s_\ell}{2}.
\label{EJ}
\end{equation}
Here $\Jvec=(J_1,\dots,J_K)$, where $J_\ell=\pm 1$, denotes the combination of $K$ spin values that is
assigned an energy penalty of $2$ \footnote{This value of the energy of a  violated constraint is an
arbitrary quantity, it is often chosen to be equal to 2 only to simplify calculations
\cite{Mezard:2002Sci,Mezard:2002PRE}.}. The argument $\Jvec$ of the cost function will be written as a
subscript unless it refers to a specific hyperedge as in Eq.~(\ref{Hcl0}). The values of \emph{disorder
variables} ${\Jvec_\evec}$ are chosen independently and uniformly at random for each constraint. The
corresponding probability distribution assigns the
 probability of $1/2^K$ to each realization of $\Jvec$:
\begin{equation}
  p(\Jvec) = \prod_{\ell=1}^K \frac{\delta(J_\ell-1)+\delta(J_\ell+1)}{2}.
\label{pJ}
\end{equation}

The energy (\ref{Hcl0}) equals twice the number of violated constraints.
When the number of constraints $M$ is sufficiently small, all of them may be
satisfied at the same time and the energy is zero. The properties of
random \K-SAT are studied in the limit when the number of variables $N$ and
constraints $M$ goes to infinity, while the constraint-to-variable
ratio $\gamma = M/N$ is kept constant. In this limit the fraction of variables
involved in $d$ constraints is Poisson with mean $K\gamma$
\begin{equation}
  f_d(K\gamma) = \frac{1}{d!}(K\gamma)^d \ee^{-K\gamma},
\label{fd}
\end{equation}
so that each variable appears in $K\gamma$ constraints on average.

It has been shown by computer studies that there exists a threshold $\gamma_c$ such that with overwhelming
probability, there exists a configuration of $N$ spins with zero energy if and only if $\gamma < \gamma_c$
(in the limit of large $N$). In the language of statistical mechanics, the random \K-SAT undergoes a
phase transition between the satisfiable (SAT) and unsatisfiable (UNSAT) phases at $\gamma=\gamma_c$. The
interaction term (\ref{EJ}) imposes a ``weak'' constraint on the spins involved in it. For this reason,
unlike the Viana-Bray model with Ising interactions, the phase transition for random \K-SAT does not
coincide with the percolation transition for the corresponding hypergraph. For $3$-SAT, the percolation
transition takes place at $\gamma\subs{perc}=1/6$, while the ``experimental'' value of the satisfiability
threshold is $\gamma_c \approx 4.2$ \cite{Hogg}. The exact value of $\gamma_c$ for
random \K-SAT for $K \geqslant 3$ is not known.

Random \K-SAT can be formulated as a statistical mechanics problem by introducing
the artificial temperature $T=1/\beta$ and writing the Gibbs free energy
\begin{equation}
  F = - \frac{1}{N \beta} \ln \sum_{\{s_i\} \in \{\pm 1\}^N} \ee^{-\beta \Hcal\subs{cl}(\{s_i\})}.
\end{equation}
The extra factor of $1/N$ ensures that this is the free energy \emph{per spin} so that
$F$ does not scale with $N$. It is related to the total internal energy $\Epsilon$ and
the total entropy $\Sigma$ via the standard identity:
\begin{equation}
  F = \frac{1}{N} (\Epsilon - T \Sigma)
\label{FETS}
\end{equation}
In the limit $T=0$ thermal fluctuations disappear and the second term in Eq.~(\ref{FETS}) vanishes. In
this limit $\Epsilon$ converges to the minimum value of energy $H\subs{cl}$. Therefore, $F=0$ for $\gamma
< \gamma_c$. Note that in random \K-SAT there is no region where the minimum number of violated
constraints is $o(N)$ except in the immediate vicinity of $\gamma_c$. For $\gamma > \gamma_c$ this number
is $O(N)$ and $F>0$.

Instance-to-instance fluctuations of $F$ are small:  $o(1)$. Therefore, with overwhelming probability a
randomly chosen instance has  free energy within $o(1)$ from $\langle F \rangle$, which is the
disorder-averaged value. This is the central quantity which is computed using the replica method. We
briefly discuss the main results obtained in \cite{Monasson:1996,Monasson:1997}. The authors
demonstrated that the disorder-averaged free energy of random \K-SAT corresponds to the
extremal value of the free energy functional
\begin{multline}
  \Fcal[P(h)] = \\
    \gamma \intxlim{-\infty}{\infty}{\!\!\!\dd h_1} \dotsi \intxlim{-\infty}{\infty}{\!\!\!\dd h_K}
    P(h_1)\dots P(h_K) \langle \Ucal_\Jvec(h_1,\dots,h_K) \rangle_\Jvec \\
    - \intxlim{-\infty}{\infty}{\dd h} |h| \intxlim{-\infty}{\infty}{\frac{\dd\omega}{2\pp}}
    \ee^{\ii\omega h} \Pbar(\omega) \left( 1 - \Pbar(\omega) \right).
\label{FPh0}
\end{multline}
Here  $\langle \dots \rangle_\Jvec$ denotes averaging over the parameters $J_\ell=\pm 1$
($\ell=1,\dots,K$) with equal weights assigned to all $2^K$ possibilities. The function
$\Ucal_\Jvec(\{h_\ell\})$ is defined as
\begin{equation}
  \Ucal_\Jvec(h_1,\dots,h_K)= 2 \min \left( 1, \left(J_1 h_1\right)_+,\dots,\left(J_K h_K\right)_+ \right).
\label{Uh0}
\end{equation}
Here and throughout the paper we use a shorthand $(\dots)_+$ which we define as follows
\begin{equation}
  (x)_+ = \begin{cases}
    x &\text{for $x > 0$,} \\
    0 &\text{for $x \leqslant 0$.}
  \end{cases}
\end{equation}
The function $\Pbar(\omega)$ in (\ref{FPh0}) is the Fourier transform of the distribution $P(h)$:
\begin{equation}
  \Pbar (\omega) = \intxlim{-\infty}{\infty}{\dd h} \ee^{-\ii \omega h} P(h).
\end{equation}

The function $P^{\ast}(h)$ is found by extremizing $\delta \Fcal[P(h)]$ subject to the constraint
$\intxlim{-\infty}{+\infty}{\dd h}P(h)=1$ a has the meaning of the histogram of effective fields $h_i$
associated with each spin. Whenever $h_i \neq 0$ spin $s_i$ takes the same value $s_i = \sgn h_i$ in all
spin configurations with the lowest energy. The absolute value $|h_i|$ is one-half of the energy cost
needed to flip it.

The fraction of frozen (s.t. $h_i \neq 0$) spins $q=\intxlim{-\infty}{-0}{\dd h}P^{\ast}(h)+
\intxlim{+0}{+\infty}{\dd h}P^{\ast}(h)$ is the order parameter associated with the satisfiability
transition. In the satisfiable phase $q=0$, corresponding to $P^{\ast}(h)=\delta(h)$, whereas the
unsatisfiable phase is described by finite $q>0$.

The simplest solution $P^{\ast}(h)$ of the extremality condition for the functional (\ref{FPh0})
is \cite{Monasson:1996}:
\begin{subequations}
\label{T0intPh}
\begin{equation}
  P(h) = \sum_{k = - \infty}^{+ \infty} \ee^{- K \gamma (q / 2)^{K-1}}
  I_{|k|} \left( K \gamma (q / 2)^{K-1} \right) \delta (h - k),
\tag{\ref{T0intPh}}
\end{equation}
where $I_k(x)$ is the modified Bessel function of first kind.
The value of $q$ may be determined self-consistently from
\begin{equation}
  1 - q = \ee^{- K \gamma (q / 2)^{K-1}} I_0 \left( K \gamma (q / 2)^{K-1}
  \right) .
\label{T0intq}
\end{equation}
\end{subequations}
For $K = 3$ and $\gamma > \gamma_d \approx 4.667$ Eq.~(\ref{T0intq}) has two stable solutions:
the trivial $q = 0$ and the non-trivial $q > 0$. The non-trivial solution does not becomes
stable until $\gamma > \gamma_c \approx 5.181$. The corresponding bound is very close to the
annealed bound of $\gamma\subs{ann}=\ln 2 / \ln(7/8) \approx 5.191$ and greatly overestimates
the ``experimental'' value of the satisfiability threshold $\gamma_{\exp} \approx 4.2$ from
computer simulations \cite{Hogg}.

A similar integer-delta-peaks solution \cite{Kanter:1987} for the order parameter in the Viana-Bray model
\cite{Viana:1985} was shown to be unstable in the longitudinal sector (i.e. within the replica-symmetric
ansatz)\cite{deAlmeida:1988}. The longitudinally stable solution exhibited a continuous part in addition
to delta-peaks. Though the appearance of the continuous component is believed to signal the breakdown of
replica symmetry, the replica-symmetric result may still be useful if regarded as a type of variational
approximation.

The incorporation of the continuous component led to an improved upper bound of the satisfiability
transition $\gamma_c \approx 4.60$ obtained numerically \cite{Monasson:1997}. This problem will be
revisited in Sec.~\ref{sec:CT0} and we will show that although the value of $\gamma_c$ had been determined
correctly, the phase transition predicted by the replica-symmetric theory is actually continuous rather
than first-order as was claimed in Ref.~\cite{Monasson:1997}.

Subsequent analysis by M.~M\'ezard and R.~Zecchina of 1-step replica symmetry breaking (RSB) in random
\K-SAT improved the bound for satisfiability threshold to $\gamma_c \approx 4.267$
\cite{Mezard:2002PRE,Mezard:2002Sci}. It is believed that this 1-step RSB solution is stable. What made
the $T=0$ RSB analysis tractable (and yet required a lot of numerical effort) was the integer-delta-peaks
ansatz for the distribution of effective fields \emph{within each pure state}. It is a daunting task to
extend 1-step RSB analysis to finite temperatures (where non-integer effective fields are certain to
exist), let alone including quantum effects. This paper only considers the replica-symmetric solution.

Using replica-symmetric analysis to study the quantum problem may have some
merit. It has been argued in the literature \cite{Chakrabarti}, based in part on results on
quantum SK model \cite{Thirumalai:1989,Ray:1989}, that effects of quantum
tunneling may stabilize the replica symmetric solution. Even if true, such symmetry must break down
for extremely small transverse fields $\Gamma=o(N)/N$ or in the limit $\Gamma / T \ll 1$.
Indeed, the purely classical limit $\Gamma = 0$ should be described by the 1-step RSB solution
obtained in Ref.~\cite{Mezard:2002PRE}.

\section{Replica solution of quantum $K$-SAT\label{sec:Replica}}

\subsection{Replica-symmetric free energy functional\label{sec:F}}

The quantum Hamiltonian given by Eq.~(\ref{H}) is a sum of two terms: the purely classical term describing
the interaction of Ising spins and the quantum term describing the coupling to the external magnetic field
applied in the transverse direction. By employing Suzuki-Trotter transformation, the problem of finding
the partition function $Z=\operatorname{Tr} \ee^{-\beta\hat{H}}$ can be reformulated as that of computing
the partition function of the purely classical model. The corresponding classical partition function is
written as a sum over all possible paths $s_i(t)$:
\begin{subequations}
\label{ZJ}
\begin{equation}
  Z(\{\Jvec\}) = \sum_{[\{s_i(t)\}]} \ee^{-\intxlim{0}{\beta}{\dd t}
  \Hcal\subs{cl}\blparen\{s_i(t)\}\brparen + \sum_i \Kcal[s_i(t)]}, \tag{\ref{ZJ}}
\end{equation}
where the functional $\Kcal[s(t)]$ is given by
\begin{multline}
  \Kcal[s(t)] = -\frac{1}{2} \ln(\tanh\Gamma\Delta t)
  \sum_{t=0,\Delta t,\dots,\beta-\Delta t} s(t) s(t+\Delta t) \\
  +{}\frac{1}{2} \ln \Bigl( \frac{1}{2} \sinh 2\Gamma\Delta t \Bigr).
\label{Kst}
\end{multline}
\end{subequations}
The time variable $t$ takes $L$ discrete values $t=k \Delta t$ ($\Delta t=\beta/L$). Periodic boundary
conditions $s_i(\beta)=s_i(0)$ are assumed.

The sum (\ref{ZJ}) is over $N \times L$ spin variables labeled by $i=1,\dots,N$ and $t \equiv k \Delta t$.
In anticipation of the limit $L \to \infty$ that will be taken eventually, we treat time as a continuous
variable. In particular, we write $\intxlim{0}{\beta}{\dd t}\dotsi$ to mean $\sum_{k=0}^{L-1} \Delta t
\dotsm$. We use square brackets for writing functionals and for indicating sets labeled by continuous
variables. Sets indexed by a discrete variable will be designated using curly braces. To avoid ambiguities
we may adorn brackets or braces with subscripts and superscripts to indicate index variables and ranges
(e.g. $\left[\{s_i(t)\}_i\right]_{t=0}^{\beta}$)

A constant in expression (\ref{Kst}) ensures proper normalization of the statistical sum (\ref{ZJ}). It
can be verified that (\ref{ZJ}) reduces to \ $Z\sups{0}=(2 \cosh \beta\Gamma)^N$ for the non-interacting
problem ($\Hcal\subs{cl} \equiv 0$).

We choose to write the classical Hamiltonian (\ref{Hcl0}) in the following form
\begin{equation}
  \Hcal\subs{cl} = \sum_{i_1<i_2<\dots<i_K} c_{i_1\dots i_K}
  E(s_{i_1},\dots,s_{i_K};\Jvec_{i_1\dots i_K}),
  \label{Hcl}
\end{equation}
where the cost function $E_\Jvec(s_1,\dots,s_K)$ for \K-SAT is given by Eq.~(\ref{EJ}).
Disorder variables $\Jvec_{i_1 \dots i_K}$ are assumed to be uniformly distributed
according to Eq.~(\ref{pJ}).

The value of $c_{i_1\dots i_K}$ is chosen to be $1$ if the instance contains a
constraint involving a set of variables $i_1,\dots,i_K$, and zero otherwise.
Random variables $c_{i_1 \dots i_K}$ are statistically independent and distributed according to
\begin{equation}
  p(c) = \Bigl( 1 - \frac{K!\gamma}{N^{K-1}} \Bigr) \delta(c)
  + \frac{K!\gamma}{N^{K-1}} \delta(c-1).
\end{equation}
In the asymptotic limit ($N \to \infty$) the number of constraints will be $M=\gamma N$.
Form (\ref{Hcl}) is preferable to (\ref{Hcl0}) because it emphasizes the long-range
character of random \K-SAT.

In this paper we will keep the derivation as general as possible. Formulae written without expanding
(\ref{EJ}) will be --- by substituting appropriate expressions for $E_\Jvec(s_1,\dots,s_K)$ and $p(\Jvec)$
--- directly generalizable to any random combinatorial optimization problem with binary variables and
\K-local interaction (e.g. \K-XOR-SAT, \K-NAE-SAT, $1$-in-$K$ SAT).

The central physical quantity of interest is the disorder averaged value of the free energy $\langle F
\rangle = - \frac{1}{N\beta} \langle \ln Z \rangle$. This is the same as the value of the free energy for
a typical realization of disorder, the free energy (in contrast to $Z$) being a self-averaging quantity.
We use the replica method to perform the disorder averaging. The average of the logarithm is rewritten
using the following identity:
\begin{equation}
  \langle \ln Z \rangle = \lim_{n\to 0} \frac{\partial}{\partial n} \langle Z^n \rangle.
\end{equation}
For integer $n$, $Z^n$ is the partition function of a system of $n$ non-interacting replicas of the
original random instance. Computing $\langle F \rangle$ will require performing the analytical
continuation in $n$. The gist of the method is that disorder averaging in the expression for $\langle Z^n
\rangle$ is done prior to performing the sum over classical spin configurations.
\begin{equation}
  \langle Z^n \rangle = \sum_{[\{s_i^a (t)\}]} \ee^{\sum_{a,i} \Kcal[s_i^a(t)]}
  \left\langle \ee^{-\sum_a \intxlim{0}{\beta}{\dd t} H\subs{cl}\blparen\{s_i^a(t)\}\brparen} \right\rangle ,
  \label{Znsum}
\end{equation}
where the replica index $a$ runs from $1$ to $n$, effectively increasing the number of spin
variables to $N \times L \times n$.

Disorder averaging couples together formerly non-interacting replicas. However, it also transforms the
dilute model with strong $O(1)$ interactions into a completely connected model with weak $O(1/N^{K-1})$
interaction. This permits the exact evaluation of the sum over the spin variables using mean field theory.
We express the mean field solution in terms of a set of order parameters: spin correlation functions
\begin{equation}
  Q_{a_1\dots a_p} (t_1,\dots,t_p) = \frac{1}{N} \sum_i s_i^{a_1}(t_1)
  s_i^{a_2}(t_2) \dotsm s_i^{a_p}(t_p). \label{Qt1tp}
\end{equation}
In the thermodynamic limit, the partition function (\ref{Znsum}) can be written in the form of a
functional integral:
\begin{equation}
  \left\langle Z^n \right\rangle = \intx{\mathcal{D}Q\mathcal{D}\lambda}
  \ee^{-N n \beta \Fcal[\{Q\},\{\lambda\}]}. \label{Znint}
\end{equation}
The argument of the exponential is (up to a factor) the free energy functional $\Fcal$ that depends on
correlation functions $\{Q_{a_1\dots a_p}(t_1,\dots,t_p)\}$ as well as Lagrange multipliers
$\{\lambda_{a_1\dots a_p}(t_1,\dots,t_p)\}$ that enforce constraints (\ref{Qt1tp}). In Eq.~(\ref{Znint})
we have suppressed indices and time arguments for conciseness; similarly $\mathcal{D}Q$ and
$\mathcal{D}\lambda$ are a shorthand for multiple functional integrals.

In the limit $N \to \infty$ the integral (\ref{Znint}) is dominated
by the saddle-point value of $\Fcal$:
\begin{equation}
  F = - \frac{1}{N n \beta} \ln \langle Z^n \rangle
    = \Fcal[\{Q^{\ast}\}, \{\lambda^{\ast}\}].
\end{equation}
The right hand side is evaluated for $\{Q^{\ast}_{\boldsymbol{a}}(\boldsymbol{t})\}$,
$\{\lambda^{\ast}_{\boldsymbol{a}}(\boldsymbol{t})\}$ that make $\Fcal$ stationary with respect to small
variations. Note that in the following we will use a calligraphic $\Fcal$ to indicate a functional and a
roman $F$ to denote its value at the saddle point.

In practice, working with an infinite set of time-dependent correlation
functions is infeasible. Instead, as often done in the analysis of quantum
spin glasses \cite{Bray:1980,Thirumalai:1989}, we resort to the static approximation.
We solve stationarity condition for the reduced set of functions
--- those that are independent of time arguments. Note that consistency requires that if $Q_{a_1\dots
a_p}(t_1,\dots, t_p)$ are replaced by their static counterparts $Q_{a_1\dots a_p}$, any time-dependence
be ignored for $\lambda_{a_1\dots a_p}(t_1,\dots,t_p)$ as well. Implemented in this form, the
static approximation may be regarded as a type of variational approximation.

Integrating out $\{\lambda_{a_1\dots a_p}\}$, we may write $\Fcal(\{Q\})$
as a function of $\{Q_{a_1\dots a_p}\}$ alone. It may be verified that
static $Q_{a_1\dots a_p}$ are the time-averaged dynamic correlation functions:
\begin{equation}
  Q_{a_1\dots a_p} = \frac{1}{\beta^p} \intx{\dd t_1 \dots \dd t_p}
  Q_{a_1\dots a_p}(t_1,\dots,t_p).
\end{equation}
We work within replica-symmetric ansatz, which posits that $Q_{a_1\dots a_p}$
at the saddle-point of $\Fcal$ are symmetric with respect to
permutations of replicas. Due to this symmetry, not all $Q_{a_1\dots a_p}$
are independent. The value of $Q_{a_1\dots a_p}$ may only depend on the set
of numbers $k_1, k_2, \dots$ which, respectively, indicate the number of
distinct replica indices that appear exactly once, twice, etc. We will write
\begin{equation}
  Q_{a_1 a_2 \dots a_{k_1} b_1 b_1 b_2 b_2 \dots b_{k_2} b_{k_2} \dots} =
  Q_{k_1 k_2 \dots},
\end{equation}
where $\{a_i\}, \{b_i\}, \dots$ are all distinct. Although for finite
integer $n$, the inequality $\sum_r k_r \leqslant n$ must hold, performing the
analytical continuation to $n \to 0$ requires the knowledge of
$\langle Z^n \rangle$ for all integer values of $n$. Thus, paradoxically, in the limit
$n \to 0$, the values $k_r$ may run from $1$ to $\infty$.

Note that in classical limit $\Gamma = 0$, only two paths [$s(t) \equiv +1$ and $s(t) \equiv -1$]
contribute to (\ref{Znsum}). Due to that, the static approximation becomes exact in this limit, and the
order parameters $Q_{\{k_r\}}$ may depend only on $p=\sum_r k_{2r+1}$ as evidenced from Eq.~(\ref{Qt1tp}).
It has been recognized in the analysis of classical Viana-Bray model by I.~Kanter and H.~Sompolinsky
\cite{Kanter:1987} that the order parameters $Q_p$ are the moments of the probability distribution $P(m)$
of average spin magnetizations. For a quantum model, $Q_{\{k_r\}}$ are related to the functional
distribution $P[h(m)]$, where functions $h(m)$ are defined on the interval $[-1;1]$:
\begin{equation}
  Q_{\{k_r\}} = \intx{[\dd h(m)]} P[h(m)] \prod_{r=1}^{\infty} \biggl(
    \frac{\intx{\dd m} \ee^{- \beta h(m)} m^r}{\intx{\dd m} \ee^{- \beta h(m)}}
  \biggr)^{k_r}. \label{Qkr}
\end{equation}
That the right hand side of (\ref{Qkr}) is a functional integral is indicated by
the use of square brackets ($\intx{[\dd h(m)]}\dotsi$). Such notation is
customary in quantum field theory (see e.g. \cite{Zinn-Justin}) and is
consistent with our practice of using square brackets to indicate sets indexed
by continuous variables. Regular multidimensional integrals will be written
using curly braces (e.g. $\intx{\{\dd m_i\}_{i=1}^k} \dotsi$). Note that
integrals over magnetizations run from $-1$ to $+1$.

We refer to functions $h(m)$ as \emph{effective fields}. It can be guessed from the formm of (\ref{Qkr}) that these
effective fields represent probability distributions of individual spin magnetizations via $p_i(m) \propto
\ee^{-\beta h(m)}$. The distribution $P[h(m)]$ is the histogram of effective fields $h_i(m)$ associated
with each spin. It may be interpreted as a probability distribution of probability distributions of
magnetizations. Such constructs appear in replica analysis of classical problems in the description of
replica symmetry breaking (RSB). As one can see, in the quantum case they are already present at the
replica-symmetric level. Note that the effective fields $h(m)$ are defined only up to a shift by an
arbitrary constant  $h(m) \to h(m) + \const$.

We express $\Fcal(\{Q_{\{k_r\}}\})$ in terms of the distribution $P[h(m)]$
as a sum of two terms, which we will call a ``quasipotential''
$\Vcal$ and a ``quasientropy'' $\Scal$; themselves dependent on $P[h(m)]$:
\begin{equation}
  \Fcal \llb P[h(m)] \rrb = \gamma \Vcal \llb P[h(m)] \rrb
  -\Scal \llb P[h(m)] \rrb.
\label{FPhm}
\end{equation}
We have used double square brackets to indicate that arguments of $\Fcal$, $\Vcal$ and $\Scal$ are
functionals. Detailed derivations are given in Appendix~\ref{app:Replica}; here we provide the
resulting expressions. For the quasipotential $\Vcal\llb P[h(m)] \rrb$ we obtain
\begin{equation}
  \Vcal = \intx{\left[\{\dd h_\ell(m)\}_{\ell=1}^K\right]}
  \prod_{\ell=1}^K P[h_\ell(m)] \times
  \langle \Ucal_\Jvec [\{h_\ell(m)\}] \rangle_\Jvec,
\label{VPhm}
\end{equation}
where $\langle \dots \rangle_\Jvec$ indicates averaging over
$2^K$ possible realizations of vector $\Jvec$. The functional
integral over $h_1(m),\dots,h_K(m)$ describes averaging over probability
distributions $P[h_\ell(m)]$ of the \emph{quasipotential density}
$\Ucal_\Jvec[h_1(m),\dots,h_K(m)]$ given by the following expression:
\begin{multline}
  \Ucal_\Jvec[\{h_\ell(m)\}] = \frac{1}{\beta} \sum_{\ell=1}^K \ln \intx{\dd m} \ee^{- \beta h_\ell(m)} \\
  - \frac{1}{\beta} \ln \intx{\{\dd m_\ell\}_{\ell=1}^K}
  \ee^{- \beta \Ehat_\Jvec(m_1,\dots,m_K) - \beta \sum_{\ell=1}^K h_\ell(m_\ell)},
\label{UJ}
\end{multline}
Integrals over magnetizations run from $-1$ to $+1$. We write
$\intx{\{\dd m_\ell\}_{\ell=1}^K} \dotsi$ to indicate the \K-dimensional
integral over magnetizations $m_1,\dots,m_K$.

The function $\Ehat_\Jvec(m_1,\dots,m_K)$ that appears in Eq.~(\ref{UJ}) is
multilinear in $m_1,\dots,m_K$ and coincides with $E_\Jvec(\dots)$
when $\{m_\ell\} \in \{\pm 1\}^K$. These two conditions determine it uniquely.
For \K-SAT the expression is obtained by formally replacing discrete spin variables
in Eq.~(\ref{EJ}) with continuous magnetizations $\{m_\ell\}$:
\begin{equation}
  \Ehat_\Jvec(m_1,\dots,m_K) = 2\frac{1+J_1m_1}{2}\dotsm\frac{1+J_Km_K}{2}.
\label{Ehat}
\end{equation}
It is easily seen that for any $\ell$ one may write $\Ehat_\Jvec(m_1,\dots,m_K) =
A_\ell + B_\ell m_\ell$, where $A_\ell$ and $B_\ell$ are independent of
$m_\ell$ but depend on $\Jvec$ and other magnetizations $\{m_{\ell'}\}_{\ell'\neq\ell}$.

For the quasientropy $\Scal\llb P[h(m)] \rrb$, we obtain the following expression:
\begin{subequations}
\label{SPhm}
\begin{equation}
  \Scal = \intx{[\dd h(m)]} \Lcal[h(m)]
  \intx{[\dd\omega(m)]} \ee^{\ii\intx{\dd m}\omega(m)h(m)} \SigmaBar[\omega(m)],
\tag{\ref{SPhm}}
\end{equation}
with $\SigmaBar[\omega(m)]$ given by
\begin{equation}
  \SigmaBar =
  \Pbar[\omega(m)] \left( 1 - \ii \intx{\dd m} \omega(m) u_0(m) -
  \ln \Pbar[\omega(m)] \right) \label{Sigmawm},
\end{equation}
\end{subequations}
which in turn is written in terms of the functional Fourier transform of $P[h(m)]$
that we denote $\Pbar[\omega(m)]$. It is implied that the normalization inside
the functional integral over $\omega(m)$ is such that the inverse Fourier
transform of $\Pbar[\omega(m)]$ reproduces $P[h(m)]$, i.e. $\intx{[\dd\omega(m)]}
\ee^{\ii \intx{\dd m} \omega(m) h(m)} \Pbar[\omega(m)] = P[h(m)]$.

The functional $\Lcal[h(m)]$ is given by the following expression:
\begin{equation}
  \Lcal[h(m)] = -\frac{1}{\beta} \intx{\dd m} \ee^{-\beta h(m)}. \label{Lhm}
\end{equation}

The function $u_0(m)$ that appears in Eq.~(\ref{Sigmawm}) is entirely due to the kinetic term $\Kcal[s(t)]$. In
the limit of continuous magnetizations $(L \to \infty)$ it can be evaluated in closed form:
\begin{multline}
  \ee^{-\beta u_0(m)} =
  \frac{\beta\Gamma}{\sqrt{1-m^2}} I_1\left(\beta\Gamma\sqrt{1-m^2}\right) \\
  + \delta(m-1) + \delta(m+1).
  \label{ebu0}
\end{multline}
Observe that in the limit $\Gamma=0$ only contributions from $m=\pm 1$ are expected. We demonstrate in
Appendix~\ref{app:Classical} that the free energy functional (\ref{FPhm}) may be re-expressed, using the
reduced order parameter $P(h)$, in the form given by Eq.~(\ref{FPh0}).

It would seem from the form of Eq.~(\ref{Znint}) that the free energy should correspond to the minimum of
the free energy functional (\ref{FPhm}). Because of the peculiar nature of the limit $n \to 0$, this is
not the case. In Appendix~\ref{app:Classical} we show that in the classical limit ($\Gamma=0$) the free
energy is a local \emph{maximum} with respect to symmetric perturbations of $P(h)$ [i.e., such that
$\delta P(-h) = \delta P(h)$] and a local \emph{minimum} with respect to antisymmetric perturbations [s.t.
$\delta P(-h) = -\delta P(h)$]. The quantum case is considerably more complex; fortunately, we only need
to make sure that $P[h(m)]$ makes the free energy functional $\Fcal$ stationary and do not care whether it
is a minimum or a maximum.

A few notes must be made about approximations made in this section. The
assumption of replica symmetry is justified for sufficiently small
connectivities $\gamma$; above the replica-symmetry-breaking transition ($\gamma >
\gamma\subs{RSB}$), it becomes an approximation. In contrast, the static
approximation is not guaranteed to be exact anywhere except $\Gamma = 0$. It
is a type of mean-field approximation, whereby fluctuating spins are replaced
by average magnetizations.

The physical interpretation of the static approximation is rather intuitive. One can define the effective
classical model with discrete spins replaced by \emph{continuous} magnetizations $m_i \in [-1;1]$. For a
specific realization of disorder
\begin{subequations}
\label{ZJmi}
\begin{equation}
  Z(\{\Jvec\}) = \intx{\{\dd m_i\}_{i=1}^N} \ee^{-\beta\Hcal\subs{eff}(\{m_i\};\{\Jvec\})},
\tag{\ref{ZJmi}}
\end{equation}
where the effective Hamiltonian $\Hcal\subs{eff}(\{m_i\};\{\Jvec\})$ is
\begin{equation}
  \Hcal\subs{eff} = \sum_{(i_1 \dots i_K)}
  \Ehat_\Jvec(m_{i_1},\dots,m_{i_K};\Jvec_{i_1\dots i_K})
  + \sum_i u_0(m_i),
\label{Hmi}
\end{equation}
\end{subequations}
where $\sum_{(i_1 \dots i_K)} \dotsi$ denotes sum over all hyperedges $c_{i_1 \dots i_K}=1$.
Magnetizations $m_i$ roughly correspond to expectation values $\langle \hat{\sigma}_i^z \rangle$.
Eq.~(\ref{ZJmi}) depends on $\Gamma$ indirectly through form of $u_0(m)$.

In Appendix~\ref{app:BP} we demonstrate that the replica-symmetric static solution is equivalent to the
Bethe-Peierls approximation \cite{Bethe:1935} of the effective classical model defined by
Eqs.~(\ref{ZJmi}).

\subsection{Stationarity condition and the Monte Carlo
method\label{sec:Stationarity}}

To complete the derivation of the replica free energy we need to find $P[h(m)]$ that makes the free energy
functional $\Fcal \llb P[h(m)] \rrb$ stationary with respect to small variations; its value will be the
desired free energy $F$, formally a function of $\beta$, $\Gamma$, and $\gamma$. The stationarity
condition may be written as follows:
\begin{equation}
  \frac{\delta \Fcal}{\delta P[h(m)]} \equiv \gamma \frac{\delta
  \Vcal}{\delta P[h(m)]} - \frac{\delta S}{\delta P[h(m)]} = \const.
  \label{dFdPhm}
\end{equation}
The arbitrary constant appears on the right hand side of Eq.~(\ref{dFdPhm}) is a Lagrange multiplier
associated with the normalization condition $\intx{\dd h(m)} P[h(m)]=1$. Substituting expressions
(\ref{VPhm}) and (\ref{SPhm}) we will formulate the equation that must be satisfied by the saddle-point
value of $P[h(m)]$. Due to a remarkable cancelation we will able to write this self-consistency equation
in a relatively simple form.

Due to the specific form of the functionals (\ref{VPhm}) and (\ref{UJ}), we may express the variation of
$\Vcal$ in the following form:
\begin{multline}
  \frac{\delta \Vcal}{\delta P[h(m)]} = \\
  K \left( \intx{[\dd u(m)]} Q[u(m)] \Lcal[h(m)+u(m)]-\Lcal[h(m)] \right).
\label{dVdPhm}
\end{multline}
This identity can be used as a definition of a new functional $Q[u(m)]$. It is necessarily normalized to unity
($\intx{[\dd u(m)]} Q[u(m)]=1$). We will see that its meaning is that of the probability distribution of
$u(m)=u_\Jvec \left(m;\left[\{h_\ell(m)\}_{\ell=2}^K\right]\right)$ where
\begin{multline}
  u_\Jvec\blparen m;[\{h_\ell(m)\}]\brparen = \frac{1}{\beta} \bigg(
  \sum_{\ell = 2}^K \ln \intx{\dd m} \ee^{-\beta h_\ell(m)}
  \\ -{} \ln \intx{\{\dd m_\ell\}_{\ell=2}^K}
  \ee^{- \beta \Ehat_\Jvec(m,m_2,\dots,m_K) - \beta \sum_{\ell = 2}^K h_\ell(m_\ell)}
  \bigg),
\label{uJ}
\end{multline}
and under the assumption that $\Jvec$ is uniformly distributed and $h_2(m),\dots,h_K(m)$ are taken
from $P[h(m)]$.

On the other hand, the variation of the quasientropy with respect to $P[h(m)]$
reads
\begin{multline}
  \frac{\delta S}{\delta P[h(m)]} = \\
  -\frac{1}{\beta} \intx{[\dd u(m)]} \Lcal[h(m)+u(m)]
  \intx{[\dd\omega(m)]} \ee^{\ii\intx{\dd m}\omega(m)u(m)} \\
  \times \left( \ln \Pbar[\omega(m)]+\ii\intx{\dd m}\omega(m)u_0(m) \right).
\label{dSdPhm}
\end{multline}

Combining Eqs.~(\ref{dVdPhm}) and (\ref{dSdPhm}) uncovers the following system
of self-consistency equations:
\begin{subequations}
\label{QumPhm}
\begin{multline}
  Q[u(m)] = \intx{\left[\{\dd h_\ell(m)\}_{\ell = 2}^K \right]}
  \prod_{\ell = 2}^K P[h_\ell(m)] \\
  \times \left\langle \delta \boldsymbol\blbracket u(m)
  - u_\Jvec\blparen m; [\{h_\ell(m)\}]\brparen \brbracket
  \right\rangle_\Jvec,
\label{Qum}
\end{multline}
\begin{multline}
  P[h(m)] = \intx{[\dd\omega(m)]}
  \ee^{\ii \intx{\dd m} \omega(m) \boldsymbol{(}h(m)-u_0(m)\boldsymbol{)}} \\
  \times \exp K \gamma \left(-1+\intx{[\dd u(m)]} \ee^{\ii\intx{\dd m}\omega(m)u(m)} Q[u(m)] \right).
\label{Phm}
\end{multline}
\end{subequations}
In (\ref{Qum}) we use a \emph{functional} generalization of the delta function, defined so that $F[x(m)] =
\intx{[\dd y(m)]} F[y(m)] \delta[x(m)-y(m)]$. Note that Eq.~(\ref{Phm}) may be written in an alternative
form by expanding the exponential in the integrand (the term corresponding to $d=0$ is $\ee^{-K\gamma}
\delta[h(m)-u_0(m)]$):
\begin{multline}
  P[h(m)] = \sum_{d=0}^{\infty} f_d(K\gamma)
  \intx{\left[\{\dd u_k(m)\}_{k=1}^d\right]} \prod_{k=1}^d Q[u_k(m)] \\
  \times \delta\biggl[ h(m)-u_0(m)-\sum_{k=1}^d u_k(m) \biggr]
  \tag{\ref{Phm}$'$}.\label{Phm2}
\end{multline}
The appearance of the Poisson distribution $f_d(\alpha) = \frac{\alpha^d}{d!} \ee^{-\alpha}$ is
intimately related to the hypergraph model that we study, as it is the distribution of the degrees (number
of incident hyperedges) of the vertices. From the form of Eqs.~(\ref{Qum}), (\ref{Phm2}) it is apparent
that $h(m)$ are properly associated with the vertices of the random hypergraph, whereas $u(m)$ correspond
to its hyperedges. This link is explained in Appendix \ref{app:BP}.

The system of equations (\ref{QumPhm}) can be solved iteratively.
Starting from some initial distribution $P\sups{0}[h(m)]$,
we may compute a sequence of $\{Q\sups{r}[u(m)]\}$ and $\{P\sups{r}[h(m)]\}$
by applying (\ref{Qum}) and (\ref{Phm}). The limiting distribution
\begin{equation}
  P^{\ast}[h(m)] = \lim_{r \to \infty} P\sups{r}[h(m)]
\end{equation}
must be a solution to the stationarity condition (\ref{dFdPhm}). The \emph{value}
of the free energy is obtained from $F = \gamma V - S$, where the quasipotential $V$
is found by substituting $P^{\ast}[h(m)]$ into (\ref{VPhm}), and the expression
for the quasientrtopy $S$ is rewritten using self-consistency equations (\ref{QumPhm}):
\begin{multline}
  S = K \gamma \intx{[\dd h(m) \dd u(m)]} P^{\ast}[h(m)] Q^{\ast}[u(m)] \\
    \times \boldsymbol{(} \Lcal[h(m)] - \Lcal[h(m)+u(m)] \boldsymbol{)} \\
  + \intx{[\dd h(m)]} P^{\ast}[h(m)] \Lcal[h(m)].
\label{SQum}
\end{multline}
The iterative procedure described above lends itself to a straightforward implementation using a Monte
Carlo method. Observe that both expressions (\ref{Qum}) and (\ref{Phm}) are written as averages over
probability distributions $P[h(m)]$ and $Q[u(m)]$ and vectors $\Jvec$. The Monte Carlo algorithm that we
describe below represents distributions $P[h(m)]$ and $Q[u(m)]$ as finite samples $\{h_i(m)\}_{i=1}^N$ and
$\{u_i(m)\}_{i=1}^N$. (Implementation details of storing \emph{functions} $h(m)$ and $u(m)$ in memory are
not discussed here; we assume that it can be done without any loss in precision). A single iteration step
can be implemented as follows:
\begin{enumerate}
  \item Compute a sample $\{u_i(m)\}$. For each $i \in 1,\dots,N$
  \begin{enumerate}
    \item Choose $h_2(m),\dots,h_K(m)$ from the set $\{h_i(m)\}$
    uniformly at random.

    \item Choose a disorder vector $\Jvec$ at random.

    \item Evaluate $u(m) = u_\Jvec\left(m; \left[\{h_\ell(m)\}_{\ell=2}^K\right]\right)$
    using Eq.~(\ref{uJ}).
  \end{enumerate}
  \item Compute an updated sample $\{h'_i(m)\}$. For each $i \in 1,\dots,N$:
  \begin{enumerate}
    \item Choose a random integer $d$ from the Poisson distribution with parameter
    $K \gamma$.
    \item If $k = 0$, let $h'(m) = u_0(m)$, otherwise

    \item Choose $u_1(m),\dots,u_d(m)$ form the set $\{u_i(m)\}$
    uniformly at random and

    \item Evaluate $h'(m)$ using
    \begin{equation}
      h'(m) = u_0(m) + \sum_{k=1}^d u_k(m). \label{hm2}
    \end{equation}
  \end{enumerate}
\end{enumerate}
The convergence criterion for the algorithm is that step-to-step fluctuations are entirely due to the
finiteness of $N$, i.e. that both the old $\{h_i(m)\}$ and the updated $\{h'_i(m)\}$ histograms sample the
same probability distribution. This can be verified by the Kolmogorov-Smirnov test for instance
\cite{Eadie}.

Self-consistency equations and their Monte Carlo implementation are related to
Thouless-Anderson-Palmer (TAP) equations discussed in Appendix~\ref{app:BP}.

\subsection{Quantum limit ($T = 0$, $\Gamma > 0$).\label{sec:QT0}}

A number of simplifications are possible in this limit. Since all integrals over magnetizations have the
form  $\intx{\dd m} \ee^{- \beta f(m)}$, in the limit $\beta \to \infty$ they are dominated by the minimum
value of $f(m)$. The replica free energy functional $\Fcal\llb P[h(m)]\rrb \equiv \gamma \Vcal\llb
P[h(m)]\rrb - \Scal\llb P[h(m)]\rrb$ retains the form given by Eqs.~(\ref{VPhm}) and (\ref{SPhm}), but
expressions (\ref{UJ}) and (\ref{Lhm}) for $\Ucal_\Jvec[\{h_\ell(m)\}]$ and $\Lcal[h(m)]$ simplify to,
respectively,
\begin{multline}
  \Ucal_\Jvec\left[\left\{h_\ell(m)\right\}\right]= \\
  \min_{\{m_\ell\}} \biggl[ \Ehat_\Jvec(\{m_\ell\}) + \sum_{\ell=1}^K h_\ell(m_\ell) \biggr]
  - \sum_{\ell=1}^K \min_m[h_\ell(m)]
\end{multline}
and
\begin{equation}
  \Lcal[h(m)] = \min_m[h(m)],
\end{equation}
while Eq.~(\ref{ebu0}) assumes the asymptotic form
\begin{equation}
  u_0(m) = - \Gamma \sqrt{1 - m^2} .
\end{equation}

Self-consistency equations retain the form of Eqs.~(\ref{QumPhm}), but the expression
for $u_\Jvec\blparen m;[\{h_\ell(m)\}]\brparen$ reduces to the following:
\begin{multline}
  u(m) = \min_{m_2,\dots,m_K} \biggl[ \Ehat_\Jvec(m, m_2,\dots,m_K)
  + \sum_{\ell = 2}^K h_\ell(m_\ell) \biggr] \\
  - \sum_{\ell = 2}^K \min_m[h_\ell(m)].
\label{uJT0}
\end{multline}
The physical meaning of the effective fields $h(m)$ is particularly evident in the limit $T=0$. The
free energy corresponds to the minimum of the effective Hamiltonian of Eq.~(\ref{ZJmi}):
\begin{multline}
  \Hcal_{T=0}(\{m_i\};\{\Jvec\}) = \\
  \sum_{(i_1 \dotsc i_K)}
  \Ehat(m_{i_1},\dots,m_{i_K};\Jvec_{i_1\dots i_K})
  - \Gamma \sum_i \sqrt{1-m_i^2}.
\label{HmiT0}
\end{multline}
In the limit $\Gamma \to \infty$ the free energy is dominated by the second term: $F=-\Gamma$, which
corresponds to a state with all spins completely polarized along the $x$ direction. In the limit $\Gamma
\to 0$ the free energy is expected to be $F \approx 0$ in the satisfiable phase and $F \gtrsim 0$ in the
unsatisfiable phase.

For each spin, $h_i(m)$ is, up to a constant, the increase in energy if the
magnetization of spin $i$ is set to $m$ (magnetizations of other spins are allowed
to adjust).

It is possible to set up a deceptively simple system of equations for magnetizations
$\{m_i^{\ast}\}$ corresponding to the minimum of (\ref{HmiT0}). Solving
$\partial \Hcal_{T=0} / \partial m_i=0$ we observe that $m_i^{\ast}$ may be represented in terms
of scalar effective fields $h_i^{\ast}$ via
\begin{equation}
  m_i^{\ast} = \frac{h_i^{\ast}}{\sqrt{\Gamma^2 + \left( h_i^{\ast}\right)^2}},
\end{equation}
while each $h_i^{\ast}$ is a sum of contributions $u_k$ from each hyperedge incident to vertex $i$. E.g. for
\K-SAT
\begin{equation}
  u_k^{\ast} = \prod_{\ell=1}^K \frac{1+J_\ell m_{k\ell}^{\ast}}{2}.
\end{equation}

The description of the problem in terms of order parameter $P(h^{\ast})$ --- the histogram
of fields $h_i^{\ast}$ --- is effective for large values of $\Gamma$ where (\ref{HmiT0})
has only one local minimum. However in the limit of small $\Gamma$ the number of local minima
becomes exponential in $N$, which is the essential reason for the introduction of the functional
order parameter.

\section{Small transverse field regime at zero temperature. \label{sec:smallG}}

For small values of the transverse field, the free energy functional can be expanded in powers of $\Gamma$
around $\Gamma=0$ corresponding to the classical limit. The limit $\Gamma=0$ has been considered in
Appendix~\ref{app:Classical}. Taking the limit $T=0$ afterwards,  expression (\ref{FPh0}) for the
classical free energy is recovered. We expect that the physically relevant value of the free energy
(unlike that of the order parameter) cannot be affected by the order in which the limits are taken. It is
instructive to verify that the same result is obtained when the limit $T=0$ is taken first, followed by
$\Gamma=0$. Even though the effective field functions $h(m)$ will be finite everywhere in the interval
$[-1;+1]$, the value of the free energy will by determined by the values $h_{\pm 1}$ attained on both ends
of the interval.

As a first step, we demonstrate that the function $u_\Jvec\blparen m;[\{h_\ell(m)\}]\brparen$ is always
convex. This convexity property is valid for arbitrary values of $\Gamma$. We evaluate $u(m)$ for some
linear combination of magnetizations $m_0$ and $m_1$. Writing $m_2^\ast,\dots,m_K^\ast$ to denote the
values of magnetization that minimize the first term on right hand side of Eq.~(\ref{uJT0}) and using the
property that $\Ehat_\Jvec(m_1,\dots,m_K)$ is a multilinear function of magnetizations, we write the
lengthy inequality proving the convexity of $u(m)$:
\begin{widetext}
\begin{equation}
\begin{split}
  u\blparen \alpha m_0 + (1-\alpha) m_1\brparen &={} \min_{m_2,\dots,m_K}
  \biggl[\Ehat_\Jvec\blparen \alpha m_0 + (1-\alpha) m_1,m_2,\dots,m_K\brparen
     + \sum_{\ell=2}^K h_\ell(m_\ell)\biggr] - \sum_{\ell = 2}^K \min_m[h_\ell(m)] \\
  &={} \alpha \Ehat_\Jvec(m_0, m_2^{\ast},\dots,m_K^{\ast})
     + \alpha \sum_{\ell=2}^K h_\ell(m_\ell^{\ast})
     + \alpha \sum_{\ell=2}^K \min_m[h_\ell(m)] \\
  &\quad+{} (1-\alpha) \Ehat_\Jvec(m_1,m_2^{\ast},\dots,m_K^{\ast})
     + (1-\alpha) \sum_{\ell=2}^K h_\ell(m_\ell^{\ast})
     + (1-\alpha) \sum_{\ell=2}^K \min_m[h_\ell(m)] \\
  &\geqslant{} \alpha u(m_0) + (1-\alpha) u(m_1).
\end{split}
\end{equation}
\end{widetext}
Using the convexity of $u(m)$, it can be established from Eq.~(\ref{Phm2}) that in the limit
$\Gamma=0$ the effective field functions $h(m)$ are also convex due to the vanishing of $u_0(m)$. The
convexity of $h(m)$ and the multilinearity of $\Ehat_\Jvec(\{m_\ell\})$, together, ensure that expressions
of the form $\Ehat_\Jvec(m_1,\dots,m_K) + h_\ell(m_\ell)$ achieve their minimum values for $m_\ell = \pm
1$. Similarly, minima of effective fields $h(m)$ can be replaced by $\min(h_{-1}, h_{+1})$ due to
convexity of $h(m)$. It follows that the value of the free energy will be unchanged if minima over the
interval $m \in [-1;+1]$ are replaced with minima over the discrete set $m \in \{-1;+1\}$. Hence, the
free energy of the quantum model in the limit $\Gamma=0$ must equal that of the classical model.

One corollary to this is that in the limit $\Gamma=0$ the functions $u(m)$ are
piecewise linear. Indeed, $\dd u / \dd m = (\partial / \partial m)
\Ehat_\Jvec(m, m_2^{\ast},\dots,m_K^{\ast})$ may depend on $m$ only indirectly
via $\{m_\ell^\ast\}_{\ell=2}^K$. Since $m_\ell^\ast \in \{-1;+1\}$ the slope
of $u(m)$ cannot change continuously; instead it assumes one of finitely many
values depending on the value of $m$.

\begin{figure}[!h]
  \includegraphics[width=3.4in]{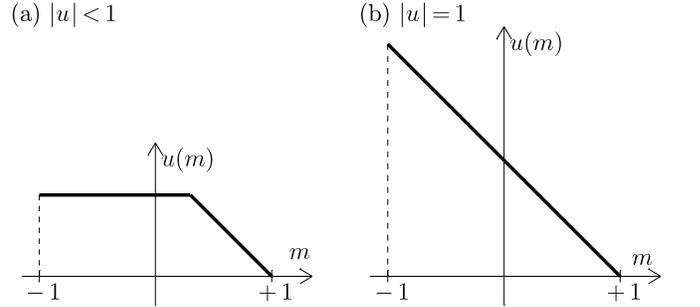}
  \caption{Form of $u(m)$ in the classical limit ($\Gamma=0$).
  Two cases are depicted: (a) $|u|<1$, $u>0$ and (b) $|u|=1$, $u>0$.
  Analogous figures for $u<0$ may be obtained by mirror reflection $m \to -m$.
\label{fig:um0}}
\end{figure}

So far we have kept the derivation as general as possible. In the following we
restrict our attention to random \K-SAT proper described by the cost function
(\ref{Ehat}). In the limit $\Gamma=0$ functions $u(m)$ (sketched in Fig.~\ref{fig:um0})
may be parametrized by a single parameter $u$ as follows:
\begin{equation}
  u(m) = \min \blparen 2,2|u|,1-(\sgn u)m\brparen. \label{umT0}
\end{equation}
Using the same letter for the function $u(m)$ and the parameter $u$ should not
lead to confusion. We will always include the magnetization argument to refer
to the function $u(m)$. The value of $u(m)$ for a particular magnetization
(e.g. $m=0$ or $m=\pm 1$) will be indicated using subscripts: i.e. $u_0$, $u_{\pm 1}$.

It can be seen from Eq.~(\ref{umT0}) that $u=\frac{1}{2}(u_{-1}-u_{+1})$. Although
$h(m)=u_0(m)+\sum_{k=1}^d u_k(m)$ does not admit a simple parametrization, we can still
define scalar $h=\frac{1}{2}(h_{-1}-h_{+1})$. This choice ensures that
$h=\sum_{k=1}^d u_k$. As expected, $u(m)$ defined by Eq.~(\ref{uJT0}) assumes the form of
Eq.~(\ref{umT0}) and depends on $\{h_\ell(m)\}$ only via $\{h_\ell\}$:
\begin{equation}
  u=\min\bigl(1,\left(J_2 h_2\right)_+,\dots,\left(J_K h_K\right)_+\bigr),
\end{equation}
with $(x)_+$ used to denote $\max(x,0)$.

This describes two different regimes. Function $u(m)$ has the form depicted in Fig.~\ref{fig:um0} (left)
whenever $\min_\ell\{(J_\ell h_\ell)_+\} < 1$, and that shown on the right if $\min_\ell\{J_\ell
(h_\ell)_+\} \geqslant 1$.

\begin{figure}[!h]
  \includegraphics[width=3.4in]{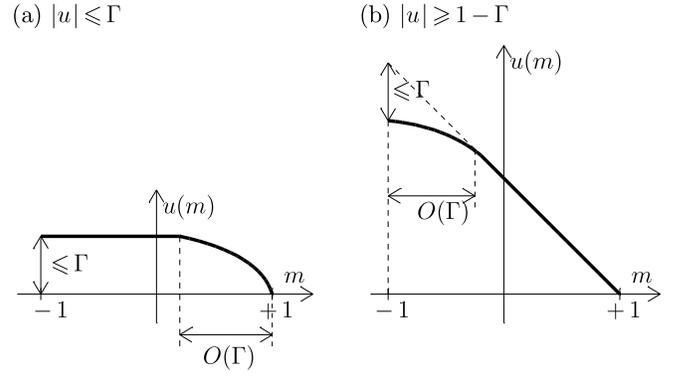}
  \caption{Possible form of $u(m)$ for finite, but small $\Gamma$. Figures depict
  two possibilities
  corresponding to $u>0$ ($u<0$ corresponds to mirror images $m \to -m$):
  (a) $|u| \leqslant \Gamma / 2$ and (b) $|u| \geqslant 1 - \Gamma / 2$.
  For $1 \pm m = O (\Gamma)$, the functions $u(m)$ are not piecewise linear.
  Together with Fig.~\ref{fig:um0}, this encompasses all possible forms of $u(m)$
  in the limit of small $\Gamma$.
\label{fig:umsmall}}
\end{figure}

The order parameter $P[h(m)]$ may be obtained by iterating Eqs.~(\ref{QumPhm}) starting from, e.g.
$P\sups{0}[h(m)]=\delta[h(m)-u_0(m)]$ corresponding to the non-interacting model \footnote{When
$\Gamma=0$, as $P\sups{0}[h(m)]=\delta[h(m)]$ could be a metastable solution. However, for finite
$\Gamma$, successive iterations will always converge to the stable solution. }. The effects of small, but
finite values of $\Gamma$ can be illustrated by performing a single iteration. Substituting
$h_2(m)=\dots=h_K(m)=u_0(m)$ into (\ref{uJT0}) gives
\begin{multline}
  u_\Jvec(m) = \min_{m_2,\dots,m_K}
  \biggl[2 \frac{1+J_1 m}{2}\prod_{\ell=2}^K\frac{1+J_\ell m_\ell}{2} \\
  -(K-1)\Gamma \sqrt{1-m^2} \biggr] + (K-1)\Gamma.
\end{multline}
This expression is neither zero (as in the classical case), nor even piecewise linear. It should be
declared in advance that we do not need the precise analytical expression for $u_\Jvec(m)$ as the free
energy will not depend on such details. It is easily seen that $u_\Jvec(m)$ is monotonic in $m$, and that
it is zero at $m=-J_1$. In addition one can demonstrate that
\begin{align}
  u_\Jvec(m) = \Gamma - o (\Gamma) && \text{when $1 + J_1 m \gg \Gamma$}.
\end{align}
When $K \geqslant 3$ this approximate identity is strengthened to $u_\Jvec(m) = \Gamma$ for $1 + J_1 m
\geqslant C \Gamma$ (for some constant $C$). This form of $u_\Jvec(m)$ is sketched in Figure
\ref{fig:umsmall} (left) for $J_1 < 0$ (in this particular case $u(0) = \Gamma$). Since we are not
concerned with the precise form of $u(m)$ it is still permissible to describe it using a single parameter
$u=\frac{1}{2}(u_{-1}-u_{+1})$ (which would equal $-J_1 \Gamma / 2$ in the present case). Expression
(\ref{umT0}) would apply everywhere on $[-1;+1]$ except for the vicinity of $m = \sgn u$, where $1-(\sgn
u)m=O(\Gamma)$. Note that if either $1-(\sgn u)m \ll \Gamma$, or $1-(\sgn u) m \gg \Gamma$, expression
(\ref{umT0}) remains valid up to $o(\Gamma)$.

By considering additional iterations of Eqs.~(\ref{QumPhm}) it is possible to classify
all possible forms of $u(m)$ that can be encountered. In addition to the piecewise
linear forms of Fig.~\ref{fig:um0}, it may have one of the forms depicted in
Fig.~\ref{fig:umsmall}. The latter form may occur only if $|u| \leqslant \Gamma/2$
(Fig.~\ref{fig:um0}, left) or $1-\Gamma/2 \leqslant |u| < 1$ (Fig.~\ref{fig:um0}, right).

Observe that Eq.~(\ref{umT0}) is approximately valid for all $m$, with the possible exception of
$1-|m| \gg \Gamma$. Recognizing that $u(m)$ is monotonic and that $|\dd u/\dd m| \leqslant 1$, we can
restate the condition in an equivalent form. We require that $\dd u / \dd m$ approximately [up to $o
(\Gamma)$] equal either $0$ or $\pm 1$ for $1 - |m| \gg \Gamma$. For values of $m$ such that $1-|m|=O(1)$
the derivative $\dd u/\dd m$ equals either $0$ or $\pm 1$ with a correction of at most $O(\Gamma^2)$.

\begin{figure}[!h]
  \includegraphics[width=2.2in]{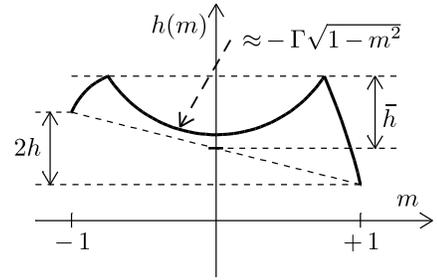}
  \caption{Typical form of the function $h(m)$, parameterized by $h$ and $\hBar$.
  In general $h=\frac{1}{2}(h_{-1}-h_{+1})$. The distance from the middle minimum
  at $m \approx 0$ to the centerpoint of the line joining $h(-1)$ and $h(+1)$ is
  $\Gamma - \hBar$. All features $O(\Gamma^2)$ have been suppressed (see discussion in text).
\label{hmform}}
\end{figure}

To investigate the qualitative form of effective fields $h(m)$ write Eq.~(\ref{hm2})
substituting the value of $u_0(m)$:
\begin{equation}
  h(m) = -\Gamma\sqrt{1-m^2} + \sum_{k=1}^d u_k(m).
\label{hmsum}
\end{equation}
Function $h(m)$ is a sum of concave and convex functions. One of possible forms of $h(m)$ is sketched in
Fig.~\ref{hmform}. All features that are $o(\Gamma)$ have been suppressed. In particular,
Fig.~\ref{hmform} fails to reflect that the locations of the local  minima at $m=\pm 1$ are shifted by
$O(\Gamma^2)$
In general, local minima of $h(m)$ away from the endpoints of the interval $[-1;+1]$ must satisfy $\dd
h/\dd m=0$. Since $\dd u_k/\dd m$ are approximately integers for $1-|m| \gg \Gamma$, such local minimum
can exist only if $\sum_k \dd u_k / \dd m |_{m = 0} \approx 0$ and can only be located at $m^{\ast}
\approx 0$ [up to $O(\Gamma)$]. Neglecting contributions of $O(\Gamma^2)$ and higher, the free energy is
determined by values of $u(m)$ and $h(m)$ at $m=0$ or $m=\pm 1$.

We will parameterize each of $u(m)$ and $h(m)$ by scalars $u$, $\uBar$ and $h$, $\hBar$ respectively. We
define
\begin{subequations}
\label{uuhh}
\begin{align}
  u &={} \frac{u_{-1}-u_{+1}}{2}, \\
  \uBar &={} \frac{u_{-1}-2u_0+u_{+1}}{2} .
\end{align}
\end{subequations}
And $h$, $\hBar$ parameterizing $h(m)$ of Eq.~(\ref{hmsum}) are chosen as follows:
\begin{subequations}
\begin{align}
  h &={} \sum_{k=1}^d u_k,  \label{huk}\\
  \hBar &={} \sum_{k=1}^d \uBar_k.  \label{huk2}
\end{align}
\end{subequations}

Note that $u$ and $\uBar$ are not independent variables, but are related by
\begin{equation}
  \uBar = \min(|u|,1-|u|). \label{ubar}
\end{equation}
Combining Eqs.~(\ref{umT0}), (\ref{hmsum}) and (\ref{uuhh}), we obtain for
values of $h(m)$ at $m=\pm 1$ and $m=0$:
\begin{subequations}
\begin{align}
  h_{\pm 1} &={} \sum_{k=1}^d |u_k| \, \pm h, \\
  h_0 &={} \sum_{k=1}^d |u_k| \, + \hBar - \Gamma.
\end{align}
\end{subequations}
The expression $\sum_{k=1}^d |u_k|$ represents a constant shift which must cancel out in the expression
for the free energy. This cancellation allows one to parameterize $h(m)$ by $h,\hBar$ alone.

It is straightforward to rewrite the self-consistency equations (\ref{QumPhm}) in terms of the reduced
distributions $P(h,\hBar)$ and $Q(u,\uBar)$. However, it is more instructive to derive self-consistency
equations from the stationarity condition for the free energy functional $\Fcal\left[P(h,\hBar)\right]$
that can be derived by using our ansatz for $h(m)$.

As before, we separate the free energy functional into two parts corresponding
to the quasipotential and the quasientropy:
$\Fcal\left[P(h,\hBar)\right] = \gamma \Vcal\left[P(h,\hBar)\right]
- \Scal\left[P(h,\hBar)\right]$. We write down without proof the expression for
the quasientropy $\Scal\left[P(h,\hBar)\right]$:
\begin{subequations}
\label{SPhh}
\begin{equation}
  \Scal = \intx{\dd h\dd\hBar} \Lcal(h,\hBar)
  \intx{\frac{\dd\omega\dd\omegaBar}{(2\pp)^2}}
  \ee^{\ii\omega h+\ii\omegaBar\hBar}
  \SigmaBar(\omega,\omegaBar),
\tag{\ref{SPhh}}
\end{equation}
where $\Lcal(h,\hBar)$ and $\SigmaBar(\omega,\omegaBar)$ are given by, respectively,
\begin{equation}
  \Lcal(h,\hBar) = \max(|h|,\Gamma-\hBar),
\end{equation}
\begin{equation}
  \SigmaBar(\omega,\omegaBar)=\Pbar(\omega,\omegaBar)\left(1-\ln\Pbar(\omega,\omegaBar)\right),
\end{equation}
\end{subequations}
with $\Pbar (\omega, \bar{\omega}) = \intx{\dd h \dd \hBar} \ee^{\ii
\omega h + \ii  \bar{\omega}  \hBar} P(h,\hBar)$ used to denote the Fourier
transform of $P(h,\hBar)$. The derivation of this expression
is straightforward and relies on the ability to replace all minima over
magnetizations in the interval $[-1;+1]$ by those over the discrete set
$m \in {0,\pm 1}$.

The derivation of the quasipotential is slightly more intricate. The minimum of
\begin{equation}
  E'_\Jvec(m_1,\dots,m_K) = 2\prod_{\ell=1}^K \frac{1+J_\ell m_\ell}{2}
  + \sum_{\ell=1}^K h_\ell(m_\ell) \label{minE}
\end{equation}
may occur only for $m_1,\dots,m_K=0,\pm 1$. It is unnecessary to consider
all $3^K$ possibilities, however. Let $m_\ell^{\ast}$ denote the location of
a global minimum of $h_\ell(m)$. The location of the global minimum of
(\ref{minE}) is such that $m_\ell=0$ or $m_\ell=-J_\ell$ for some
$\ell$, while all other magnetizations are $m_{\ell'} = m_{\ell'}^{\ast}$. It
is never advantageous to have more than one magnetization different from
$m_\ell^{\ast}$ as long as $\Gamma<1$.

\begin{figure}[!h]
  \includegraphics[width=3.4in]{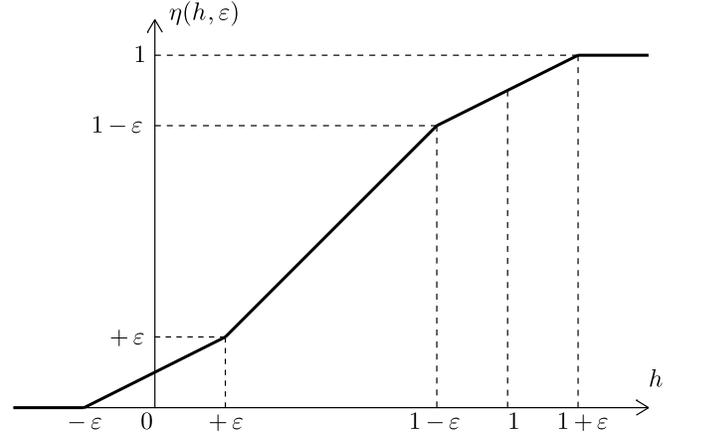}
  \caption{The form of the function $\eta=\eta (h, \varepsilon)$ defined in Eq.~(\ref{etahe}).
\label{fig:eta}}
\end{figure}

Therefore, $E'(m_1,\dots,m_K)$ may be written as a minimum over just
$K$ distinct possibilities. After some algebra we obtain the following
expression for the quasipotential $\Vcal\left[P(h,\hBar)\right]$:
\begin{subequations}
\label{VPhh}
\begin{equation}
  \Vcal = \intx{\left\{\dd h_\ell \dd\hBar_\ell\right\}}
  \prod_{\ell=1}^K P(h_\ell,\hBar_\ell) \times\langle\Ucal_\Jvec(h,\hBar)\rangle_\Jvec,
\tag{\ref{VPhh}}
\end{equation}
with $\Ucal_\Jvec(h,\hBar)$ given by
\begin{equation}
  \Ucal_\Jvec(h,\hBar) = 2 \min_{\ell=1,\dots,K}\left\{\eta\left(J_\ell
  h_\ell,\Gamma-\hBar_\ell\right)\right\},
\end{equation}
and the definition of $\eta(h,\varepsilon)$ is
\begin{equation}
  \eta (h,\varepsilon) = \min\blparen 1,\left(h\right)_+ \brparen
  + \frac{1}{2}(\varepsilon-|h|)_+ - \frac{1}{2}(\varepsilon-|h-1|)_+
\label{etahe}
\end{equation}
\end{subequations}
(the auxiliary function $\eta(h,\varepsilon)$ is sketched in Fig.~\ref{fig:eta} for illustrative
purposes). Note that for $\varepsilon \leqslant 0$, Eq.~(\ref{etahe}) reduces to $\eta(h,\varepsilon) =
\min\blparen 1,(h)_+\brparen$.

It is immediately seen that in the limit $\Gamma=0$, Eqs.~(\ref{SPhh}) and (\ref{VPhh}),
rewritten in terms of $P(h)=\intx{\dd\hBar} P(h,\hBar)$, coincide with classical $T=0$
expressions for the quasientropy and the quasipotential respectively.

The stationarity condition is $\delta (\gamma \Vcal-\Scal) / \delta P(h,\hBar) = \const$.
It should come as no surprise that the following identity holds:
\begin{multline}
  \frac{\delta\Vcal}{\delta P(h,\hBar)} = \\
  K \left(
  \intx{\dd u\dd\uBar} Q(u,\uBar) \Lcal(h+u,\hBar+\uBar)
  - \Lcal(h,\hBar)
  \right) + \const,
\end{multline}
where $Q(u,\uBar)$ is effectively a distribution of just one parameter $u$:
\begin{subequations}
\begin{equation}
  Q(u,\uBar) = Q(u)\delta\blparen \uBar-\min(|u|,1-|u|)\brparen,
\end{equation}
\begin{equation}
  Q (u) = \intx{\left\{\dd h_\ell\dd\hBar_\ell\right\}}
 \prod_{\ell=2}^K P(h_\ell,\hBar_\ell) \times
 u_{\Gamma;\Jvec}\left(\{h_\ell,\hBar_\ell\}_{\ell=2}^K\right),
\end{equation}
with $u_{\Gamma;\Jvec}(h_2,\hBar_2;\dots;h_K,\hBar_K)$ given by
\begin{equation}
  u_{\Gamma;\Jvec}(\{h_\ell,\hBar_\ell\}) = \min_{\ell=2,\dots,K} \left\{
 \eta\left(J_\ell h_\ell,\Gamma-\hBar_\ell\right)
 \right\}.
\end{equation}
\label{Quu}
\end{subequations}
Solving the stationarity condition reveals the following relationship between
$P(h,\hBar)$ and $Q(u,\uBar)$:
\begin{equation}
  P(h,\hBar) = \intx{\dd h \dd \hBar} \ee^{\ii \omega h + \ii
  \bar{\omega}  \hBar} \exp K \gamma \Qbar (\omega, \bar{\omega}),
  \label{Phh}
\end{equation}
where $\Qbar (\omega, \bar{\omega})$ is the Fourier transform of $Q (u,
\uBar)$. An alternative form of (\ref{Phh}) is
\begin{multline}
  P(h,\hBar) = \sum_d f_d(K\gamma) \intx{\{\dd u_k \dd \uBar_k \}}
  \prod_{k=1}^d Q(u_k,\uBar_k) \\
  \times \delta \biggl( h - \sum_{k=1}^d u_k \biggr) \delta
  \biggl( \hBar - \sum_{k=1}^d \uBar_k \biggr). \tag{\ref{Phh}$'$}
\label{Phh2}
\end{multline}
The order parameter can be found by solving Eqs.~(\ref{Quu}) and (\ref{Phh})
self-consistently. It is straightforward to write down $\Gamma \ll 1$ TAP equations for
a particular disorder realization by ``reverse-engineering'' these
relations, interpreting $P(h,\hBar)$ and $Q(u, \uBar)$ as the histograms of
effective fields associated with vertices and hyperedges of the random hypergraph.

\section{Classical zero-temperature solution revisited\label{sec:CT0}}

\subsection{Scale-invariant replica-symmetric solution\label{sec:ModelO}}

The analysis of Ref.~\cite{Monasson:1997} presents a classical picture of the first-order phase
transition: as a competition between two locally stable solutions: the trivial $P(h) = \delta (h)$ and the
non-trivial $P(h)$. However, a Monte Carlo study reveals that the non-trivial solution is not stable for
any $\gamma < \gamma_c \approx 4.60$ under iterations of the self-consistency equations for $P(h)$ and
$Q(h)$. This casts doubt on the picture of competition between two local maxima (the free energy must be
maximized) or the prediction of the dynamic transition at $\gamma_d \approx 4.43$. We claim that the phase
transition at $\gamma_c \approx 4.60$ is in fact continuous. While the value of $\gamma_c$ has been
determined correctly, the discontinuity of the order parameter is an artifact of the discretization used
in the numerical procedure (values of $(\Delta h)^{-1}$ up to $30$ have been used in
\cite{Monasson:1997}).

In this section we consider the model described by the free energy
functional (\ref{FPh0}), but with the expression (\ref{Uh0}) modified to
\begin{equation}
  \Ucal\sups{O}_\Jvec(h_1,\dots,h_K) = 2 \min_{\ell=1,\dots,K} \left\{(J_\ell h_\ell)_+\right\}.
\end{equation}
We will refer to this modified version as Model~O. The original version will
be called Model~A.

The distinguishing feature of Model~O is the absence of any explicit scale.
The free energy functional becomes covariant with respect to scaling transformation
(rescaling of effective fields by a factor of $\lambda$):
\begin{equation}
  \Fcal\sups{O}[\lambda P(h/\lambda)] = \lambda \Fcal\sups{O}[P(h)]. \label{scalel}
\end{equation}
The immediate consequence is that the maximum value of $\Fcal\sups{O}[P(h)]$ can be either $0$
or $+\infty$, depending on the value of $\gamma$. We can still formally write self-consistency equations
satisfied by $P(h)$ and $Q(u)$:
\begin{subequations}
\label{SelfQP}
\begin{align}
  Q(u) &= \intx{\{\dd h_\ell\}}
          \prod_{\ell=2}^K P(h_\ell)
          \times \delta \blparen u-u_\Jvec(h_2,\dots,h_K)\brparen, \label{SelfQ}\\
  P(h) &= \sum_d f_d(K\gamma) \intx{\{\dd u_k\}}
          \prod_{k=1}^d Q(u_k)
          \times \delta\biggl(h-\sum_k u_k\biggr), \label{SelfP}
\end{align}
\end{subequations}
however $u_\Jvec(h_2,\dots,h_K)$ becomes linear in $h_2,\dots,h_K$:
\begin{equation}
  u_\Jvec\sups{O} (h_2,\dots,h_K) = - J_1 \min_{\ell = 2,\dots,K}
  \left\{ \left( J_\ell h_\ell \right)_+ \right\}. \label{uJO}
\end{equation}
Under successive iterations of self-consistency equations (\ref{uJO},
the distribution quickly converges to a universal
form, with any subsequent iterations merely rescaling effective fields by a
factor of $\lambda$ that depends on the value of $\gamma$:
\begin{equation}
  P\sups{r+1} (h) = \lambda P\sups{r} (h/\lambda).
\end{equation}
It is convenient to introduce the simplified order parameter: the width
$\Delta$ of distribution $P(h)$. One possible choice for the definition of $\Delta$ is
\begin{equation}
  \Delta = \int{\dd h}P(h) |h|.
\label{Delta}
\end{equation}
Successive iterations rescale the value of $\Delta$ by a factor of $\lambda$
so that it flows towards one of two fixed points: $\Delta_{\ast} = 0$ or $\Delta_{\ast}
= + \infty$, depending on the value of $\lambda$. Since $\lambda(\gamma)$ is
expected to be monotonically increasing, we expect that $\Delta_{\ast} = 0$
whenever $\gamma < \gamma_c$ and $\Delta_{\ast} = + \infty$ for $\gamma >
\gamma_c$; the value of $\gamma_c$ being a solution to
\begin{equation}
  \lambda (\gamma_c)=1.
\end{equation}
It is easy to verify that the condition $\lambda=1$ is equivalent to
$\Fcal[P^{\ast}(h)] = 0$, where the free energy is evaluated for the
limiting distribution $P^{\ast}(h)$ rescaled so that it has finite non-zero
weight. This universal form of $P(h)$ contains a delta-function peak at $h=0$
as well as a continuous part.
\begin{equation}
  P(h) = (1 - q) \delta (h) + q \rho (h), \label{Prhoh}
\end{equation}
where $\rho (h)$ has been normalized to unity. From Eq.~(\ref{uJO}) we can
deduce the general form of $Q (u)$, which we separate into a delta function peak
at $u = 0$ and a continuous part $\chi (u)$:
\begin{equation}
  Q(u) = \left(1 - \left(\frac{q}{2}\right)^{K-1} \right) \delta(u) +
  \left(\frac{q}{2}\right)^{K-1} \chi(u).
\end{equation}
Because an iterated convolution of $\chi(u)$ is also continuous we can write
self-consistency equation for the singular part of $P(h)$ and $Q(u)$ only:
\begin{equation}
  q=1 - \exp \Bigl(-\frac{K\gamma}{2^{K-1}}q^{K-1} \Bigr).
  \label{hyper}
\end{equation}
A similar equation appears in the analysis of the leaf-removal algorithm for random \K-XOR-SAT
\cite{Mezard:2003Ricci}. The correspondence becomes exact with the replacement $\gamma / 2^{K-1} \to \gamma$.
Equation (\ref{hyper}) admits two solutions: a trivial solution $q = 0$ as well as a non-trivial solution
$q>0$ which appears discontinuously above some threshold. For $K = 3$ the critical value of $\gamma$ is
$\gamma_q \approx 4 \times 0.818 \approx 3.272$. This threshold is irrelevant for our problem, because the
corresponding $\lambda(\gamma_q)<1$ and $P(h)=\delta(h)$ maximizes the free energy.

\begin{figure}[!h]
  \includegraphics[width=3.4in]{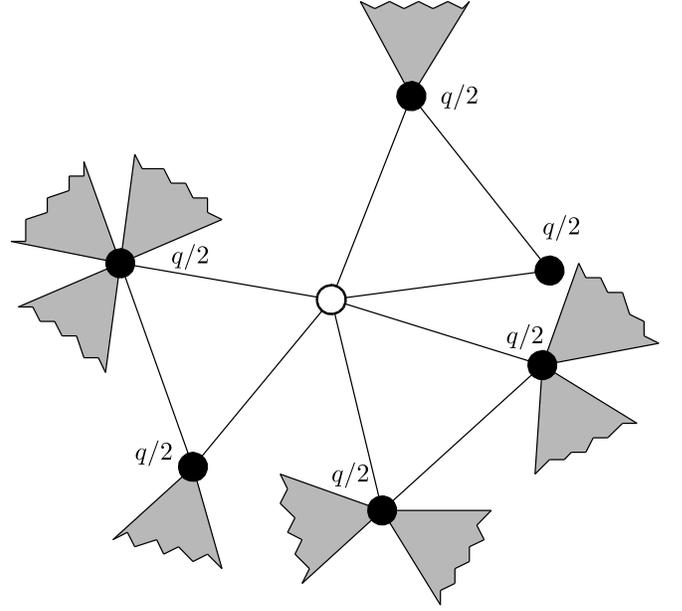}
  \caption{Hypergraph with $K=3$. Cavity vertex (white circle) has $d=3$
  (as shown in this picture) hyperedges incident to it. Each hyperedge connects the
  cavity vertex to $K-1=2$ neighboring vertices (black circles).
  In the absence of a  cavity vertex (and incident hyperedges), each of
  the neighboring vertices would be \emph{almost} frozen with probability $q/2$.
  The cavity vertex is almost frozen with probability $q$. The self-consistency
  condition on $q$ that takes into account the Poisson distribution of degrees $d$
  is given by Eq.~(\ref{hyper}).
\label{fig:hyper}}
\end{figure}

Equation (\ref{hyper}) has the following interpretation. We identify $q$ with the fraction of
\emph{almost} frozen variables: variables that take the same value for all configurations with the lowest
energy except for an exponentially small fraction. We randomly choose a spin variable and the
corresponding vertex in associated hypergraph (call it a cavity vertex). The degree of this vertex (the
number of hyperedges incident to it) is Poisson-distributed with mean $K\gamma$. Each hyperedge connects
the cavity vertex to $K-1$ neighbors (see Fig.~\ref{fig:hyper}). Each of these vertices corresponds to
another almost frozen spin with probability $q$; since we expect that spins are equally likely to be
frozen to $+1$ and $-1$, with probability $(q/2)^{K-1}$ the effect of the corresponding constraint is to
force the cavity spin to have a value of $-J_1$ in all but an exponentially small fraction of
configurations with the lowest energy. The cavity spin will be almost frozen if this happens for at least
one hyperedge. Since the cavity spin has been chosen randomly this probability equals $q$, which leads to
the self-consistency condition (\ref{hyper}). Observe that our insistence on variable being \emph{almost}
frozen rather than \emph{completely} frozen is crucial. It may happen that two or more constraints satisfy
the condition described above, and exactly half of constraints have $J_1=+1$ while the other half have
$J_1=-1$. The net effect is that the cavity variable remains unfrozen as a result. When we speak of almost
frozen variables, we assume that in the thermodynamic limit it is unlikely that an exact cancellation takes place,
i.e. that the number of configurations with $s_0=+1$ and $s_0=-1$ is roughly the same (neither is
exponentially smaller than the other). This simplified picture should work if the number of completely
frozen variables is much smaller than the number of almost frozen variables.

\begin{figure}[!h]
  \includegraphics{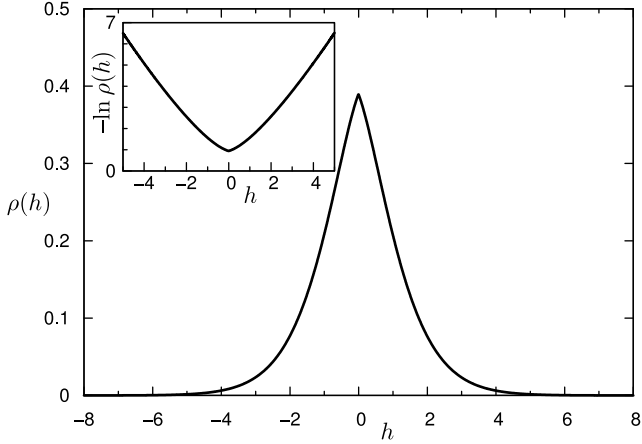}
  \caption{Scale-free solution $\rho^{\ast}(h)$ of equations (\ref{chirho}),
  for $\tilde{\gamma} = \tilde{\gamma}_c$. Inset shows $\ln \rho (h)$ to
  illustrate the approximately exponential decrease
  of $\rho (h)$ as $h \to \infty$.
\label{fig:rho}}
\end{figure}

Once the nontrivial solution $q > 0$ to $(\ref{hyper})$ is found (there will be two solutions $q > 0$, but
only the larger one is stable), we can write the self-consistency equations for the continuous parts
$\rho(h)$ and $\chi(u)$:
\begin{subequations}
\label{chirho}
\begin{align}
  \chi(u) &= (K-1)\rho(u) \biggl( \intxlim{|u|}{+\infty}{\dd h} \rho(h) \biggr)^{K-2},
  \label{chi} \\
  \rho(h/\tilde{\lambda}) &= \sum_{d=1}^{\infty} \frac{\tilde{\gamma}^d}{\ee^{\tilde{\gamma}}-1}
     \intx{\{\dd u_k \}} \prod_{k=1}^d \chi(u_k) \times \delta \biggl(h-\sum_k u_k\biggr)
  \label{rho}
\end{align}
\end{subequations}
(the ``renormalized'' connectivity is $\tilde{\gamma} = K \gamma q^{K-1}$).
These equations can be solved iteratively. The value of
$\tilde{\lambda}$ is estimated after each iteration by rescaling $\rho(h)$ so that $\int{\dd h} \rho (h)
|h|=1$. The critical value $\gamma_c$ is found from $\tilde{\lambda}(\tilde{\gamma}_c)=1$. We obtain
$\tilde{\gamma}_c \approx 3.1650$  which translates into $\gamma_c \approx 4.6002$. This agrees with the
threshold value reported in Ref.~\cite{Monasson:1997}. In Fig.~\ref{fig:rho} we plot $\rho^{\ast}(h)$
at the critical value of $\gamma$.

\subsection{Continuous phase transition in classical $T=0$ \K-SAT\label{sec:ModelA}}

Armed with the solution to Model~O obtained in the previous section, we can make qualitative predictions
about the solution to classical $T = 0$ \K-SAT (Model~A). The self-consistency equations for Model~O and
Model~A differ only in the expression for $u_\Jvec (h_2,\dots,h_K)$:
\begin{equation}
  u_\Jvec\sups{A} (h_2,\dots,h_K) = - J_1 \min \bigl( 1,
    \bigl| u_\Jvec\sups{O} (h_2,\dots,h_K) \bigr| \bigr).
\end{equation}
In the limit $\Delta \ll 1$ both expressions are nearly equal, since $\Delta$
defines the scale of $h_2,\dots,h_K$. As a result, we expect that whenever
$\gamma < \gamma_c\sups{O}$, successive iterations of self-consistency equations
for Model~A flow toward $\Delta \to 0$. The fact that $\bigl|u_\Jvec\sups{A}\bigr|
< \bigl|u_\Jvec\sups{O}\bigr|$ will only accelerate the process. Conversely, when
$\gamma > \gamma_c\sups{O}$, the fact that $\bigl|u_\Jvec\sups{A}\bigr| < 1$
prevents the divergence of $\Delta$, which will stabilize at $\Delta=O(1)$.
The value of $\gamma = \gamma_c\sups{O}$ is the boundary between the satisfiable
($\Delta=0$) and the unsatisfiable ($\Delta>0$) phases in Model~A.

The dependence of the order parameter $\Delta$ on connectivity $\gamma$ in the
vicinity of phase transition $|\gamma-\gamma_c| \ll 1$ can be estimated
variationally. We expect that $\Delta$ increases continuously from the value
of $0$ at $\gamma = \gamma_c$. Right above the transition, in the limit
$\Delta \ll 1$, Model~O and Model~A are essentially equivalent.
For the variational ansatz for $P(h)$, we choose the scale-invariant solution
of Model~O, $P^{\ast}_{\Delta}(h)$, corresponding to $\gamma=\gamma_c$.
The width $\Delta$ of the distribution appears explicitly and is the adjustable
parameter. Exactly at $\gamma=\gamma_c$, the free energy of Model~O is degenerate
($F\sups{O}=0$). For Model~A, this degeneracy is lifted, and we can obtain the
width of distribution $\Delta$ by optimizing
\begin{equation}
\begin{split}
  V\subs{var}\sups{A}(\Delta) &= \intxlim{0}{1}{\dd h}
    \left( \intxlim{h}{\infty}{\dd h'} P^{\ast}_\Delta(h') \right)^K \\
  &= \Vcal\sups{O}[P_\Delta(h)] - \intxlim{1}{\infty}{\dd h}
    \left( \intxlim{h}{\infty}{\dd h'} P^{\ast}_\Delta(h') \right)^K.
\end{split}
\label{Vvar}
\end{equation}
The quasientropy is independent of the choice of the model:
$S\subs{var}\sups{A}(\Delta) = \Scal\sups{O}[P^{\ast}_\Delta(h)]$.
Observe that
\begin{equation}
  \Scal\sups{O}[P^{\ast}_\Delta(h)] = \gamma_c \Vcal\sups{O}[P^{\ast}_\Delta(h)].
\end{equation}
The asymptotic form of $P^{\ast}_\Delta(h)$ is related to that of $\rho^{\ast}(h)$
(see Fig.~\ref{fig:rho}, inset):
\begin{equation}
  \rho^{\ast}(h) \propto \ee^{- \mu(h) |h|},
\end{equation}
where $\mu (h)$ is a function of very slow growth. In particular, it grows slower than iterated
logarithm of $|h|$. Therefore, in the limit $\Delta \ll 1$, the correction term in (\ref{Vvar}) scales as
$\ee^{-\mu(1/\Delta) / \Delta}$.

The variational free energy may be written as follows:
\begin{equation}
  F\subs{var}(\Delta) \approx \alpha (\gamma - \gamma_c) \Delta - \ee^{-\mu(1/\Delta) / \Delta}.
\end{equation}
Solving $\dd F\subs{var} / \dd \Delta = 0$ with respect to $\Delta$ yields
\begin{equation}
  \gamma - \gamma_c \propto \frac{1}{\Delta^2} \ee^{-\mu(1/\Delta) / \Delta}.
\label{gammaDelta}
\end{equation}
With some abuse of notation (we write $x \sim y$ to mean that $x$ is asymptotically
proportional to $y$ with the coefficient of proportionality being an extremely
slow-varying function of $y$), the dependence of the order parameter $\Delta$ on
connectivity $\gamma > \gamma_c$ may be written as follows:
\begin{equation}
  \Delta \sim \frac{1}{\lvert \ln(\gamma-\gamma_c) \rvert}.
\end{equation}
Given the extremely singular character of this function, it is not surprising that the transition
looks like a first order transition in numerical simulations. Critical exponents $\alpha=1$, $\beta = 0$
are precisely those expected for the first order transition (the scaling exponent associated with the
logarithm is zero). In the vicinity of the phase transition, just above it, the behavior of the free energy
is
\begin{equation}
  F \sim \frac{\gamma-\gamma_c}{\lvert \ln (\gamma-\gamma_c) \rvert}.
\label{Fsim}
\end{equation}

\begin{figure}[!h]
  \includegraphics[width=3.4in]{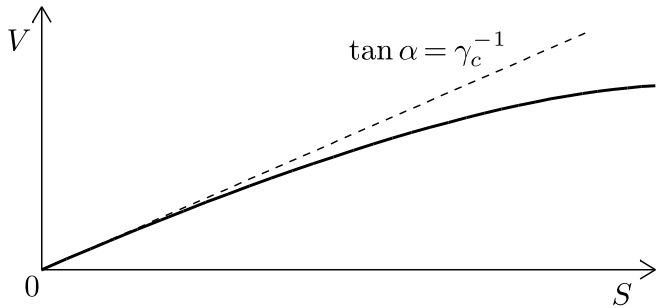}
  \caption{Quasipotential as a funtion of quasientropy in Model A. Tangentials to the curve
  have local slope $\gamma^{-1}$. There are no tangentials with slopes corresponding
  to $\gamma < \gamma_c$, hence $S=V=0$ in this region.
\label{fig:VS}}
\end{figure}

Let us briefly describe the mechanism of this continuous phase transition. In the classical case, the free
energy is the maximum of the free energy functional [over symmetric distributions $P(h)$] as demonstrated
in Appendix~\ref{app:Classical}:
\begin{equation}
  F(\gamma) = \max_{P(h)} \bigl\llb \gamma \Vcal[P(h)] - \Scal[P(h)] \bigr\rrb.
\end{equation}
It is convenient to regard the reciprocal of the connectivity $1/\gamma$ as a Lagrange multiplier for
$\Scal[P(h)]$. We consider the quasipotential $V$ as a formal function of the quasientropy $S$:
\begin{equation}
  V(S) = \max_{P(h)} \bigl\llb \Vcal[P(h)] \big| \Scal[P(h)]=S \bigr\rrb.
\end{equation}
In Fig.~\ref{fig:VS} we sketch the function $V(S)$, which should be convex.

Via newly defined $V(S)$, the free energy $F(\gamma)$ can be expressed as follows:
\begin{equation}
  F(\gamma) = \max_S[\gamma V(S) - S]. \label{FVS}
\end{equation}
The value of $S$ that maximizes the right hand side of (\ref{FVS}) is a solution to
\begin{equation}
  \gamma^{-1} = \dd V(S) / \dd S, \label{recipgamma}
\end{equation}
for $\gamma \geqslant \gamma_c$. When $\gamma < \gamma_c$, Eq.~(\ref{recipgamma})
has no solutions and the r.h.s. of (\ref{FVS}) is maximized by $S=V=0$.

Iterating self-consistency equations (\ref{SelfQP}) for various
values of connectivity $\gamma$ becomes extremely time-consuming in the
vicinity of the phase transition. We have verified the general trend that
iterations converge to $\Delta=0$ for $\gamma<\gamma_c$ and to a finite
value of $\Delta$ for $\gamma>\gamma_c$. To find the equation for the order
parameter $\Delta(\gamma)$ numerically, we took a different route. Instead of
fixing the value of $\gamma$ and iterating equations until convergence, we fix
the width $\Delta$ of probability distribution $P(h)$ and choose the value of
$\gamma$ at each iteration step so that $P(h)$ has the desired width
$\Delta$. In fact, by making an appropriate choice for the somewhat arbitrary
definition of $\Delta$, the corresponding value of $\gamma$ may be obtained at
every step with just a single arithmetical operation. One such choice ---
$\tilde{\Delta} = \intx{\dd h}P(h) h^2$ --- exploits the identity
\begin{equation}
  \intx{\dd h} P(h)h^2 = K\gamma \intx{\dd u} Q(u)u^2.
\end{equation}
The above-described approach results in tremendous speed-up. While the number of
iterations required for convergence for fixed $\gamma$ increases to infinity
as $\gamma$ approaches $\gamma_c$, for fixed $\Delta$ a complete convergence
(to machine precision limit) is achieved within 20 iterations.

Additionally, we utilize a Quasi Monte Carlo (QMC) method \cite{Niederreiter}
that we have formulated specifically for the
problems involving probability distributions. Under ideal conditions, the expected error for Quasi Monte
Carlo (QMC) is $O\bigl( \log N \big/ N \bigr)$ compared to $O\bigl( 1 \big/ \sqrt{N} \bigr)$ for standard
Monte Carlo (MC). In practice, discontinuities and singularities worsen the error estimate, however the
asymptotic behavior of QMC is always better. For values of $N$ as small as $N = 8192$ the error in the
value $\gamma$ obtained using standard Monte Carlo was $1\%$, while Quasi Monte Carlo produced an error of
$0.05\%$. The advantage of QMC over MC only increases with increasing $N$. We have used values of $N$ up
to $N = 2^{25} = 33554432$. An overview of our method is presented in Appendix~\ref{app:QMC}.
\begin{figure}[!h]
  \includegraphics{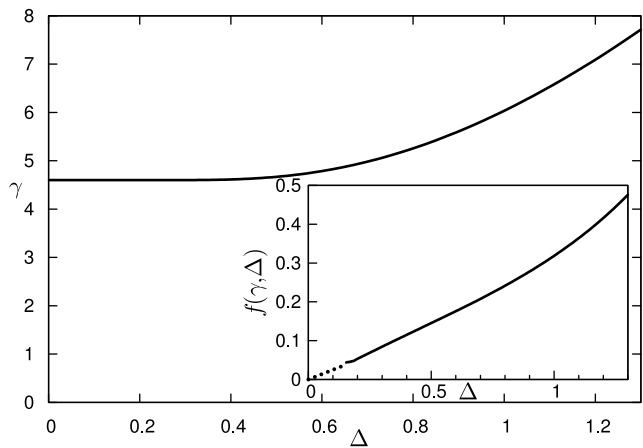}
  \caption{Connectivity $\gamma$ vs. width of the distribution of effective
  fields $\Delta$ in Model A.
  We predict that $\gamma>\gamma_c$ for any $\Delta>0$, so $\Delta(\gamma)$ has
  no discontinuities. The inset replots the data to illustrate the asymptotic relation
  (\ref{gammaDelta}). The $\hat y$ axis corresponds to
  $f(\gamma,\Delta)=[3.5-\ln \Delta^2(\gamma-\gamma_c)]^{-1}$,
  which is asymptotically linear in $\Delta$ as $\Delta \to 0$.
  Dotted line is the result of extrapolation to small values of $(\gamma-\gamma_c)$
  where the numerical error is too large.
\label{fig:A1}}
\end{figure}

\begin{figure}[!h]
  \includegraphics{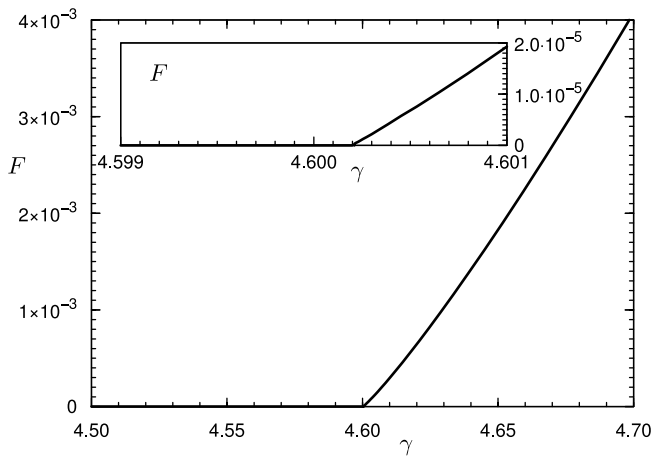}
  \caption{Free energy $F$ vs. connectivity $\gamma$. Contrary to visual perception,
  $F(\gamma)$ is \emph{not} linear for $\gamma >~ \gamma_c$. The inset zooms
  in on the transition, keeping the aspect ratio the same. The apparent slope of
  $F(\gamma)$ in the inset is smaller. This apparent slope will tend to zero as
  progressively higher zoom ratios are used.
\label{fig:A2}}
\end{figure}

In Fig.~\ref{fig:A1}, we plot the function $\gamma(\Delta)$ obtained numerically.
To establish that the phase transition is continuous, we must convince ourselves
that $\gamma(\Delta)$ is a strictly increasing function of $\Delta$, i.e. that
$\gamma(\Delta) > \gamma(0)$ for arbitrarily small $\Delta$. The inset shows the roughly
linear dependence of $f(\gamma,\Delta)=[C-\ln \Delta^2(\gamma-\gamma_c)]^{-1}$
on $\Delta$ (cf. Eq.~\ref{gammaDelta}) that, when extrapolated, predicts the vanishing of $\Delta$
as $\gamma \to \gamma_c$.

In Fig.~\ref{fig:A2}, we plot the dependence of the free energy $F$ on connectivity
$\gamma$ in the vicinity of $\gamma_c$. Contrary to visual perception the slope of
$F(\gamma)$ at $\gamma = \gamma_c + 0$ is zero from Eq.~(\ref{Fsim}). The ``apparent''
slope $\Delta F/\Delta\gamma$ decreases as a function of $\Delta$ as can be seen
by comparing the main figure with the inset in Fig.~\ref{fig:A2}.

\section{Numerical results\label{sec:Numerical}}

\subsection{Classical regime ($\Gamma = 0$, $T > 0$)\label{sec:Classical}}

In addition to Model~O and Model~A described in Sec.~\ref{sec:CT0}, we
introduce two new models: Classical Model~B and Classical Model~AB.
Classical Model~AB is precisely the finite temperature classical random \K-SAT.
The distinguishing feature of Model~B is the absence of explicit temperature.
It is defined using Eqs.~(\ref{VPhm}), (\ref{SPhm}), but with the following choice
for $\Ucal_\Jvec(\{h_\ell\})$ and $\Lcal(h)$ respectively:
\begin{subequations}
\label{ULTL}
\begin{equation}
  \Ucal\sups{B} = - \ln \left( 1 - \frac{1}{(1+\ee^{-2 J_1 h_1})\dotsb(1+\ee^{-2 J_K h_K})} \right),
\end{equation}
\begin{equation}
  \Lcal\sups{B} = \ln(2\cosh h).
\end{equation}
\end{subequations}
All four models (O,A,B and AB) can be described by a single form of the free energy functional that
depends explicitly on two parameters: the temperature $T=1 / \beta$ and the energy scale parameter
$\Lambda$. This common model can be defined using the following expression for $\Ucal_\Jvec(\{h_\ell\})$
and $\Lcal(h)$:
\begin{subequations}
\begin{equation}
  \Ucal_{T,\Lambda} = -T \ln
    \left( 1 - \frac{1-\ee^{-2\Lambda/T}}{\prod_{\ell=1}^{K} \bigl(1+\ee^{-2J_\ell h_\ell/T}\bigr)}
    \right),
\end{equation}
\begin{equation}
  \Lcal_{T,\Lambda} = T \ln \left( 2\cosh\frac{h}{T} \right).
\end{equation}
\end{subequations}
We summarize the values of $T$ and $\Lambda$ for the four models we have introduced in
Table~\ref{tab:ModelsC}.

\begin{table}[!h]
  \caption{Four different models defined by the values taken by parameters
  $T$, $\Lambda$. Statistical properties of Models O, A, and B should be
  similar, since $T/\Lambda = 0$ in all three cases. Model~AB is the
classical
  finite-temperature random $K$-SAT.\label{tab:ModelsC}}
\begin{ruledtabular}
  \begin{tabular}{lll}
   Type of Model & Temperature & Scale $\Lambda$\\
    \hline
    Model~O & $T=0$ & $\Lambda=\infty$\\
    Model~A & $T=0$ & $\Lambda=1$\\
    Model~B (classical) & $T=1$ & $\Lambda=\infty$\\
    Model~AB (classical) & $T>0$ & $\Lambda=1$\\
    \end{tabular}
  \end{ruledtabular}
\end{table}

This common model with explicit dependence on $T$,$\Lambda$ satisfies the following scaling relations:
\begin{subequations}
\begin{align}
  \Fcal_{T,\Lambda}[P(h)] &= T \Fcal_{T=1,\Lambda}[T P(h/T)], \\
  &= \Lambda \Fcal_{T,\Lambda=1}[\Lambda P(h/\Lambda)].
\end{align}
\end{subequations}
The implication is that the statistical-mechanical properties of this model
depend on the ratio of two scales $T/\Lambda$. In particular, we expect
that Model~O, Model~A and Model~B undergo phase transition at the same value
of the critical connectivity, since $T/\Lambda=0$ in all three models.
We have previously established that $\gamma_c\sups{A}=\gamma_c\sups{O}$.

\begin{figure}[!h]
  \includegraphics{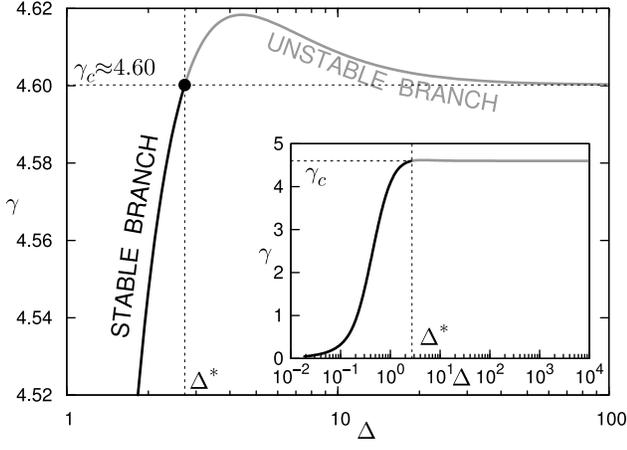}
  \caption{Connectivity $\gamma$ vs. width of distribution of effective fields
  $\Delta$ in Classical Model~B. The main figure shows $\gamma(\Delta)$ in the region close
  to critical. The inset shows $\gamma(\Delta)$ in a wider range.
  The branch $\Delta > \Delta^{\ast}$ is unstable, hence $\Delta=+\infty$ as soon
  as $\gamma>\gamma_c$. The unstable branch is drawn with the gray solid line.
\label{fig:CB1}}
\end{figure}

\begin{figure}[!h]
  \includegraphics{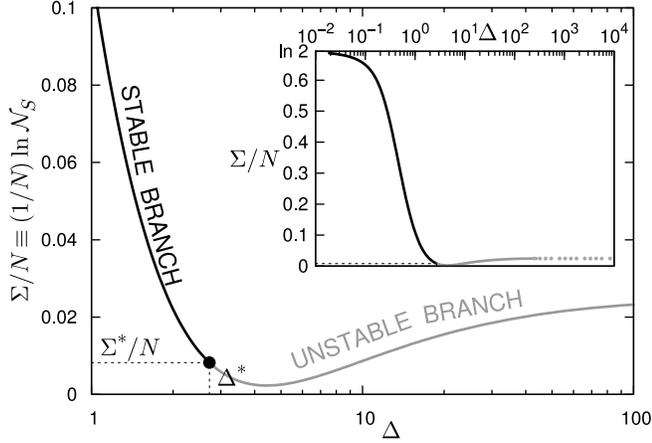}
  \caption{Specific entropy $\Sigma/N$ (logarithm of the number of solutions)
  as a function of $\Delta$. The main figure shows the dependence in the critical
  region, while a wider range of $\Delta$ is used for the inset.
  The unstable branch ($\Delta > \Delta^{\ast}$) is drawn in gray.
  The entropy decreases to $\Sigma^{\ast}$ as $\gamma$ approaches $\gamma_c$,
  but jumps to $-\infty$ (corresponding to zero solutions) for $\gamma>\gamma_c$.
  Entropy corresponding to very large values of $\Delta$ in the unstable branch
  could not be determined with good precision. Results of extrapolation are
  indicated using the dotted gray line.
\label{fig:CB2}}
\end{figure}

\begin{figure}[!h]
  \includegraphics[width=3.4in]{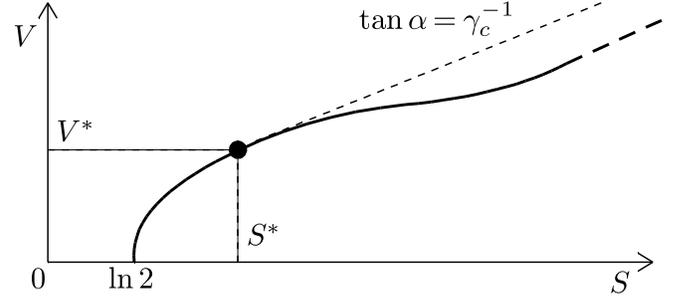}
  \caption{Illustration of the non-convex behavior of $V(S)$ for Classical Model~B.
  Slopes of tangentials to the curve determine values of $\gamma$ via Eq.~(\ref{recipgamma}).
  The largest attainable values of the quasipotential and the quasientropy,
  $V^{\ast}$ and $S^{\ast}$ correspond to $\gamma=\gamma_c$.
\label{fig:non-convex}}
\end{figure}

The numerical results for the Classical Model B are presented in Figs.~\ref{fig:CB1} and \ref{fig:CB2}. To
obtain the numerical solution, we adopted the same strategy as for Model~A. We computed $\gamma$ as well
as a number of other quantities for each value of $\Delta$. An interesting feature of Model~B is that the
function $\gamma(\Delta)$ plotted in Fig.~\ref{fig:CB1} is non-monotonic and cannot be inverted
unambiguously for $\gamma > \gamma\sups{B}(+\infty)$. Although not reflected in the figure; formally there
exists another solution corresponding to $\Delta=+\infty$, with the free energy $F=+\infty(-\infty)$ for
$\gamma>\gamma_c\sups{O}(<\gamma_c\sups{O})$. Since a branch with the higher free energy must be chosen
(see Appendix~\ref{app:Classical}), the branch $\Delta^{\ast} < \Delta < +\infty$ is unstable and
$F\sups{B} = +\infty$ for $\gamma>\gamma_c\sups{O}$. This behavior may be understood in terms of the form
of function $V(S)$. The discontinuous transition occurs because it has non-convex form as sketched in
Fig.~\ref{fig:non-convex}

Whereas the free energy $F\sups{A}$ of Model~A (see Sec.~\ref{sec:MZ}) corresponds to the internal energy
$\Epsilon$ of Eq.~(\ref{FETS}) (or the number of violated constraints); the free energy of Model~B is
related to entropy $\Sigma$:
\begin{equation}
-N F\sups{B} = \langle \Sigma \rangle = \langle \ln \Ncal_S \rangle,
\end{equation}
where $\Ncal_S$ represents the number of solutions that satisfy all constraints. The divergence of
$F\sups{B}$ signals the transition to the unsatisfiable phase ($\Ncal_S=0$).

In Fig.~\ref{fig:CB2} we plot the specific entropy (i.e. the negative of the free energy) as a function of
$\Delta$. Note that since the entropy is finite at $\gamma=\gamma_c$, the number of solutions, just prior
to the satisfiability transition, is exponentially large. It is the expected behavior: for the associated
hypergraph,  random graph theory \cite{Bollobas} predicts that there are $O(N)$ vertices that are either
isolated or belong to small isolated clusters. These make a finite contribution to the entropy but do not
affect the overall satisfiability of the random instance.

\begin{figure}[!h]
  \includegraphics{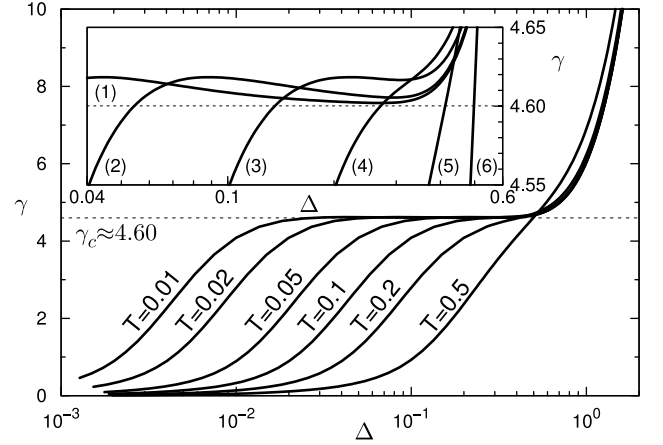}
  \caption{The dependence $\gamma(\Delta)$ in Classical Model~AB for a range
  of temperatures. Curves labeled (1)--(6) in the inset correspond
  to temperatures $T=0.01$, $0.02$, $0.05$, $0.1$, $0.2$, and $0.5$ respectively.
  The temperatures are labeled explicitly in the main figure.
  The curves smoothly interpolate between the regime of Model~B [$\Delta=O(T)$]
  and that of Model~A [$\Delta=O(1)$]. The inset shows that $\gamma(\Delta)$ is not
  monotonic for sufficiently small temperatures.
\label{fig:CAB1}}
\end{figure}

Based on results for Model~B we expect that the non-monotonic behavior of $\gamma(\Delta)$ persists for
some sufficiently small but finite temperatures. In Fig.~\ref{fig:CAB1} we plot the functions $\gamma(\Delta)$
for the classical Model~AB for a range of temperatures from $T=0.01$ to $T=0.5$. It is seen that far away
from $\gamma=\gamma_c$ Model~AB interpolates between the regimes of Model~B with $\Delta=O(T)$ and Model~A
with $\Delta=O(1)$. For small temperatures $\gamma(\Delta)$ is non-monotonic, which gives rise to the
first-order phase transition.

\begin{figure}[!h]
  \includegraphics{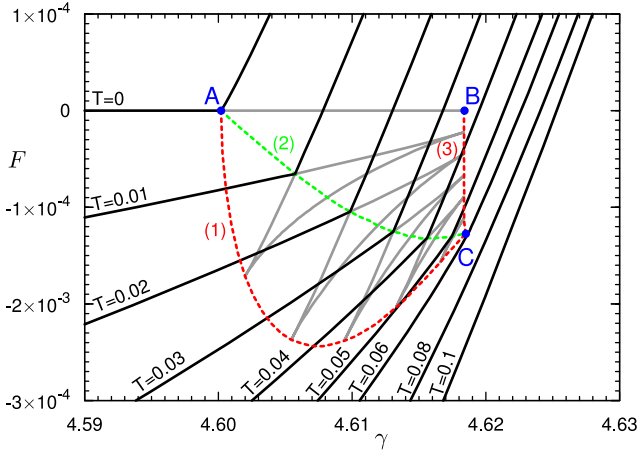}
  \caption{
Two-dimensional parametric plots $\blparen \gamma(\Delta),F(\Delta)\brparen$ for the Classical Model AB
(finite temperature $K$-SAT). Different lines correspond to different temperatures. Black solid lines
correspond to the stable branches of the free energy; gray lines correspond to unstable solutions.
Switching between stable branches occurs along the green dashed line (2). Along this line, the first
derivative $\dd F / \dd \gamma$ of the free energy has a discontinuity. Red dashed lines (1) and (3) are
the spinodals $\dd V / \dd S = 0$. Points A and B along $T=0$ line correspond, respectively, to the
critical threshold in Model A and the metastable solution $\gamma \approx 4.6184$ of Model B. Lines
(1),(2), and (3) meet at a critical point C, corresponding to $T^{\ast} \approx 0.05864$. Note that the line BC
is \emph{nearly} vertical, with corrections that are $\sim \exp(-1/T)$.
 \label{fig:CAB2}}
\end{figure}
We make two dimensional parametric plots $\blparen \gamma(\Delta),F(\Delta)\brparen$ for a range
of temperatures $T$ (see Fig.~\ref{fig:CAB2}). For $T<T^{\ast}$ these curves are self-intersecting. Stable
branches (black solid lines) have a discontinuous slope at the point of self-intersection which leads to
the discontinuity of the order parameter. The  dashed green line (marked $(2)$ in the figure) is the line
of singularities between $(\gamma_c,0)$ and $(\gamma^{\ast},F^{\ast})$ terminating at the critical point.
In the space of variables $(\gamma,T,F$)  stable and unstable branches of $F$ form a dovetail
  singularity. It should be recalled that for $T=0$ the derivative $\dd F/\dd \gamma$ has no discontinuity, although it
is difficult to see from the figure.

\begin{figure}[!h]
  \includegraphics{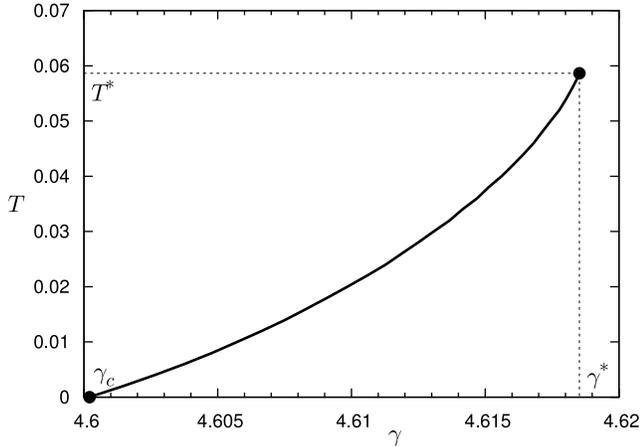}
  \caption{Numerical phase digram of the Classical Model~AB
  (finite temperature $K$-SAT). The phase boundary (phase transition line)
  starts from $T=0$, critical connectivity $\gamma_c$ and terminates at the
  critical point with $\gamma^{\ast} \approx 4.6185$ and $T^{\ast} \approx 0.05864$.
\label{fig:CABpd}}
\end{figure}

Finally, in Fig.~\ref{fig:CABpd}, we show the numerical phase diagram in the plane $(\gamma,T)$.
The discontinuity of the order parameter becomes zero at both ends of the phase boundary between
$(\gamma_c,0)$ and $(\gamma^{\ast},T^{\ast})$.

\subsection{Quantum regime ($\Gamma>0$, $T=0$)\label{sec:Quantum}}

We introduce Quantum Model~B and Quantum Model~AB as follows. We will keep the definition (\ref{SPhh}) for the
quasientropy, but in the definition (\ref{VPhh}) for the quasipotential the function $\eta(h,\varepsilon)$
will be replaced with
\begin{equation}
  \eta_{\Lambda}(h,\varepsilon) = \min \blparen\Lambda,(h)_+\brparen
  + \frac{1}{2}(\varepsilon-|h|)_+ - \frac{1}{2} (\varepsilon - |h-\Lambda|)_+
\end{equation}
so that the free energy functional will contain a characteristic scale of
the effective fields $\Lambda$ explicitly.
By choosing the values of $\Gamma$ and $\Lambda$ according to Table~\ref{tab:ModelsQ}
we define the two quantum models: Model B and Model AB. The purely classical models
--- Model~O and Model~A --- correspond to the limit $\Gamma=0$. \begin{table}[!h]
  \caption{Four different models defined by the values taken by parameters
  $\Gamma$, $\Lambda$. Statistical properties of Models O, A, and B
should be
  similar, since $\Gamma / \Lambda = 0$ in all three cases. Model~AB is
  quantum random $K$-SAT; we study the limit $\Gamma \ll
1$.\label{tab:ModelsQ}}
\begin{ruledtabular}
  \begin{tabular}{lll}
    Type of Model & Transverse Field& Scale $\Lambda$\\
    \hline
    Model~O & $\Gamma = 0$ & $\Lambda = \infty$\\
    Model~A & $\Gamma = 0$ & $\Lambda=1$ \\
    Model~B (quantum) & $\Gamma=1$ &$\Lambda = \infty$\\
    Model~AB (quantum) & $\Gamma > 0$ &$\Lambda=1$\\
    \end{tabular}
  \end{ruledtabular}
\end{table}

Ordinarily, the free energy of the quantum model corresponds to the smallest eigenvalue of the
Hamiltonian. However, in the limit $\Lambda = +\infty$, where $\Lambda$ defines the energy scale of the
classical Hamiltonian, the contribution  from the states with energy $\Epsilon>0$ vanishes. The value
$F\sups{B}$ of the free energy for an instance of Model~B may be evaluated using the degenerate perturbation
theory. In the limit $\Lambda=+\infty$, this free energy is proportional to $\Gamma$, which can also be
seen from scaling analysis. We choose $\Gamma=1$; the free energy $F\sups{B}$ is directly related to a
property of the space of solutions $\boldsymbol\sigma$.

Consider a graph $\Gcal$ having $\Ncal_S$ vertices corresponding to
spin configurations that satisfy all constraints. We draw edges between
vertices of $\Gcal$ corresponding to configurations $\boldsymbol\sigma,\boldsymbol{\sigma'}$
that differ by a single spin-flip. Let $\Acal$ denote the adjacency matrix for this graph, i.e.
\begin{equation}
  \Acal_{\boldsymbol\sigma\boldsymbol{\sigma'}} = \begin{cases}
    1 &\text{if $d(\boldsymbol\sigma,\boldsymbol{\sigma'})=1$,} \\
    0 &\text{if $d(\boldsymbol\sigma,\boldsymbol{\sigma'})=0$
             or $d(\boldsymbol\sigma,\boldsymbol{\sigma'})\geqslant 2$,}
  \end{cases}
\label{Acal}
\end{equation}
where $d(\boldsymbol\sigma,\boldsymbol{\sigma'})$ denotes the Hamming distance
between spin configurations $\boldsymbol\sigma$ and $\boldsymbol{\sigma'}$.
The free energy of Model~B will be related to the norm of matrix $\Acal$:
\begin{equation}
  -N F\sups{B} = \langle \| \Acal \| \rangle =
  \langle \lambda\subs{max}(\Acal_{\boldsymbol\sigma\boldsymbol{\sigma'}}) \rangle
\end{equation}
(the spectrum of $\Acal_{\boldsymbol\sigma\boldsymbol{\sigma'}}$ is symmetric).

The expression for the free energy may be simplified in the limit $\Lambda=+\infty$.
Since in this limit, $\uBar = |u|$, and $u$ is equally likely to be positive or
negative, we may express joint probability distribution $Q(u,\uBar)$ in terms
of the probability distribution of $\uBar$ denoted $Q_+(\uBar)$:
\begin{equation}
  Q(u,\uBar) = \frac{1}{2} \blparen
    Q_+(\uBar) \delta(u-\uBar) + Q_+(\uBar) \delta(u+\uBar) \brparen.
\label{Qplus}
\end{equation}
Substituting this into Eq.~(\ref{Phh}), we ontain the following factorization of the joint distribution
$P(h,\hBar)$:
\begin{subequations}
\label{Pplus}
\begin{gather}
  P(h,\hBar) = P_+\Bigl( \frac{\hBar+h}{2} \Bigr) P_+\Bigl( \frac{\hBar - h}{2} \Bigr), \\
  P_+(\hBar) = \intx{\dd\hBar} \ee^{\ii\omegaBar\hBar}
    \exp \frac{K\gamma}{2} \left( \Qbar_+(\omegaBar)-1 \right) .
\end{gather}
\end{subequations}
We can use relations (\ref{Qplus}) and (\ref{Pplus}) to write the free energy as a functional of
$P_+(\hBar)$ alone. The resulting expression for $\Fcal[P_+(h)]$ is
\begin{subequations}
\label{FBPh}
\begin{multline}
  \Fcal\sups{B}[P_+(h)] = \gamma \intx{\{\dd\eta_\ell\}} \prod_{\ell=1}^K R(\eta_\ell)
                  \times 2 \min_{\ell=1,\dots,K} \{\eta_\ell\} \\
                 -{} \intx{\dd h_1 \dd h_2} \Lcal(h_1,h_2)
                  \intx{\frac{\dd\omega_1\dd\omega_2}{2\pp}}
                     \ee^{\ii \omega_1 h_1 + \ii \omega_2 h_2} \SigmaBar(\omega_1,\omega_2),
\tag{\ref{FBPh}}
\end{multline}
where the expressions for $\Lcal(h_1,h_2)$ and $\SigmaBar(\omega_1,\omega_2)$ are
\begin{gather}
  \Lcal(h_1,h_2) = \max(|h_1-h_2|,\Gamma-\hBar), \\
  \SigmaBar(\omega_1,\omega_2) = \Pbar_+(\omega_1) \Pbar_+(\omega_2)
    \left( 1 - \ln \Pbar_+(\omega_1) - \ln  \Pbar_+(\omega_2) \right)
\end{gather}
\end{subequations}
[$\Pbar(\omega)$, as usual, denotes the Fourier transform of $P(h)$].

\begin{figure}[!h]
  \includegraphics{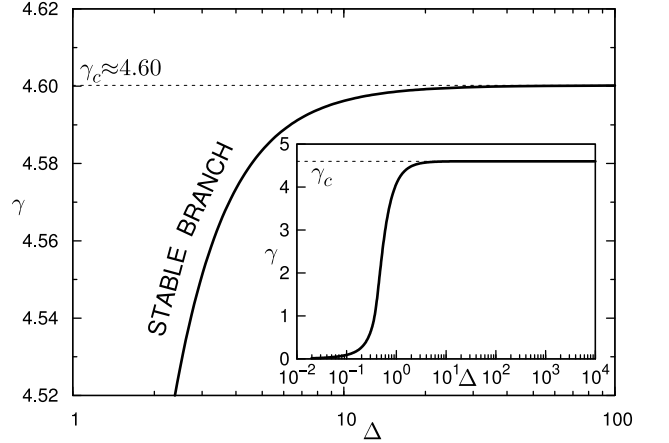}
  \caption{Connectivity $\gamma$ vs. width of distribution of effective fields
  $\Delta$ in Quantum Model~B. The main figure shows $\gamma(\Delta)$ in the
  region close to critical. Since $\gamma(\Delta)$ is monotonic, there is only a
  stable branch. The inset shows $\gamma(\Delta)$ in a wider range.
  As $\gamma$ approaches $\gamma_c$ the value of $\Delta$ increases to infinity
  conntinuously.
\label{fig:QB1}}
\end{figure}

\begin{figure}[!h]
  \includegraphics{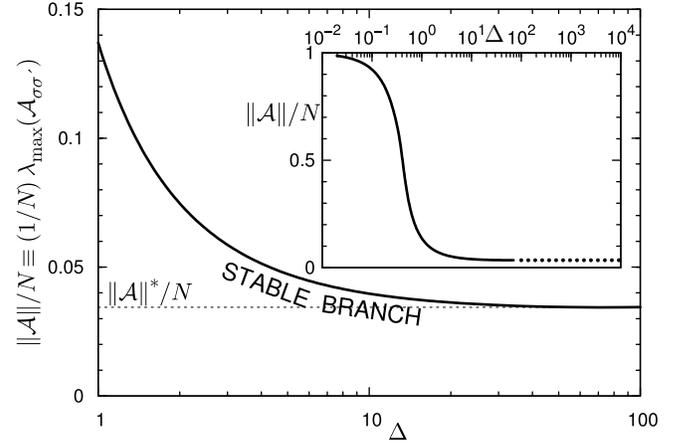}
  \caption{The norm (largest eigenvalue) of matrix $\Acal$, defined by Eq.~(\ref{Acal}),
   as a function of $\Delta$. The main figure shows the dependence in the critical
  region, while a wider range of $\Delta$ is used for the inset. There is only a stable
  branch. Values of $\|\Acal\|$ that could not be computed reliably for very large
  $\Delta$ are estimated using extrapolation, which is indicated by the use of
  black dotted line.
\label{fig:QB2}}
\end{figure}

The distribution $R(\eta)$ that enters on the r.h.s. of Eq.~(\ref{FBPh}) may
be related to $P_+(h)$ as follows:
\begin{multline}
  R(\eta) = \intx{\dd h_1 \dd h_2} P_+(h_1) P_+(h_2) \\
  \times \delta\bigl( \eta - \blparen \max(\Gamma/2,h_1) - h_2 \brparen_+ \bigr).
\label{Reta}
\end{multline}
The numerical results for Quantum Model~B are presented in Figures \ref{fig:QB1} and \ref{fig:QB2}. In
contrast to Classical Model~B, $\gamma (\Delta)$ is a monotonically increasing function of $\Delta$ (See
Fig.~\ref{fig:QB1}). Its inverse $\Delta(\gamma)$ is a single-valued function exhibiting no
discontinuities. It diverges as $\gamma$ approaches $\gamma_c$. It is fortunate that there is a single
branch, as the stability analysis is more complicated in the quantum case.

In Fig.~\ref{fig:QB2}, we plot $\| \Acal \| / N$: the norm of the matrix describing the connectivity of
solutions. It is seen that this quantity does not go to zero as $\gamma \to \gamma_c$. This can be
explained by the effect of small clusters in a random hypergraph associated with an instance. This
hypergraph is a collection of isolated clusters: a giant cluster of size $O(N)$ and a large [$O(N)$]
number of small clusters. Each cluster may be used to define a subinstance of the problem. The space of
solution of the large instance is a Cartesian product of spaces of solutions of subinstances. It can be
shown that the norm $\| \Acal \|$ for the full instance may be written as a sum of norms $\| \Acal_k \|$
for all the subinstances corresponding to isolated clusters. The large number of small clusters
contributes to the finite value of $\|\Acal\|$ as $\gamma \to \gamma_c$.

We should mention that the computed value of $F\sups{B}$ is not quantitatively correct even in
the regime where the replica-symmetric solution is stable. This is due to our making a static
approximation. Although Quantum Model~B describes the limit $\Gamma \to 0$, the static approximation
requires a stronger condition $\beta \Gamma \to 0$ in order to be exact. We, however, work in the opposite
limit $\beta\Gamma \to \infty$.
\begin{figure}[!h]
  \includegraphics{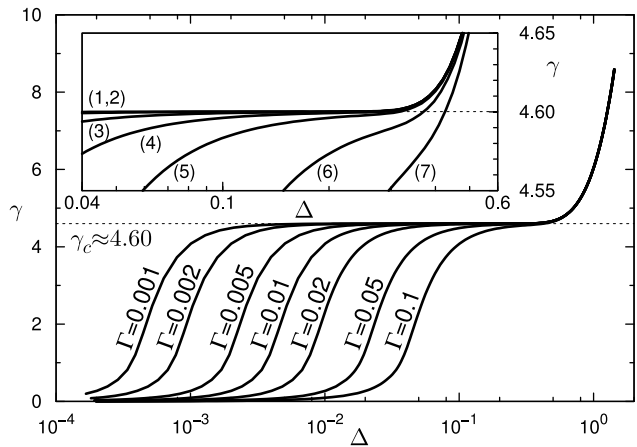}
  \caption{The dependence $\gamma(\Delta)$ in Quantum Model~AB for a range
  of transverse magnetic fields. Curves labeled (1)--(7) in the inset correspond
  to transverse fields $\Gamma=0.001$, $0.002$, $0.005$, $0.01$, $0.02$, $0.05$, and $0.1$ respectively.
  The curves in the main figure are explicitly labeled with values of $\Gamma$.
  The curves smoothly interpolate between the regime of Model~B [$\Delta=O(\Gamma)$]
  and that of Model~A [$\Delta=O(1)$]. In contrast to the classical case, functions
  $\gamma(\Delta)$ are monotonic and free of singularities.
\label{fig:QAB1}}
\end{figure}

\begin{figure}[!h]
  \includegraphics[width=3.4in]{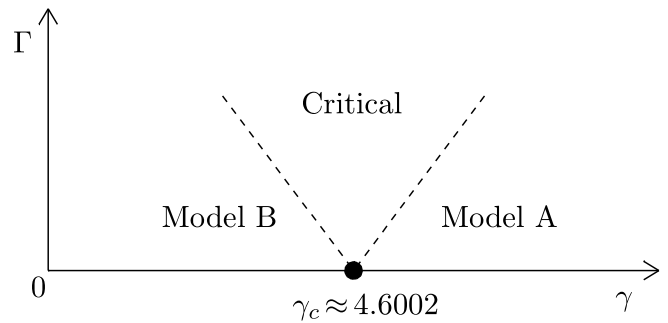}
  \caption{Illustration of crossover transition for quantum $K$-SAT. The sharp phase
  transition predicted in classical $K$-SAT is the critical point at $\Gamma = 0$.
  For small but non-zero values of $\Gamma$ it is replaced by a smooth
  crossover transition of finite width between the underconstrained ($\gamma < \gamma_c$)
  regime described by Model~B and the overconstrained ($\gamma > \gamma_c$)
  regime described by Model~A. The width of the critical region decreases as $\Gamma \to 0$.
\label{fig:crossover}}
\end{figure}

\begin{figure}[!h]
  \includegraphics{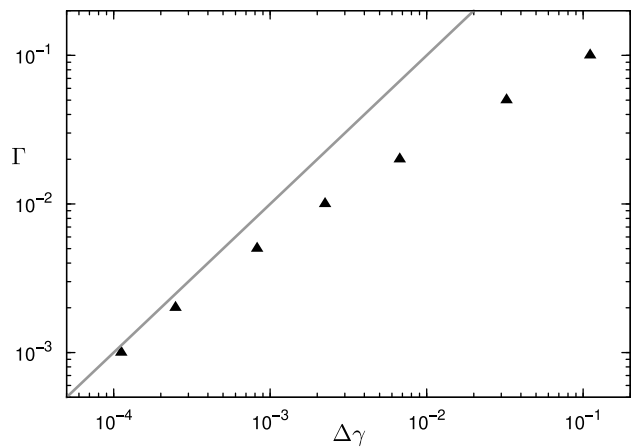}
  \caption{Transverse field $\Gamma$ vs. the width of the transition $\Delta\gamma$.
  Log-log scale is used to obtain a power-law fit between the width of the transition
  and the transverse field $\Gamma = (\Delta\gamma)^z$. Filled triangles correspond to
  numerical estimates of $\Delta\gamma$ for the following values of $\Gamma$:
  $0.001$, $0.002$, $0.005$, $0.01$, $0.02$, $0.05$, and $0.1$.
  The slope of the gray solid line corresponds to $z=1$.
\label{fig:QABsc}}
\end{figure}

Numerical results for Quantum Model~AB are presented in Fig.~\ref{fig:QAB1}.
We plot $\gamma(\Delta)$ for transverse field $\Gamma$ ranging from
$\Gamma = 0.001$ to $\Gamma = 0.1$. It can be seen that
functions $\gamma(\Delta)$ are always monotonic. In contrast to Classical
Model~AB, the free energy does not exhibit non-analytic behavior.
The continuous phase transition present for $\Gamma = 0$ disappears and is
instead replaced by a smooth crossover for arbitrarily small
$\Gamma > 0$ as depicted in Fig.~\ref{fig:crossover}.
The effect of the critical point $(\gamma=\gamma_c,\Gamma=0)$ is that the width
of the transition $\Delta\gamma$ goes to zero together with $\Gamma$.
We conjecture, by analogy with quantum phase transitions in physical systems, that the
characteristic width of the transition scales as some power of~$\Gamma$,
\begin{equation}
  \Delta\gamma \propto \Gamma^{1/z}.
\label{powerlaw}
\end{equation}
where the width of the transition has been formally defined as follows:
\begin{equation}
  \Delta\gamma = \max_\gamma \Bigl[ \frac{1}{\Delta} \frac{\dd\gamma}{\dd\Delta} \Bigr].
\end{equation}
This power law may be verified by plotting points ($\Delta\gamma$ and $\Gamma$) on a
log-log plot (see Fig.~\ref{fig:QABsc}).
For small $\Gamma$, the data seems to converge to power-law scaling with scaling exponent
$z=1$ (the slope corresponding to $z=1$ is indicated with the gray solid line).
However, we have not studied this scaling dependence analytically and cannot completely
rule out the possibility that the dependence of the width of the transition on $\Gamma$
is more complex and cannot be described by a simple power law.

\section{Conclusion\label{sec:Conclusion}}

The main result of this paper is that the thermodynamic phase transition
between SAT and UNSAT phases in classical random \K-Satisfiability problem
does not survive when quantum effects are incorporated via coupling to
the external transverse magnetic field. We have studied the free energy as a
function of connectivity $\gamma$ for different values of transverse field $\Gamma$.
The case $\Gamma=0$ corresponds to the purely classical limit,
and there exists a phase transition when $\gamma$ is crossing its
critical value. We have demonstrated that for any small value of $\Gamma$
the free energy becomes analytic and the sharp phase transition at $\Gamma=0$
is replaced by a smooth crossover transition. This stands in contrast to
classical \K-Satisfiability  model at finite temperatures, where we have
found a first-order phase transition line on the temperature-connectivity
plane approaching continuously a zero temperature limit.
However, it is not inconceivable that the seeming difference between the classical
and quantum case is an artifact of the replica-symmetric approximation.
The RSB analysis of dilute antiferromagnetic Potts glass at finite temperature indicates
that analogous zero-temperature static transition becomes a smooth crossover at finite
temperature \cite{Krzakala:2008}. Whether the inclusion of RSB in $T>0$ classical \K-SAT
will similarly lead to the smoothing of the static transition is open to investigation.

We believe the above-mentioned phenomenon is not universal among dilute long range
spin glasses. We expect that in models with Ising or \p-spin interactions
the phase transition at $\gamma = \gamma_c$ is not affected by transverse fields
below a certain threshold ($\Gamma<\Gamma_c$) and that the phase boundary has the
form of line labeled (1) in Fig.~\ref{fig:quantum}. We attribute the difference to
the fact that constraints involved in Viana-Bray and dilute \p-spin models are
``stronger'' (i.e. a greater number of spin combinations are penalized) than
those in \K-Satisfiability.

The limitations of our approach are the assumptions of the replica symmetry and the failure to include
the time-dependence of correlation functions. The latter approximation has been justified on the ground
that we work in the limit of small transverse fields, as we establish the absence of the phase transition
line on ($\gamma,\Gamma$) plane. It is known that the replica-symmetric approximation can capture the
existence of the thermodynamic transition in the classical \K-Satisfiability model while providing an
overestimated value for the transition point (critical connectivity).

Previously, we attempted to analyze the $O(\Gamma)$ corrections to the free energy of the $\Gamma=0$
(classical) \K-SAT problem along similar lines \cite{Knysh:2006}. In contrast to the present analysis, we
computed corrections to the integer-delta-peaks solution (\ref{T0intPh}) of zero-temperature classical
\K-SAT developed in Ref. \cite{Monasson:1997}. Although the assumption of integer values of effective
fields is more natural for $T=0$, $\Gamma=0$ and has been used to construct an RSB theory of $T=0$
\K-SAT \cite{Mezard:2002PRE}, at the replica-symmetric level it does not give a truly stable solution for
either $T>0$ or $\Gamma>0$. At the same time, the solution derived in the present paper is globally
stable in the RS sector at finite temperatures and transverse fields.

To obtain a correct location of the phase transition one needs to take into account the spontaneous
breaking of the replica symmetry \cite{Mezard:2002PRE}. However, we hope that our main result --- the
smoothing of the phase transition at finite values of the transverse field --- will be immune to the
effects of the replica symmetry breaking. On the other hand, it is well known that the replica symmetric
approximation fails to account for the dynamic transition and the complex structure of local minima
existing in the classical \K-Satisfiability model at connectivity that is smaller than that of the static
transition. From the perspective of quantum adiabatic algorithm, the likely conclusion is that the
bottleneck of QAA may be in this dynamic transition rather than the static transition.
Recent results on \K-SAT show the presence of another transition: so-called condensation transition
(which coincides with the dynamic transition for $K=3$) \cite{Krzakala:2007,Montanari:2008}. It is believed that
the crossover to the exponential complexity of classical algoritm happens at the condensation transition.
It may also be relevant for the performance of quantum adiabatic family of algorithms.

Since the non-analytic behavior of the free energy associated with the static transition is the isolated
singularity, it should be irrelevant to the complexity of QAA except when $\gamma=\gamma_c$ precisely.
It is conceivable that the complexity of QAA will be subexponential for $\gamma \neq \gamma_c$ if
singularities of the free energy associated with e.g. condensation transition are weaker than that
of static (satisfiability) transition.

The analysis of a  quantum version of \K-Satisfiability problem lead us to the re-examination of the
static phase transition in zero-temperature classical limit. In Ref. \cite{Monasson:1997} it has
been predicted, using a replica-symmetric analysis, that this phase transition is of a random
first-order (discontinuous) type. We have  found that the transition is, in fact, of  second-order
(continuous). In the vicinity of the phase transition, the functional order parameter is given by the new
scale-free solution that we described in Sec.~\ref{sec:ModelO}. This second-order transition is of a
peculiar nature, as it possesses critical exponents typically associated with first-order phase
transitions. Consequently, numerical studies must be performed with care as finite-size effects can make
this phase transition indistinguishable from a first-order transition.

We have found that in the vicinity of the transition the singular
component of the free energy is $F=\frac{t}{\ln t}$ where $t=\gamma-\gamma_c$.
The logarithmic correction is sufficient to make the first derivative
$\dd F/\dd \gamma$ continuous, therefore there is no associated ``latent heat''.
Ordinarily, finite ``latent heat'' must be a continuous function of thermodynamic
variables, which ensures that the phase transition persists for finite
$\Gamma$, at least up to the critical point. Conversely, when it is zero, it is
plausible that the phase transition might disappear for arbitrarily small $\Gamma$
as we claim.

Throughout the paper we have attempted to keep the discussion as general as possible.
All formulae derived using the replica method can be applied to a host of spin glass
models defined on random hypergraphs. In the analysis presented in the paper it will
usually involve the replacement of the cost function $E_{\Jvec}(s_1,\ldots,s_K)$ by
a suitable expression and performing disorder averages $\left\langle \ldots \right\rangle_{\Jvec}$
appropriately. Quantum analogues of dilute \p-spin \cite{Mezard:2003Ricci}, \K-NAE-SAT
\cite{Achlioptas:2001}, the Exact Cover \cite{Knysh:2004,Kalapala:2005,Raymond:2007},
and the Vertex Cover \cite{Weigt:2000} problems can be studied using this method.

We have also devised a new method, of Quasi Monte Carlo variety, for the
numerical determination of the functional order parameter. Since it significantly
outperforms standard Monte Carlo, it can be used to improve the accuracy in
the numerical studies of 1-step replica symmetry breaking, which so far required
significant numerical effort \cite{Mertens:2006}.

For future work it is of interest to investigate the stability region of the replica symmetric solution on
the plane ($\gamma,\Gamma$) in a quantum regime corresponding to finite values of  $\Gamma$. In the
classical case, $\Gamma=0$, the  replica symmetric solution loses the stability at the point of the
dynamic (replica-symmetry breaking) transition $\gamma=\gamma_d$. Beyond this point the energy landscape
is characterized by a proliferation of an exponentially large (in $N$) number of deep local minima in the
energy landscape, which traps classical annealing algorithms. It is of interest to explore how this
picture is modified for finite values of $\Gamma$. The structure of the free energy landscape will have
implications for the scaling of the minimum gap in the Quantum Adiabatic Algorithm.

The effective classical Hamiltonian (\ref{HmiT0}) may be used as a starting point for performing RSB analysis.
Although it reflects static approximation, the disorder dependence is explicit. The replica-symmetric
ansatz that we made corresponds to the assumption the distributions of magnetizations on different sites
are not correlated. In the limit of small $\Gamma$ relevant local minima correspond to integer values of
magnetizations: $m \approx \pm 1$ and $m \approx 0$. Although local minima with intermediate values
of $m$ exist, our analysis indicates that corresponding free energies correspond to excitations
with energies much larger than the typical $O(\Gamma)$. In this limit, continuous magnetizations
may be replaced by discrete variables taking 3 possible values. The third possibility ($m=0$) makes the
problem distinctly different from the classical \K-SAT.

\subsection{Acknowledgments}
We acknowledge the financial support of the United States National Security Agency's Laboratory for
Physical Sciences.
We would also like to thank Lenka Zdeborov\'a from the Laboratoire de Physique Th\'eorique et Modeles
Statistiques, Univ. Paris-Sud for interesting and useful comments.

\appendix

\begin{widetext}

\section{Details of Replica Calculations.\label{app:Replica}}

\subsection{The free energy functional.\label{app:F}}

Substitution of the classical Hamiltonian of random \K-SAT (\ref{Hcl}) into Eq.~(\ref{Znsum}) leads
to the following expression for the disorder-averaged $n$-replica partition function:
\begin{equation}
  \left\langle Z^n \right\rangle = \sum_{[\{s_i^a (t)\}]} \ee^{\sum_{a, i}
  \Kcal[s_i^a (t)]} \left\langle  \prod_{i_1 < \dots < i_K} \exp \biggl(
  - c_{i_1 \dots i_K}  \sum_a \intxlim{0}{\beta}{\dd t} E\left(s_{i_1}^a (t), \dots,
  s_{i_K}^a (t) ; \Jvec_{i_1 \dots i_K}\right) \biggr) \right\rangle .
  \label{Znsum1}
\end{equation}
To perform the disorder average over $c_{i_1 \dots i_K}$, we expand the exponential and exploit the
fact that averages $\left\langle c_{i_1 \dots i_K}^p \right\rangle = K! \gamma / N^{K-1}$ are
independent of $p \geqslant 1$. The resulting expression is
\begin{equation}
  \langle Z^n \rangle = \sum_{[\{s_i^a(t)\}]} \ee^{-N\Hcal\subs{avg}[\{s_i^a(t)\}]
  + \sum_{a,i} \Kcal[s_i^a(t)]},
\end{equation}
where we the term $\exp(-N\Hcal\subs{avg}[\dots])$ results from disorder averaging.
\begin{equation}
  \Hcal\subs{avg}[\{s_i^a(t)\}] = \frac{K!\gamma}{N^K}
  \sum_{i_1<\dots<i_K} \sum_{p=1}^{\infty} \frac{(-1)^{p-1}}{p!}
  \biggl\langle \biggl(
  \sum_{a,t} \intxlim{0}{\beta}{\dd t} E_\Jvec\left(s_{i_1}^a(t),\dots,s_{i_K}^a(t)\right)\biggr)^p
  \biggr\rangle_\Jvec.
\label{Havg}
\end{equation}
Like everywhere else, the disorder dependence of the cost function $E_\Jvec(\{s_\ell\})$ is indicated
with a subscript whenever $\Jvec$ is a dummy variable and does not refer to a specific constraint
(cf. $J_{i_1 \dots i_K}$ in Eq.~(\ref{Znsum1})). In this particular case, an average over $2^K$ possible
realizations of $\Jvec$ is performed, which is indicated by the use of angle brackets
$\langle \dots \rangle_\Jvec$. Similar to \K-SAT, for many other combinatorial problems defined on random
hypergraphs that are described with the present formalism, this notation is used to denote averaging over
a ``non-geometric'' component of the disorder. For instance, for random \K-XOR-SAT problem
\cite{Mezard:2003Ricci}
the non-geometric  component of disorder is a scalar $J$, taking values $\pm 1$ with probability of $1/2$.
However, for random Exact Cover \cite{Farhi:2001} the disorder is purely geometric and the application of
$\langle \dots \rangle_\Jvec$ produces no effect.

As the next step we rewrite Eq.~(\ref{Havg}) as a sum of terms with Ising-like interactions among spins
(i.e. so that terms only involve products of spin variables). We observe that any function of $K$ binary
variables can be written as a polynomial in $s_1 \dots s_K$.
\begin{equation}
  E_\Jvec(s_1, \dots, s_K) = \sum_{\{n_i \} \in \{0, 1\}^K} \Ebar_\Jvec\sups{\{n_i \}}
  s_1^{n_1} \dotsm s_K^{n_K} = \Ebar_\Jvec\sups{00 \dots 0} + \Ebar_\Jvec\sups{10 \dots
  0} s_1 + \Ebar_\Jvec\sups{01 \dots 0} s_2 + \dots + \Ebar_\Jvec\sups{11 \dots 1} s_1
  s_2 \dots s_K . \label{Epoly}
\end{equation}
In particular, for \K-SAT, Eq.~(\ref{Epoly}) holds with $\Ebar_\Jvec\sups{n_1 \dots n_K} =
\frac{1}{2^{K-1}} J_1^{n_1} \dots J_K^{n_K}$. Using this transformation and inverting the orders of
summation and integration we can rewrite Eq.~(\ref{Havg}) as follows:
\begin{equation}
  \Hcal\subs{avg} = K! \gamma \sum_{p=1}^{\infty}
  \frac{(-1)^{p-1}}{p!} \sum_{a_1 \dots a_p} \intx{\dd t_1 \dotsm \dd t_p}
  \sum_{\{n_{r\ell} \} \in \{0,1\}^{p\times K}}
  \biggl\langle \prod_{r=1}^p \Ebar_\Jvec\sups{n_{r1} \dots n_{rK}} \biggr\rangle_\Jvec
  \frac{1}{N^K} \sum_{i_1 < \dots < i_K}
  \prod_{\ell=1}^{K} \prod_{\{r|n_{r\ell}=1\}} s_{i_\ell}^{a_r}(t_r).
\label{Havg2}
\end{equation}
From the form of Eq.~(\ref{Havg2}), it is obvious that the partition function (\ref{Znsum1}) may be
evaluated using mean field theory. The expression for $\Hcal\subs{avg} [\{s_i^a (t)\}]$ can be written
entirely in terms of $Q_{a_1 \dots a_p} (t_1, \dots, t_p)$ defined by Eq.~(\ref{Qt1tp}). We may write
$\Hcal\subs{avg}[\{s_i^a (t)\}] = \gamma n \beta \Vcal[\{Q_{a_1 \dots a_p} (t_1, \dots, t_p)]$, where the
newly introduced ``quasipotential'' is expressed in terms of spin correlation functions:
\begin{equation}
  \Vcal = \frac{1}{n \beta} \sum_{p=1}^{\infty} \frac{1}{p!}
  \sum_{\{n_{r1} \}, \dots, \{n_{rK} \} \in \{0, 1\}^p} \sum_{a_1 \dots
  a_p} \intx{\dd t_1 \dotsm \dd t_p}
  \biggl\langle \prod_{r=1}^p \Ebar_{n_{r1} \dots n_{rK}}
  (\Jvec) \biggr\rangle_\Jvec
  \prod_{\ell=1}^K Q_{\{a_r|n_{r\ell}=1\}} (\{t_r|n_{r\ell}=1\}). \label{VQ}
\end{equation}
The only other term that contributes to the replica free energy, is of an entropic nature, as it appears
from the summation over all spin configurations subject to the constraints (\ref{Qt1tp}) on correlation
functions. This term will be referred to as ``quasientropy'' and denoted $\Scal$. In order to compute the
quasientropy, we introduce auxiliary variables $\lambda_{a_1 \dots a_p} (t_1, \dots, t_p)$. Each
constraint of the form $Q = \frac{1}{N} \sum_i x_i$ is enforced by inserting the integral $\int_{- \ii
\infty}^{+ \ii \infty} \dd \lambda \,\ee^{- N \lambda Q} \,\prod_i \ee^{\lambda x_i}$. We may write
\begin{equation}
  \Scal = \frac{1}{Nn \beta} \ln \intx{\mathcal{D} \lambda} \ee^{- N
  \sum_p \sum_{a_1 \dots a_p} \intx{\dd t_1 \dotsm \dd t_p} \lambda_{a_1 \dots
  a_p} (t_1, \dots, t_p) Q_{a_1 \dots a_p} (t_1, \dots, t_p)}
  \Zcal_1^N [\{\lambda_{a_1 \dots a_p} (t_1, \dots, t_p)\}],
  \label{SintL}
\end{equation}
where $\mathcal{D} \lambda$ denotes a functional integration over the set of $\lambda_{a_1 \dots a_p}
(t_1, \dots, t_p)$, and $\Zcal_1$ is the ``single-site'' partition function
\begin{equation}
  \Zcal_1 [\{\lambda\}] = \sum_{[\{s_a(t)\}]} \exp \biggl( \sum_{p=1}^{\infty}
  \sum_{a_1 \dots a_p} \intx{\dd t_1 \dotsm \dd t_p} \lambda_{a_1
  \dots a_p}(t_1, \dots, t_p) s_{a_1}(t_1) \dots s_{a_p}(t_p) + \sum_a
  \Kcal[s_a(t)] \biggr).
\label{Z1}
\end{equation}
The partition function may be written in the form of the functional integral over both
$Q_{a_1 \dots a_p}(t_1,\dots,t_p)$ and $\lambda_{a_1 \dots a_p}(t_1,\dots,t_p)$ as
seen in Eq.~(\ref{Znint}). This functional integral is in the limit $N \to \infty$ dominated by its
saddle-point value corresponding to the extremum of the free energy functional
\begin{equation}
  \Fcal[\{Q\},\{\lambda\}] = \gamma \Vcal[\{Q\}] - S[\{Q\},\{\lambda\}].
\label{FQl}
\end{equation}
Normalization constants have been chosen so that the real free energy
$F=(-1/N n \beta) \ln \langle Z^n \rangle$ corresponds to the extremal value of (\ref{FQl}).

\subsection{Static Approximation\label{app:Static}}

The saddle-point solution may be obtained by varying the free energy functional $\Fcal[\{Q\},\{\lambda\}]$
with respect to its time dependent arguments; and self-consistency equations for
saddle-point values of $Q_{a_1 \dots a_p}(t_1,\dots,t_p)$ and
$\lambda_{a_1 \dots a_p}(t_1,\dots,t_p)$ may be written down.
Unfortunately, the resulting infinite number of equations cannot be reduced to the closed-form
expression for the free energy. To have a closed-form solution is important since an analytic
continuation to $n \to 0$ must be performed eventually.

We resort to an approximation of variational type, which restricts the functional integration in
(\ref{Znint}) to only time-independent $Q_{a_1 \dots a_p}$ and $\lambda_{a_1 \dots a_p}$. In other words,
the  functional integral is replaced by a regular, albeit infinite-dimensional integral. Under this static
ansatz, the saddle-point condition for the integral over $\lambda_{a_1 \dots a_p}$ becomes
\begin{equation}
  0 = \frac{\partial \Scal}{\partial \lambda_{a_1 \dots a_p}}
  = -\frac{1}{n \beta} Q_{a_1 \dots a_p} + \frac{1}{n \beta}
  \frac{\partial}{\partial \lambda_{a_1 \dots a_p}} \Zcal_1 \left(
  \left\{ \lambda_{a_1 \dots a_p} \right\} \right),
\label{dSdL}
\end{equation}
where the static single-site partition function is rewritten in the following form:
\begin{equation}
  \Zcal_1 \left( \left\{ \lambda_{a_1 \dots a_p} \right\} \right) =
  \sum_{[\{s_a(t)\}]} \exp \biggl( \sum_{p=1}^{\infty} \sum_{a_1 \dots a_p}
  \lambda_{a_1 \dots a_p} \intx{\dd t_1 \dotsm \dd t_p} s_{a_1}(t_1) \dotsm
  s_{a_p}(t_p) + \sum_a \Kcal[s_a (t)] \biggr).
\label{Z1static}
\end{equation}
Stationarity condition (\ref{dSdL}) shows that the saddle-point values
of the static order parameter $Q_{a_1 \dots a_p}$ are the time-averaged correlation functions
\begin{equation}
  Q_{a_1 \dots a_p} = \left\langle \frac{1}{\beta^p} \intx{\dd t_1 \dotsm \dd t_p}
  s_{a_1}(t_1) \dotsm s_{a_p}(t_p) \right\rangle_{\Zcal_1}.
\end{equation}
Here $\langle \dots \rangle_{\Zcal_1}$ represents an average taken with respect to
the Gibbs distribution corresponding to the single-site partition function (\ref{Z1static}):
\begin{equation}
  p[\{s_a(t)\}] = \frac{1}{\Zcal_1}
  \ee^{\sum_p \sum_{a_1 \dots a_p} \lambda_{a_1 \dots a_p} \intx{\dd t_1 \dotsm \dd t_p}
  s_{a_1}(t_1) \dotsm s_{a_p}(t_p)}.
\label{pZ1static}
\end{equation}

Once time-dependence is ignored, all expressions may be rewritten in terms of the mean magnetizations
$m_a = \frac{1}{\beta} \intxlim{0}{\beta}{\dd t} s_a(t)$. In particular, we introduce the marginal
$p(\{m_a\})$ of the Gibbs distribution (\ref{pZ1static})
\begin{equation}
  p(\{m_a\}) = \sum_{[\{s_a (t)\}]} p [\{s_a (t)\}]
  \prod_a \delta \biggl( m_a - \frac{1}{\beta} \intxlim{0}{\beta}{\dd t} s_a(t) \biggr).
\end{equation}
Note that for finite $L$ the delta-function is the Kronecker's delta; the normalization has been
chosen so that $\intx{\{\dd m_a\}} p(\{m_a\})=1$. We replace sum over paths $[s(t)]$ by integrals
over magnetizations. Eq.~(\ref{Z1static}) may be rewritten as follows:
\begin{equation}
  \Zcal_1 = \intx{\{\dd m_a\}} \ee^{\sum_p \beta^p \sum_{a_1 \dots a_p}
  \lambda_{a_1 \dots a_p} m_{a_1} \dotsm m_{a_p} - \beta \sum_a u_0(m_a)},
\end{equation}
where the factor $\ee^{- \beta u_0(m)}$ comes from the summation over paths of average
magnetization $m$
\begin{equation}
  \ee^{- \beta u_0(m)} = \sum_{[s(t)]} \ee^{\Kcal[s(t)]} \delta
  \biggl( m - \frac{1}{\beta} \intxlim{0}{\beta}{\dd t} s_a(t) \biggr).
\end{equation}
For a finite value of $L$ it may be written as a sum over kinks of $s(t)$. In the limit
$L \to \infty$ the series may be expressed in the analytical form given by Eq.~(\ref{ebu0}).

We substitute $Q_{a_1 \dots a_p} = \intx{\{\dd m_a\}} p(\{m_a\}) m_{a_1} \dots m_{a_p}$ into Eq.
(\ref{VQ}) and gather all terms in an infinite sum to form an exponential. The resulting expression
for the quasipotential reads
\begin{equation}
  \Vcal[p (\{m_a \})] = \frac{1}{n \beta} \intx{\{\dd m_a\sups{\ell} \}}
  p(\{m_a\sups{1}\}) \dotsm p(\{m_a\sups{K} \}) \left\langle 1 - \ee^{- \beta
  \sum_a \Ehat_\Jvec\blparen\{m_a\sups{\ell} \}\brparen}
  \right\rangle_\Jvec, \label{Vpma}
\end{equation}
where $\Ehat_\Jvec(m_1,\dots,m_K) = \sum_{\{n_i\} \in \{0,1\}^K}
\Ebar_\Jvec\sups{\{n_i\}} m_1^{n_1} \dotsm m_K^{n_K}$ is the
generalization of $E_\Jvec(s_1,\dots,s_K)$ to continuous magnetizations.

The expression for the quasientropy, expressed in terms of $p(\{m_a\})$, is
\begin{equation}
  \Scal[p(\{m_a \})] = - \intx{\{\dd m_a\}} p(\{m_a\}) \Bigl(
  \ln p(\{m_a\}) + \beta \sum_a u_0(m_a) \Bigr).
\label{entS}
\end{equation}

\subsection{Replica-symmetric ansatz and the functional order
parameter\label{app:OrderParameter}}

Now that the free energy functional corresponding to the static approximation
has been written down, we must perform analytical continuation in $n$ and take
the limit $n \to 0$.  To accomplish that we employ the method developed by R.~Monasson
for systems with binary variables \cite{Monasson:1996,Monasson:1997,Monasson:1998}
and extend it to continuous magnetizations.

Replica symmetric ansatz assumes that the saddle point of the free energy functional,
the distribution $p(\{m_a\})$, is a symmetric function of its arguments. We make the
following ansatz for $p(\{m_a\})$:
\begin{equation}
  p(\{m_a\}) = \intx{[\dd h(m)]} P[h(m)] \prod_a
  \frac{\ee^{- \beta h(m_a)}}{\intx{\dd m} \ee^{-\beta h(m)}}.
\label{pmaPhm}
\end{equation}
Here $P[h(m)]$ is the probability distribution of functions $h(m)$ --- the
``effective fields'' that define Boltzmann distribution of magnetizations.
It is evident that $p(\{m_a\})$ is symmetric and normalized to unity,
provided that $\intx{[\dd h(m)]} P[h(m)] = 1$.

Substituting (\ref{pmaPhm}) into (\ref{Vpma}) we obtain for the quasipotential
\begin{equation}
  \Vcal = \frac{1}{n \beta} \intx{\left[ \{ \dd h_\ell(m)\}_{\ell= 1}^K \right]}
  P[h_1(m)] \dotsm P[h_K(m)] \prod_a \intx{\{\dd m_a\sups{\ell}\}_{\ell=1}^K}
  \frac{\ee^{- \beta \sum_{\ell=1}^K h_\ell\blparen m_a\sups{\ell}\brparen}}
  {\prod_{\ell=1}^K \intx{\dd m} \ee^{- \beta h_\ell(m)} }
  \left\langle 1 - \ee^{- \beta \Ehat_\Jvec\blparen m_a\sups{1},\dots,m_a\sups{K}\brparen}
  \right\rangle_\Jvec .
\end{equation}
Product over $n$ replicas appearing inside the functional integral may be replaced by its argument raised
to $n$-th power. Upon taking the limit $n \to 0$ we use $x^n \approx 1 + n \ln x$ to obtain the
expression given by Eq.~(\ref{VPhm}).

For evaluating the quasientropy (\ref{entS}) we use the following trick described in
Ref.~\cite{Monasson:1998}:
\begin{equation}
  \intx{\{\dd m_a\}} p(\{m_a\}) \ln p(\{m_a\}) = \lim_{r \to 1}
  \frac{\partial}{\partial r} \intx{\{\dd m_a\}} \blparen p(\{m_a\}) \brparen^r,
\end{equation}
where $r$-th power of $p(\{m_a\})$ for a positive integer $r$ can be written
in the form
\begin{equation}
  \blparen p(\{m_a \})\brparen^r = \int{[\dd h_1(m) \dotsm \dd h_r (m)]} P [h_1(m)] \dotsm P [h_r (m)]
  \prod_a \frac{\ee^{- \beta \blparen h_1 (m_a) + \dots + h_r(m_a) \brparen}}
  {\left( \intx{\dd m} \ee^{- \beta h_1 (m)} \right) \dotsm
  \left( \intx{\dd m} \ee^{- \beta h_r (m)} \right)} .
\label{pmar}
\end{equation}
In the limit $n \to 0$ products over $n$ replicas may be replaced by the logarithm so that
the r.h.s of Eq.~(\ref{pmar}) splits into several additive terms. It may be further further
simplified to
\begin{equation}
  \intx{\{\dd m_a\}} \blparen p(\{m_a \})\brparen^r \approx n \intx{[\dd h(m)]}
  \left( P^{[\ast r]}[h(m)] - P[h(m)] \right) \ln \intx{\dd m} \ee^{- \beta h(m)}.
\label{dmpmr}
\end{equation}
We write $P^{[\ast r]}[h(m)]$ to denote $r$-th functional convolution of $P[h(m)]$ with itself:
\begin{equation}
  P^{[\ast r]}[h(m)] = (\underbrace{P \ast P \ast \dots \ast P}_\text{$r$ times})[h(m)],
\end{equation}
where the convolution of two functionals $A[h(m)$ and $B[h(m)]$ is defined as follows:
\begin{equation}
  (A \ast B)[h(m)] = \intx{[\dd u(m)]} A[u(m)] B[h(m)-u(m)].
\end{equation}
The convolution $P^{[\ast r]}(m)$ can be written as an inverse Fourier transform of
$\bigl( \Pbar [\omega (m)] \bigr)^r$, where
\begin{equation}
  \Pbar[\omega(m)] = \intx{[\dd h(m)]} \ee^{- \ii \intx{\dd m \omega(m) h(m)}} P[h(m)].
\end{equation}
Once Eq.~(\ref{dmpmr}) is expressed in termos of $\bigl(\Pbar[\omega(m)]\bigr)^r$, it is
straightforward to perform an analytical continuation in $r$.

We now turn to the remaining term in (\ref{entS}):
\begin{equation}
\begin{split}
  \intx{\{\dd m_a\}} p(\{m_a\}) \sum_a u_0 (m_a) &= n \intx{[\dd h(m)]}
  P[h(m)] \frac{\intx{\dd m} \ee^{- \beta h(m)} u_0(m)}{\intx{\dd m} \ee^{- \beta
  h (m)}}  \\
  &= n \intx{[\dd h(m)]} \frac{\intx{\dd m} \ee^{- \beta h(m)} u_0(m)}{\intx{\dd m}
  \ee^{- \beta h (m)}}  \intx{[\dd\omega(m)]} \ee^{- \ii \intx{\dd m}
  \omega (m) h (m)}  \Pbar [\omega (m)].
\end{split}
\label{dmpmu0}
\end{equation}
We substitute
\begin{equation}
 \frac{ \ee^{- \beta h(m)}}{\intx{\dd m'}  \ee^{- \beta h(m')}} = - \frac{1}{\beta}
  \frac{\delta}{\delta h(m)} \ln \intx{\dd m'} \ee^{- \beta h(m')}
\end{equation}
into (\ref{dmpmu0}) and integrate by parts.

Putting together (\ref{dmpmr}) and (\ref{dmpmu0}), expressed entirely in terms of
$\Pbar[\omega(m)]$, we obtain the expression (\ref{SPhm}) for the quasientropy.

\subsection{Classical limit ($\Gamma = 0$) and longitudinal
stability\label{app:Classical}}

In the limit $\Gamma = 0$ the expression (\ref{ebu0}) reduces to
\begin{equation}
  \ee^{- \beta u_0 (m)} = \delta (m-1) + \delta (m+1).
\end{equation}
Since all integrations over magnetizations necessarily involve a factor of
$\ee^{- \beta u_0 (m)}$, they can be replaced by sums over $m = \pm 1$.
Specifically, we will write
\begin{equation}
  \ee^{- \beta h(m)} = \ee^{- \beta h_{+1}} \delta (m-1) + \ee^{-\beta
  h_{-1}} \delta (m+1). \label{ebhmT0}
\end{equation}
The free energy functional $\Fcal \llb P [h(m)] \rrb$ can be expressed in terms of a joint
probability distribution of $h_{+ 1}$ and $h_{- 1}$. It is convenient to make a change of variables
$h_{\pm 1} = g \mp h$. Expressed in terms of the probability distribution $P_2(h,g)$, the free energy
functional has the form
\begin{multline}
  \Fcal = \frac{\gamma}{\beta} \intx{\{\dd h_\ell \dd g_\ell \}}
  \prod_{\ell=1}^K P_2 (h_\ell, g_\ell) \times
  \biggl( \sum_\ell \ln \left( \ee^{- \beta (g_\ell -
  h_\ell)} + \ee^{- \beta (g_\ell + h_\ell)} \right) - \ln
  \sum_{\{s_\ell \}} \ee^{- \beta \Ehat_\Jvec(s_1,\dots,s_K)
  - \beta \sum_\ell g_\ell + \beta \sum_\ell h_\ell s_\ell} \biggr) \\
  -{} \frac{1}{\beta} \intx{\dd h \dd g} \ln \left( \ee^{- \beta (g - h)} +
  \ee^{- \beta (g + h)} \right) \intx{\frac{\dd \omega \dd \nu}{(2 \pp)^2}}
  \ee^{\ii \omega h + \ii \nu g}  \Pbar_2 (\omega, \nu) \left( 1 - \ln
  \Pbar_2 (\omega, \nu) \right),
\end{multline}
where $\Pbar_2(\omega,\nu)$ is the Fourier transform of $P_2(h,g)$. After some simplifications
it becomes
\begin{equation}
\begin{split}
  \Fcal =& \gamma \intx{\{ \dd h_\ell \}} \prod_{\ell=1}^K P
  (h_\ell) \times \biggl( \sum_\ell \ln (2 \cosh \beta h_\ell) - \ln
  \sum_{\{s_\ell \}} \ee^{- \beta E_\Jvec(s_1,\dots,s_K) +
  \beta \sum_\ell h_\ell s_\ell} \biggr)\\
  & -{} \frac{1}{\beta} \intx{\dd h} |h| \intx{\frac{\dd \omega}{2 \pp}}
  \ee^{\ii \omega h}  \Pbar (\omega) \left( 1 - \ln \Pbar (\omega)
  \right) + \Delta \Fcal[ \Rbar (\nu)],
\end{split}
\label{FPhRg}
\end{equation}
where $P (h) = \intx{\dd g} P_2 (h, g)$, the corresponding Fourier transform is $\Pbar (\omega)$,
and the last term is a functional of $\Rbar (\nu) \equiv \Pbar_2 (0, \nu)$:
\begin{equation}
  \Delta \Fcal[ \Rbar (\nu)] = \frac{1}{\beta} \intx{\dd g} g \intx{\frac{\dd \nu}{2 \pp}}
  \ee^{\ii \nu g}  \Rbar (\nu) \left( 1 - \ln
  \Rbar (\nu) \right).
\end{equation}
This contribution is zero, provided that the inverse Fourier transform of $\ln \Rbar (\nu)$ is
non-negative everywhere except the origin. This will happen for the saddle-point of $\Fcal$;
incidentally, it ensures that $R(g) = \intx{\dd h} P_2(h, g) \geqslant 0$, as should be expected
for a probability distribution. Without the last term, Eq.~(\ref{FPhRg}) coincides with the free energy
functional obtained in Ref.~\cite{Monasson:1997}.

\subsubsection{Longitudinal stability\label{app:Stability}}

We can express the condition for the longitudinal stability (that is, the stability \emph{within}
the replica-symmetric sector) of $P(h)$ that makes (\ref{FPhRg}) stationary as the requirement that
eigenvalues of the Hessian matrix $\partial^2 \Fcal/ \partial Q_{a_1 \dots a_p} \partial Q_{b_1
\dots b_q}$ associated with replica-symmetric eigenvectors be positive. In the classical limit ($\Gamma =
0$) the number of independent order parameters $\{Q_{\{k_r \}} \}$ is greatly reduced because $Q_{\{k_r
\}} = Q_p$ for $p = \sum_r k_{2 r+1}$. As a consequence, the free energy may be expressed entirely in
terms of the parameters $Q_{k_1 00 \dots}$, which means that we may assume that $a_1, \dots, a_p$ are
all different (as well as $b_1, \dots, b_q$).

The condition that the Hessian is semi-positive definite can be written as a condition that for an
arbitrary $\{u_p \}_{p=1}^{\infty}$
\begin{equation}
  \frac{\sum_{p, q} u_p u_q \sum_{a_1 \dots a_p} \sum_{b_1 \dots b_q}
  \partial^2 (n\Fcal) / \partial Q_{a_1 \dots a_p} \partial Q_{b_1
  \dots b_q}}{\sum_p u_p^2  \sum_{a_1 \dots a_p} 1} \geqslant 0.
  \label{Hessian}
\end{equation}
Note that retaining the denominator in (\ref{Hessian}) is mandatory. This is peculiar to the
replica theory, as in the limit $n \to 0$ the denominator may become negative. We can choose
$\{u_p\}$ in the following general form:
\begin{equation}
  u_p = \intx{\dd h} \frac{\delta Q_p}{\delta P(h)} u(h).
\end{equation}
Substituting this expression along with the degeneracy factor $\sum_{a_1
\dots a_p} 1 = \binom{n}{p}$ into (\ref{Hessian}), we obtain
\begin{equation}
  \intx{\dd h \dd h'} u(h) u(h') \frac{\delta^2 \Fcal}{\delta P(h)
  \delta P(h')} \bigg/ \frac{1}{n} \sum_p \binom{n}{p} u_p^2 \geqslant 0.
\end{equation}
The denominator includes only terms with even $p$, provided that $u(-h)=u(h)$,
and, conversely, only terms with odd $p$ when $u(-h)=-u(h)$. In the
limit $n \to 0$ the binomial $\binom{n}{p}\approx(-1)^{p-1}n/p$,
i.e. the denominator is negative for even $u(h)$ and positive for odd $u(h)$.

Therefore, the stability condition is that the free energy functional $\Fcal[P(h)]$ is a local
\emph{maximum} with respect to even perturbations and a local \emph{minimum} with respect
to odd perturbations. In the cases where  the symmetry of the problem at hand permits one
to restrict the space of $P(h)$ to even functions, as \K-SAT does, one can safely maximize
the free energy within this restricted subspace.

In a quantum case, the stability condition may not be expressed in such a simple form.
Fortunately, for \K-SAT in a quantum limit, the stationarity condition determines
saddle point uniquely, obviating the need for analysis of the longitudinal stability.

\section{Cavity method and the Bethe-Peierls approximation\label{app:BP}}

The drawback of the method of replicas is the lack of intuitiveness. However, all results obtained using
replica theory can alternatively be derived using the so-called \emph{cavity method} that does not rely on
analytic continuation in $n$. In fact, many physical properties are more easily derived using the cavity
method. It was recently used by M.~M\'ezard and R.~Zecchina to obtain the 1-step RSB solution to classical
zero-temperature random \K-SAT \cite{Mezard:2002Sci,Mezard:2002PRE}. An excellent description of the
method is given in Ref.~\cite{Mezard:2003Parisi}. In contrast to the replica method, cavity equations
\footnote{Also known as the Thouless-Anderson-Palmer (TAP) equations. See Ref. \cite{Mezard}} are written
for a specific realization of disorder; the disorder averaging is performed as a last step. At the level
of replica symmetry, cavity equations are identical to belief propagation (BP) equations as was
demonstrated in \cite{Mezard:2002PRE}.

We generalize BP equations to include quantum degrees of freedom. Although we are limited to static
approximation, this excercise allows us to establish the physical meaning of the former. By examining the
form of the self-consistency equations (\ref{QumPhm}) for the order parameter, we conjectured a particular
form of the effective Hamiltonian given by Eq.~(\ref{Hmi}). The static ansatz is prominent in that the
effect of individual spins in the effective Hamitonian is only via their magnetizations, viz.
imaginary-time averages of the spin. We corroborate the conjecture by demonstrating that we rediscover the
results obtained in Sec.~\ref{sec:Replica} using replica-symmetric static ansatz.

We introduce the Gibbs distribution $p(m_1,\dots,m_N)$ associated with partition function (\ref{ZJmi})
\begin{equation}
  p(\{m_i\}) = \frac{1}{Z(\{\Jvec\})}  \ee^{- \beta \Hcal\subs{eff}(\{m_i\};\{\Jvec\})}.
\label{pmi}
\end{equation}
The probability distribution of magnetization of a particular spin $p_i(m)$ may be written as a
marginal of $p(\{m_i\})$, where all magnetizations other than $m_i$ have been integrated over:
\begin{equation}
  p_i(m) = \intx{\{\dd m_{i'}\}_{i' \neq i}} p(\{m_{i'}|m_i=m\}).
\end{equation}
We write $p(\{m_{i'}|m_i=m\})$ to indicate that the $i$-th argument of
$p(m_1,\dots,m_N)$ should be set to $m$.

We work with the ensemble of random hypergraphs, where each hyperedge may appear with probability $K!
\gamma / N^{K-1}$. Randomly chosen vertex $i$ will have on average $K \gamma$ hyperedges incident to it.
The degree $d_i$ of the vertex is Poisson-distributed with mean $K\gamma$ (see Eq.~(\ref{fd})). We
temporarily relabel vertices as follows (see Fig.~\ref{fig:cavity}): The randomly chosen vertex shall be
referred to as the cavity vertex and shall be labeled by number $0$. Its immediate neighbors shall be
labeled by two integers: $k=1,2,\dots,d \equiv d_i$, to refer to one of the incident hyperedge, and
$\ell=2,\dots,K$, to label other vertices connected to the cavity vertex via this hyperedge. The
particular order of variables for each hyperedge or the ordering of hyperedges is unimportant due to
symmetry; however, we must remember  to reorder components of $\Jvec_{i_1 \dots i_K}$ appropriately. We
shall write $\Jvec_k$ to denote disorder variables associated with $k$-th ($k=1,\dots,d$) hyperedge; the
reordering of components of $\Jvec_k$ is such that the combination of variables with $s_0 = J\sups{1}_k$
and  $s_{k\ell} = J_k\sups{\ell}$ for $\ell = 2,\dots,K$, is penalized.

\begin{figure}[!h]
  \includegraphics{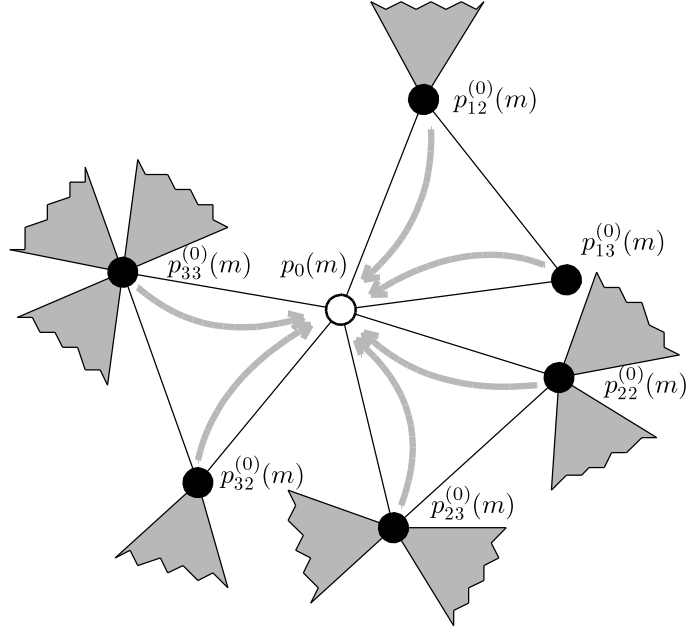}
  \caption{Demonstration of the cavity method. The white circle is a cavity
  spin, and the white triangles are the hyperedges incident to it. The black circles
  are the neighbors of the cavity spin. $p_{k\ell}\sups{0}(m)$ denotes the probability
  distribution of magnetizations of neighbors realized if the cavity vertex
  and hyperedges incident to it are removed. The probability distribution of
  magnetizations of the cavity vertex $p_0(m)$ is completely determined by
  $\{p_{k\ell}\sups{0}(m)\}$ as we indicate with gray arrows.\label{fig:cavity}}
\end{figure}
The following is the description of the Onsager's cavity method. We use $Z\sups{0}$ to denote the
partition function of a system with $N-1$ spins obtained by removal of a cavity spin. The partition
function $Z$ of the original system can be related to $Z\sups{0}$ via
\begin{equation}
  Z = Z\sups{0} \intx{\dd m \{\dd m_{k\ell}\}} \ee^{- \beta u_0(m) - \beta
  \sum_{k=1}^d \Ehat(m, m_{k2},\dots,m_{kK};\Jvec_k)}.
  \label{Z0}
\end{equation}
Since the cavity vertex has been chosen at random, the structure remaining after the removal
of the vertex and its $d$ incident hyperedges is a random graph of the same average connectivity
$K\gamma$ with $N-1$ vertices. Moreover, locations of the vertices labeled by $k$ and $\ell$ are
random and uncorrelated in the absence of a cavity spin. It is known from the theory of random graphs
that randomly chosen vertices are either unconnected to each other or separated by at least
$O(\log N)$ edges. As a result, distributions $p_{k\ell}\sups{0}(m)$ realized in absence of cavity spin
are uncorrelated for different $k,\ell$,  unless a long-range order is present in the system. The absence
of this long-range order is equivalent to the assumption of replica symmetry.

Using (\ref{Z0}) we can relate the distribution $p_0(m)$ of magnetization of the cavity spin to a set of
$p_{k\ell}\sups{0}(m)$:
\begin{equation}
\begin{split}
  p_0 (m) &= \const \times \intx{\{ \dd m_{k \ell} \}} \ee^{- \beta u_0
  (m) - \beta \sum_{k=1}^d \Ehat (m, m_{k 2}, \dots, m_{k K} ;
  \Jvec_k)} \prod_{k, \ell} p_{k \ell}\sups{0} (m_{k \ell}) \\
  &= \const \times \ee^{- \beta u_0 (m)} \prod_{k=1}^d \intx{\{ \dd m_\ell \}}
  \ee^{- \beta \Ehat (m, m_2, \dots, m_K ; \Jvec_k)}
  \prod_{\ell=2}^K p_{k\ell}\sups{0}(m_\ell).
\end{split}
\label{p0m}
\end{equation}
We perform the change of variables $h_i(m) = \const - \frac{1}{\beta} \ln p_i (m)$ to rewrite
(\ref{p0m}) in the form
\begin{subequations}
\begin{align}
  h_0 (m) &= u_0 (m) + \sum_{k=1}^d u_k(m) + \const, \\
  u_k (m) &= u_{\Jvec_k}\bigl( m; \bigl[ h\sups{0}_{k2}(m),\dots,h\sups{0}_kK(m) \bigr] \bigr),
\end{align}
\end{subequations}
where the function $u_\Jvec\blparen m;[\{h_\ell (m)\}]\brparen$ has been defined in (\ref{uJ}).

To write a closed system of equations we employ the fact that the probability distribution of a disorder
in an \N-spin system factorizes. The number $d$ of hyperdedges incident to cavity vertex and vectors
$\{\Jvec_k\}_{k=1}^d$ are not correlated with the disorder  distribution of the remaining $(N-1)$-spin
system. To each disorder realization there corresponds an effective field $h_0(m)$ associated
with the cavity vertex. Accordingly, we may speak of the probability distribution of effective fields
$P\sups{N}[h(m)]$. Absent long range order, effective fields $h_{k\ell}\sups{0}(m)$ associated with
neighbor vertices are independent and drawn from the distribution $P\sups{N-1}[h(m)]$ for a system with
$N-1$ spins. With the aid of auxiliary distribution $Q^{N-1}[u(m)]$, we can relate $P\sups{N}[h(m)]$
to $P\sups{N-1}[h(m)]$:
\begin{subequations}
\begin{align}
  Q\sups{N-1}[u(m)] & = \intx{\{ \dd h_\ell (m)\}} \prod_{\ell = 2}^K
  P\sups{N-1}[h(m)] \times \langle \delta \blbracket u(m) -
  u_\Jvec \blparen m; [h_2(m),\dots,h_K(m)] \brparen
  \brbracket \rangle_\Jvec ,\\
  P\sups{N} [h (m)] & = \sum_{d = 0}^{\infty} f_d(K\gamma)
  \intx{\{\dd h_k(m)\}} \prod_{k=1}^d P\sups{N-1}[h_k (m)] \times
  \delta \biggl[ h(m)-u_0(m)-\sum_{k=1}^d u_k(m)\biggr].
\end{align}
\end{subequations}
In the asymptotic limit $N \to \infty$ we may replace $P\sups{N}[h(m)]$
and $P\sups{N-1}[h(m)]$ by the same limiting distribution $P[h(m)]$
recovering self-consistency equations (\ref{Qum}), (\ref{Phm2}).

The free energy for the concrete realization of disorder can be obtained using Bethe-Peierls
approximation \cite{Yedidia:2002}. This approximation is exact below the percolation threshold and
should be asymptotically correct everywhere in the replica-symmetric phase. Bethe free energy takes
the form of a functional that depends on the probability distribution $\{p_i(m)\}_{i=1}^N$ as well
as the joint probability distributions $p_{i_1 \dots i_K}(m_1,\dots,m_K)$ associated with hyperedges.

For the remainder of this section we shall write $\sum_{(i_1 \dots i_K)}\dots$ to indicate sum over
all hyperedges, alternatively written as $\sum_{i_1<\dots<i_K}c_{i_1 \dots i_K} \times \dotsm$.
For example, the degree of vertex $i$ can be written as $d_i = \sum_{(i_1 \dots i_K)} \sum_\ell
\delta_{i_\ell, i}$. The Bethe free energy functional associated with static Hamiltonian (\ref{Hmi})
can be written as follows:
\begin{multline}
  \Fcal\subs{Bethe}[\{p_i(m)\},\{p_{i_1 \dots i_K}(m_1,\dots,m_K)\}] = \\
  \sum_{(i_1 \dots i_K)} \intx{\{ \dd m_\ell \}} p_{i_1
  \dots i_K} (\{m_\ell \}) \Bigl( \frac{1}{\beta} \ln p_{i_1 \dots i_K}
  (\{m_\ell \}) + \Ehat (\{m_\ell \}; \Jvec_{i_1 \dots i_K}) \Bigr)
  - \sum_i (d_i-1) \intx{\dd m} p_i(m) \Bigl( \frac{1}{\beta} \ln p_i(m) - u_0(m) \Bigr) .
\label{FBP}
\end{multline}
This free energy is to be minimized, subject to the constraint
\begin{equation}
  p_{i_\ell} (m) = \intx{\{ \dd m_{\ell'} \}_{\ell' \neq \ell}}
  p_{i_1 \dots i_K} (\{m_{\ell'}|m_\ell=m \}), \label{constrh}
\end{equation}
as well as normalization conditions $\intx{\{\dd m_\ell\}} p_{i_1 \dots i_K}(m_1,\dots,m_K)
=\intx{\dd m}p_i(m)=1$. Introducing $K \times M$ Lagrange multipliers $h_{i_1 \dots i_K}\sups{\ell}$
associated with constraint (\ref{constrh}) and extremizing the Bethe free energy, we obtain the following
equations to be solved self-consistently:
\begin{subequations}
\label{BP}
\begin{align}
  u_{i_1 \dots i_K}\sups{\ell}(m) &= \ln \intx{\{ \dd m_{\ell'} \}_{\ell'
  \neq \ell}}  \ee^{- \beta \Ehat (\{m_{\ell'}|m_\ell=m \};
  \Jvec_{i_1 \dots i_K}) - \beta \sum_{\ell' \neq \ell} h_{i_1 \dots
  i_K}\sups{\ell'}(m)} ,  \label{BPu}\\
  H_i (m) &= u_0 (m) + \sum_{(i_1 \dots i_K)} \sum_{\ell =
  1}^K \delta_{i_\ell, i} u_{i_1 \dots i_K}\sups{\ell}(m) + \const,  \label{BPH}\\
  h_{i_1 \dots i_K}\sups{\ell} (m) &= H_{i_\ell} (m) - u_{i_1 \dots
  i_K}\sups{\ell} (m) .  \label{BPh}
\end{align}
\end{subequations}
The right-hand-side (r.h.s.) of Eq.~(\ref{BPH}) is the sum of $u_0 (m)$ and contribution from all $d_i$
hyperedges incident to vertex $i$; r.h.s. of Eq.~(\ref{BPh}) is equivalent to the sum of $u_0 (m)$
and $d_{i_\ell}-1$ hyperedges incident to vertex $i_\ell$ with the exception of the hyperedge $(i_1,
\dots, i_K)$ (see Fig.~\ref{fig:BP} for the illustration).

\begin{figure}[!h]
  \includegraphics{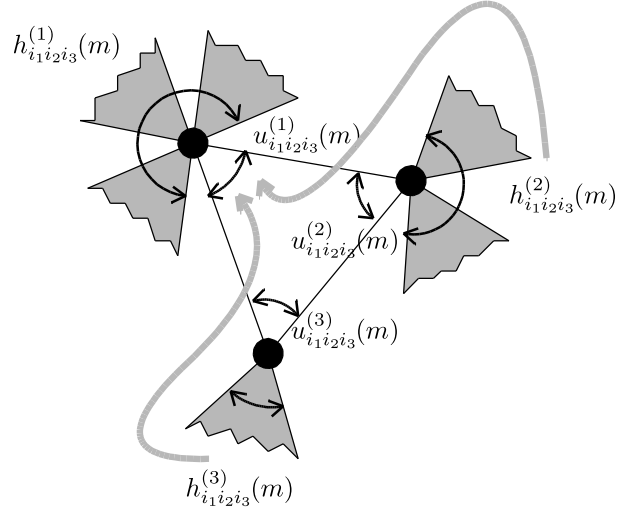}
  \caption{Illustration of the BP equations for $(K=3)$. The white triangle is a hyperedge
  $(i_1 i_2 i_3)$. With this hyperedge we associate 3 functions (beliefs)
  $u_{i_1 i_2 i_3}\sups{1}(m)$, $u_{i_1 i_2 i_3}\sups{2}(m)$, and
  $u_{i_1 i_2 i_3}\sups{3}(m)$. Beliefs $\bigl\{h_{i_1 i_2 i_3}\sups{\ell}\bigr\}$
  correspond to sum over all hyperedges incident to $i_\ell$ other than
  $(i_1 i_2 i_3)$. Grey arrows indicate that $u_{i_1 i_2 i_3}\sups{1} (m)$ is
  determined by $h_{i_1 i_2 i_3}\sups{2} (m)$ and $h_{i_1 i_2 i_3}\sups{3} (m)$.
  Similarly $u_{i_1 i_2 i_3}\sups{2} (m)$ is expressed in terms of $h_{i_1 i_2
  i_3}\sups{1} (m)$ and $h_{i_1 i_2 i_3}\sups{3} (m)$, and so on.\label{fig:BP}}
\end{figure}

These self-consistency equations obtained from Bethe-Peierls free energy functional are known as Belief
Propagation (BP) equations in statistics \cite{Yedidia:2002}, or as
Thouless-Anderson-Palmer (TAP) equations in spin glass theory \cite{Mezard}. The BP algorithm updates
local ``beliefs'' $\{u_{i_1 \dots i_K}\sups{\ell}(m)\}$,
$\{h_{i_1 \dots i_K}\sups{\ell}\}$ until convergence.
These beliefs describe distributions $p_i(m)$, $p_{i_1 \dots i_K}(m_1,\dots,m_K)$:
\begin{subequations}
\begin{align}
  p_i (m) &= \const \times \ee^{- \beta H_i(m)},  \label{p1}\\
  p_{i_1 \dots i_K}(m_1,\dots,m_K) &= \const \times \ee^{- \beta
  \Ehat(m_1,\dots,m_K;\Jvec_{i_1 \dots i_K}) - \beta
  \sum_{\ell=1}^K h_{i_1 \dots i_K}\sups{\ell}(m_\ell)}. \label{pK}
\end{align}
\end{subequations}
In the regime of the stability of the replica symmmetry, solutions
to BP equations (\ref{BP}) should be unique with high probability.

The disorder average of the Bethe free energy can be derived in terms of $P[h(m)]$
and $Q[u(m)]$ --- the histograms of $\{h_{i_1 \dots i_K}\sups{\ell} \}$
and $\{u_{i_1 \dots i_K}\sups{\ell} \}$. That $P [h (m)]$ and $Q [u (m)]$ must
satisfy self-consistency equations (\ref{QumPhm}) follows directly from (\ref{BPu}) and
(\ref{BPh}). The histogram of $H_i(m)$ is also $P[h(m)]$.

We shall require another distribution: the histogram of $\{H_i(m)\}$ but with each vertex
weighted in a proportion to its degree $d_i$. With $P_d[H(m)]$ denoting histograms of
$H_i(m)$ for vertices of a degree $d$, the desired distribution is
\begin{equation}
  P'[H(m)] = \frac{1}{K\gamma} \sum_d d\, f_d(K\gamma) P_d[H(m)].
\label{Pprime}
\end{equation}
Due to the fact that the distribution of degrees is Poisson, Eq.~(\ref{Pprime}) may
be rewritten as follows:
\begin{equation}
  P'[H(m)] = \intx{[\dd u(m)]} Q[u(m)] P[H(m)-u(m)].
\end{equation}
Substituting (\ref{p1}) and (\ref{pK}) into (\ref{FBP}) and performing the disorder average, we obtain
\begin{equation}
\begin{split}
  \left\langle F\subs{Bethe} \right\rangle &= \frac{\gamma}{\beta}
  \intx{[\{\dd h_\ell (m)\}]} \prod_{\ell=1}^K P [h_\ell (m)] \times \left\langle
  \ln \intx{\{ \dd m_\ell \}} \ee^{- \beta \Ehat (\{m_\ell \};
  \Jvec) - \beta \sum_{\ell=1}^K h_\ell (m_\ell)}
  \right\rangle_\Jvec \\
  &+{} \frac{K \gamma}{\beta} \intx{[\dd h(m) \dd u(m)]} P[h(m)] Q[u(m)]
  \ln \intx{\dd m} \ee^{- \beta h (m) - \beta u (m)}
  - \frac{1}{\beta} \intx{[\dd h(m)]} P[h(m)] \ln \intx{\dd m} \ee^{-\beta h(m)}.
\end{split}
\end{equation}
This coincides with the expression obtained using the replica method [see Eqs.~(\ref{VPhm}) and
(\ref{SQum})].

\section{Quasi Monte Carlo Implementation\label{app:QMC}}

The Quasi Monte Carlo (QMC) method of evaluating integrals replaces the random sequences of standard Monte
Carlo (MC) algorithms with  deterministic minimum discrepancy sequences \cite{Niederreiter}. For
example, a two-dimensional integral of a function $f(x, y)$ on $[0;1]^2$ is approximated by
\begin{equation}
  \intxlim{[0;1]^2}{}{\dd x \dd y} f (x, y) \approx \frac{1}{N-1} \sum_{k=1}^{N-1}
  f \Bigl( \frac{i_k}{N}, \frac{j_k}{N} \Bigr), \label{QMC2D}
\end{equation}
where $\{i_k, j_k \}_{k=1}^{N-1}$ is the minimum discrepancy sequence (since it is a finite,
we will call it a minimum discrepancy set). We use Sobol sequences \cite{Press} and choose $N$ to be an
integer power of $2$ for best results. Error estimate for the two-dimensional integral (\ref{QMC2D}) is $O (\log N / N)$
for continuous functions $f (x, y)$ and $O (1/N^{2/3})$ for discontinuous functions $f (x, y)$. This
compares favorably to the expected error of $O(1/N^{1/2})$ in standard Monte Carlo.

We adapt QMC to integrals involving univariate probability distributions. The probability distributions
will be represented internally as a finite-size sample. In contrast to the standard MC, we will ensure
that these samples are as uniform as possible. With each univariate distribution $p (x)$ we associate a
function $X (p)$ defined on the interval $[0 ; 1]$ and satisfying the condition
\begin{equation}
  X \left( \intxlim{-\infty}{x}{\dd x'} p(x') \right) = x. \label{Xp}
\end{equation}
Internally, it will be represented by a set of $\{X_k \}_{k=1}^{N-1}$,
where $X_k = X (k / N)$.

For an arbitrary function $\lambda (x)$, its expectation value may be approximated by
\begin{equation}
  \intx{\dd x} p(x) \lambda(x) = \frac{1}{N-1} \sum \lambda (X_i).
\end{equation}
The flow of the computation shall consist of a sequence of transformations of probability distributions.
The elementary operation is finding the distribution of a variable $z = f (x, y)$, given distributions of
variables $x$ and $y$
\begin{equation}
  p(z) = \intx{\dd x \dd y} p(x) p(y) \delta\blparen z-f(x,y)\brparen. \label{zxy}
\end{equation}
That is, we need to find a uniform sample $\{Z_k \}$ of a distribution $p (z)$ from uniform samples $\{X_k
\}$ and $\{Y_k\}$ of distributions $p (x)$ and $p (y)$. What the appropriate sample should be, can be
assessed indirectly by considering the expectation value of an arbitrary function $\lambda(z)$:
\begin{equation}
\begin{split}
  \frac{1}{N-1} \sum_{k=1}^{N-1} \lambda (Z_k) &= \intx{\dd z} p(z)
  \lambda(z) \\
  &= \intx{\dd x \dd y} p(x) p(y) \lambda\blparen f(x, y)\brparen \\
  &= \intxlim{[0 ; 1]^2}{}{\dd p_1 \dd p_2} \lambda\bigl( f\blparen X(p_1), Y(p_2)\brparen \bigr).
  \label{expL}
\end{split}
\end{equation}
Estimating the integral over $[0;1]^2$ in (\ref{expL}) and using (\ref{QMC2D}), we may write
\begin{equation}
  \frac{1}{N-1} \sum_k \lambda (Z_k) = \frac{1}{N-1} \sum_k \lambda \blparen f(X_{i_k}, Y_{j_k}) \brparen,
\label{ZXY}
\end{equation}
where $\{i_k,j_k\}_{k=1}^{N-1}$ is the Sobol set. The choice of the sample $\{Z_k \}$ satisfying
(\ref{ZXY}) for any $\lambda (z)$ is unique. Algorithmically, it is computed as follows:
\begin{enumerate}
  \item For $k=1, \dots, N-1$ evaluate $Z_k = f (X_{i_k}, Y_{j_k})$,
  where $\{i_k,j_k\}_{k=1}^{N-1}$ is the Sobol set.
  \item Sort the resulting vector $\{Z_k\}_{k=1}^{N-1}$ in increasing order.
\end{enumerate}
The last step is to ensure that $Z_k<Z_{k+1}$, which is required by
definition (\ref{Xp}).

\subsection{Application to $T=0$ classical 3-SAT}

Let us briefly describe how this idea can be applied to solving
self-consistency equations. For $K = 3$, Eq.~(\ref{SelfQ}) already has the
form of (\ref{zxy}),
\begin{equation}
  Q (u) = \intx{\dd h_2  \dd h_3} P (h_2) P (h_3) \delta \bigl( u + J_1 \min
  \blparen 1, ( J_2 h_2 )_+, ( J_3 h_3 )_+ \brparen \bigr),
\end{equation}
enabling one to compute $\{u_k \}$ from $\{h_k \}$. Eq.~(\ref{SelfP}) can be
rewritten as follows:
\begin{subequations}
\begin{align}
  P (h) &= \sum_{d = 0}^{\infty} f_d(3\gamma)
  P_d (h),  \label{PhPdh}\\
  P_d (h) &= \intx{\{ \dd u_k \}} \prod_{k=1}^d Q (u_k) \times \delta
  \biggl( h - \sum_k u_k \biggr) .
\end{align}
\end{subequations}
A set of $\{P_d(h)\}$ is computed using the following recursive definition
having the desired form of (\ref{zxy}):
\begin{equation}
  P_d (h) = \intx{\dd h'  \dd u} P_{d-1} (h') Q (u) \delta (h - h' - u),
\end{equation}
together with the condition $P_0(h) = \delta(h)$. The distribution $P_0(h)$ is represented by
a vector of $N-1$ zeros. Computing a sample of $P (h)$ from a set of samples of $P_d(h)$
via (\ref{PhPdh}) means that we have to select $N-1$ values from the larger set of
$(d\subs{max}+1) \times (N-1)$ values that represent distributions $P_0(h)$ through $P_{d\subs{max}}(h)$.

The elegant way to accomplish it is the following. We formally introduce the
function $H(t,p)$ defined on $[0;1]^2$. For any fixed value of $t$, viewed
as a function of one argument $p$, $H(t,p)$ represents the distribution
$P_{d(t)}(h)$:
\begin{equation}
  H \biggl( \intxlim{-\infty}{h}{\dd h'} P_{d(t)}(h') \biggr) = p,
\end{equation}
and $d(t)$ is a step-wise function of $t$ such that
\begin{equation}
  \sum_{k=0}^{d(t)} f_k(3\gamma) \leqslant t <
  \sum_{k=0}^{d(t)+1} f_k(3\gamma).
\end{equation}
The expectation value of an arbitrary function $\lambda (h)$ can be written as
\begin{equation}
  \intx{\dd h} P(h) \lambda(h) = \intxlim{[0;1]^2}{}{\dd t \dd p} \lambda\blparen H(t,p)\brparen.
\end{equation}
Applying (\ref{QMC2D}) to the integral, we construct the sample for $P(h)$
from $\bigl\{ h^{\blparen d(i_k) \brparen}_{j_k} \bigr\}$, where $\bigl\{h^{(d)}_k\bigr\}$
represents the sample for $P_d(h)$.

Memory requirements for each iteration step can be kept at $O(N)$, whereas the time
complexity is $O(N \log^2 N)$, which is the product of $d\subs{max}=O(\log N)$ and
the $O(N \log N)$ complexity of sorting.

The procedure described above is trivially extended to $K \geqslant 4$ and to
finite temperatures $T > 0$. The extension to finite temperatures merely changes the
form of $u_\Jvec(h_2,\dots,h_K)$. Distribution $Q(u)$ may be computed using either
a single $(K-1)$-dimensional integral, or as a sequence of $K-2$ two-dimensional integrals.
The latter approach is possible because for any $K \geqslant 4$, the function
$u_\Jvec(h_2,\dots,h_K)$ may be written in terms of compositions of functions
of two variables.

Instead of iterating self-consistency equations for a fixed value of $\gamma$, we achieve accelerated
convergence by specifying the desired width $\Delta$ of the distribution $P(h)$ and adjusting the value of
$\gamma$ at each iteration step to satisfy this constraint. As a result, we observe exponentially fast
convergence and avoid the effects of the critical slowing down in the vicinity of the phase transition.

\subsection{Application to quantum $K$-SAT}

The case of quantum \K-SAT in the limit $\Gamma \ll 1$ is slightly more involved. The order parameter is
the joint probability distribution (j.p.d.) $P (h, \hBar)$. For Quantum Model~B, it
is possible to parameterize this j.p.d. by a univariate distribution $P_+ (h)$ and apply
the method described previously. For Quantum Model~AB, however, no such parametrization is possible.

The workaround is to work with univariate distributions $Q(u)$ and $R(\eta)$ exclusively. It is
straightforward to compute $Q(u)$ from $R(\eta)$ by evaluating the $(K-1)$-dimensional integral
(\ref{Quu}) that can be further reduced to the sequence of two-dimensional integrals for $K \geqslant 4$.
The non-trivial part is the computation $R(\eta)$ from $Q(u)$, which we describe below.

It is easy to see that $Q(u)$ is symmetric [i.e. $Q(u)=Q(-u)$] and that $|u| \leqslant 1$.
Let $\xi_{1/2}$ denote the probability that $|u|>1/2$:
\begin{equation}
  \xi_{1/2} = \intxlim{|u|>1/2}{}{\dd u} Q(u).
\end{equation}
We introduce the reduced distributions $\Qhat_<(u)$ and $\Qhat_>(u)$ defined on intervals
$[0;1/2]$ and $[-1;-1/2]$ respectively:
\begin{subequations}
\begin{align}
  \Qhat_<(u) &= \frac{2}{1-\xi_{1/2}} Q(u) \theta(u) \theta \Bigl(\frac{1}{2}-u\Bigr), \\
  \Qhat_>(u) &= \frac{2}{\xi_{1/2}} Q(u) \theta(-u) \theta \Bigl(-\frac{1}{2}-u\Bigr),
\end{align}
\end{subequations}
where $\theta(x)$ is the Heaviside function
$\theta(x)=\begin{cases} 1& \text{for $x>0$}, \\ 0& \text{for $x \leqslant 0$}. \end{cases}$.
The factors $2/(1-\xi_{1/2})$, $2/\xi_{1/2}$ ensure that $\Qhat_<(u)$, $\Qhat_>(u)$ are normalized to unity.

We also define
\begin{equation}
  \Phat_0 (h) = \sum_{d = 0}^{\infty} f_d\Bigl( \frac{K \gamma}{2} (1-\xi_{1/2}) \Bigr)
  \intx{\{\dd u_k\}} \prod_{k=1}^d \Qhat_<(u_k) \times \delta \biggl( h-\sum_k u_k \biggr),
\end{equation}
as well as the sequence $\{ \Phat_k (h)\}$: the sequence of successive convolutions of $\Phat_0(h)$
with $\Qhat_>(u)$. It is computed via the recurrence relation
\begin{equation}
  \Phat_k (h) = \intx{\dd u} \Qhat_>(u) \Phat_{k-1} (h - u).
\end{equation}
The distribution $\Phat_k(h)$ gives the contribution from vertices that have arbitrary number
of incident hyperedges with $u \in [0;1/2]$ and precisely $k$ hyperedges with $u \in [-1;-1/2]$.
In view of a relation (\ref{ubar}) between $\uBar$ and $u$, to each $h$ in the distribution $\Phat_k(h)$
there corresponds $\hBar = k + h$.

Including contributions from mirror image regions $u \in [-1/2;0]$ and $u \in [1/2;1]$,
the distribution $P(h,\hBar)$ may be written in terms of $\{ \Phat_k (h)\}$:
\begin{equation}
  P(h,\hBar) = \sum_{k_+, k_- \geqslant 0}
  f_{k_+}\Bigl( \frac{K\gamma}{2} \xi_{1/2} \Bigr)
  f_{k_-}\Bigl( \frac{K\gamma}{2} \xi_{1/2} \Bigr)
  \intx{\dd h_+  \dd h_-} \Phat_{k_+}(h_+) \Phat_{k_-}(h_-) \delta(h - h_+ + h_-)
  \delta(\hBar - k_+ - k_- - h_+ - h_-),
\end{equation}
where $f_k(\alpha)$ denotes the Poisson distribution with mean $\alpha$ as usual.
It follows that the distribution $R(\eta)$ given by Eq.~(\ref{Reta}) may be written in the following
general form:
\begin{equation}
  R (\eta) = \intx{\dd t_+  \dd t_-  \dd h_+  \dd h_-} \Phat_{k (t_+)} (h_+)
  \Phat_{k (t_-)} (h_-) \delta \left( \eta - f_{k (t_+), k (t_-)} (h_+, h_-) \right),
\end{equation}
where $f_{k_+, k_-} (h_+, h_-) = \eta(h_+ - h_-, \Gamma - k_+ - k_- - h_+ - h_- )$
and we have defined a step-wise function $k(t)$ chosen to satisfy
\begin{equation}
  \sum_{r=0}^{k(t)} f_\ell \Bigl( \frac{K\gamma}{2} \xi_{1/2} \Bigr)
  \leqslant t <
  \sum_{r=0}^{k(t)+1} f_\ell \Bigl( \frac{K\gamma}{2} \xi_{1/2} \Bigr).
\end{equation}
Once $R(\eta)$ has been expressed as a 4-dimensional integral, values $\{\eta_k\}_{k=1}^{N-1}$ may be
sampled using Sobol sets. Memory and time requirements of this procedure remain $O(N)$ and $O(N \log^2 N)$,
respectively.

\end{widetext}

\end{document}